\documentclass{article}
\usepackage[utf8]{inputenc}

\title{EOS Manual}
\author{Li Zeng, Stein B. Jacobsen, Dimitar D. Sasselov, and Michail I. Petaev}
\date{\today}

\usepackage{commath}
\usepackage{graphicx}
\usepackage{amsmath}
\usepackage{amssymb}
\usepackage{amsfonts}
\usepackage{epsfig}
\usepackage[numbers]{natbib}
\usepackage{mathtools}
\usepackage{multirow}
\usepackage{rotating}
\usepackage{longtable}
\usepackage{verbatim}
\usepackage{url}
\usepackage{hyperref}
\usepackage{caption}
\usepackage{subcaption}
\usepackage{CJK}
\usepackage{extpfeil}
\usepackage{dirtytalk}
\usepackage{csquotes}
\usepackage{bm}
\usepackage{epigraph}

\usepackage{empheq}
\usepackage{eqparbox}
\newcommand\ddfrac[2]{\frac{\displaystyle #1}{\displaystyle #2}}
\newcommand{\overbar}[1]{\mkern 1.5mu\overline{\mkern-1.5mu#1\mkern-1.5mu}\mkern 1.5mu}
\newcommand{\Lagr}{\mathcal{L}}
\newcommand{\Celcius}{$^{\circ}$C}

\newcommand{\D}{\mathrm{d}}
\newcommand{\dhyd}[2]{\frac{\D#1}{\D#2}}
\newcommand{\dnhyd}[3][]{\frac{\D^{#1}#2}{\D #3^{#1}}}

\begin{document}

\maketitle

\section{Introduction}

\setlength\epigraphwidth{1.0\textwidth}
\epigraph{Theories of the known, which are described by different physical ideas, may be equivalent in all their predictions and hence scientifically indistinguishable. However, they are not psychologically identical when trying to move from that base into the unknown. For different views suggest different kinds of modifications which might be made and hence are not equivalent in the hypotheses one generates from them in one's attempt to understand what is not yet understood. }{\textit{R. P. Feynman [1966]}}

There is a write-up that provides the steps to turn an equation-of-state (EOS) into a mass-radius relation of exoplanets. It must be \emph{visualized}~\citep{Needham1997}!

%\subsection{A Parable}
%Here, we draw a similarity between a planet and a cooking pot: 
\begin{comment}
\begin{figure}[h!]
\centering
\includegraphics[scale=0.3]{cookingpots.pdf}
\caption{A planet can be viewed as a cooking pot.  }
\label{fig:cartoon_cooking_pots}
\end{figure}
\end{comment}

\subsection{Three Modes of Compression}

%This section deals with the path of Rankine-Hugoniot curve in the T-$\rho$ phase diagram, useful for predictions of high-pressure shock-wave experiments. 

Compression can be classified by its \emph{timescale} in the following way: 
\begin{itemize}
    \item \textbf{Supersonic}: adiabatic irreversible (Rankine-Hugoniot)
    \item \textbf{Sonic}: adiabatic reversible (isentropic)
    \item \textbf{Static}: isothermal
\end{itemize}

Shock wave, by definition, travels faster than the sonic wave in the matter. Thus, matter is compressed before the sonic signal of compression arrives.

\subsection{Three Modes of Energy Transport}

Energy transport can be classified by its \emph{physical mechanism} into three modes: 
\begin{itemize}
    \item \textbf{Convection}: large-scale flow due to internal heat source and gravitational buoyancy
    \item \textbf{Conduction}: propagation of vibrational (thermal) energy of neighboring atoms
    \item \textbf{Radiation}: absorption, transfer, reflection, and propagation of photons
\end{itemize}

All of them occur on a planet, but in different parts, and in different ways. 
Convection often occurs throughout the bulk interior of a planet. Conduction often occurs within a thin boundary layer. Radiation often occurs on or near the planet surface by interaction with host stellar radiation.

\subsection{Hydrostatic Equilibrium}
A planetary body, to a good approximation, is a spherical body with concentric layering with continuous or discontinuous physical and chemical properties changing along the radius (or equivalently, depth) in the planet interior. Here, we first consider a planetary body with uniform chemical composition. Thus, we leave aside the chemical change for now, and focus on the physical changes occurring in a planet interior. The physical changes include (1) continuous change, such as the continuous increase of pressure, temperature, and density, along an isentrope, and (2) discontinuous or abrupt change, such as a phase transition from fluid to solid. 
We will consider both cases in our calculation as follows. 

The most important physical parameter which plays in the mass-radius relation of any object, is its density ($\rho$). As often cited in high-school mathematics textbook, it was probably Archimedes who first discovered the mathematical relationship between the radius ($r$) of a sphere, and its surface area ($A =4 \pi r^2$), and its volume ($V = \frac{4 \pi}{3} r^3$). Then, volume is turned into mass by multiplying with a physical parameter ($\rho$). A good example is liquid water under room temperature and sea-level pressure (1 bar) has an approximate density of 1 gram per cubic centimeter (g/cc). In contrast, the air under room temperature and sea-level pressure (1 bar) has an approximate density of 0.001 gram per cubic centimeter (g/cc), which is one-thousand times less. 

For self-gravitating object that is larger than a few hundred kilometers, $\rho$ will not stay constant but will increase towards the center. 
Let's focus on $\rho$ in between 0.01 and 10 g/cc. 
Below 0.01 g/cc, Hydrogen can be well approximated as Ideal Gas (diatomic molecules $H_2$): 

\begin{equation}
    P (\text{pressure in bar}) = \frac{\rho (\text{density in g/cc})}{\mu (\text{molecular weight in g/mol})} \cdot 83.14 \cdot (T (\text{Kelvin}))
\end{equation}

That means even for the coldest case of $\sim$100 K, at $\sim$ 100 bar, $H_2$ can still be treated as ideal gas ($\rho \sim$ 0.01 g/cc ).

Above 10 g/cc, Hydrogen can be well approximated as a \emph{degenerate} condensed matter, with the following EOS using density $\rho$ and Temperature T as independent variables: 

\begin{equation}
    P_{\text{total}} = P_0 + P_{th} + P_{el}
\end{equation}
$P_{\text{total}}$ is the total pressure corresponding to a given pair of ($\rho$,T), for a specific composition.
$P_{th}$ is the thermal correction from (lattice) vibration of atoms.
$P_{el}$ is the electronic contribution if (partially) ionized or (partially) metallic.

In between 0.01 g/cc and 10 g/cc, there is no simple analytic expression of the EOS. This range is where different methods of Quantum Molecular Dynamics (QMD) Simulations play an important role. 
This range of density happens to coincide with the average bulk densities of many planets. 
For convenience of future calculations, we express pressure ($P$) as a function of two independent variables: density ($\rho$) and temperature ($T$). We will later explain the reasons behind this choice of independent variables.

\subsection{Differential Equations}

Now, for planetary interiors, we treat every variable as dependent on $m$ (the mass enclosed by the sphere of radius $r$ inside the planet). One can think of $m$ as a natural coordinate. For example, if a planet cools and contracts over time, then $r$ would decrease but mass $m$ is conserved. 
Therefore, we write density at a particular point in the planet as $\rho[m]$ instead of $\rho[r]$. 
Likewise, we write temperature at a particular point in the planet as $T[m]$ instead of $T[r]$. Even radius $r$ is considered a function of $m$, written as $r[m]$. This choice of dependent and independent variables will help us set up the differential equations. 

Now, let's turn out attention back to EOS for a moment. Since the density typically ranges over three orders of magnitude or more inside a planet (from center to surface, so does temperature, and pressure), it is convenience to the express the EOS in logarithmic relation. That is, to take the logarithmic of every variable ($T$,$\rho$, and $P$), and then express the EOS as the relationship among $lgT$, $lg\rho$, and $lgP$, etc. Take logarithmic on these variables also helps non-dimensionalize them, and facilitate dimensional analysis. 

For example, the ideal gas law turns into a simple linear relation after this manipulation: 

\begin{equation}
    lgP = lg\rho + lgT + \text{const}
\end{equation}

Since for mass-radius relation, the crucial parameter is density $\rho$. We write the hydro-static equilibrium equation and the mass-conservation equation into the following forms, which are explicit in $lg\rho$ but implicit in $lgP$ and $lgT$. 

mass conservation:
\begin{equation}
    \rho[m] \cdot (r[m])^2 \cdot r'[m] = 1.83843
\end{equation}

hydro-static equilibrium: 
\begin{equation}
    (r[m])^4 \cdot \frac{D}{Dm}[10.^{\bf{lgp}[lg\rho[m],lgT[m]]}]  = -115.06 \cdot m 
\end{equation}

The two constants on the righthand side are introduced for the simplicity of unit conversion so that the mass and radius are expressed in Earth units, and pressure is expressed in GPa ($10^9$ Pascal). 

\begin{equation}
    1.83843 == \frac{(M_{\oplus}=5.9742*10^{24} kg)}{{4\pi \cdot 10^3 \cdot (R_{\oplus}=6.371*10^6 m)^3}}
\end{equation}

\begin{equation}
    115.06 == \frac{(G = 6.67428*10^{-11}) \cdot (10^9) \cdot (M_{\oplus}=5.9742*10^{24} kg)^2}{{4\pi \cdot (R_{\oplus}=6.371*10^6 m)^4}}
\end{equation}

In order to solve this set of two equations, we need to supply two more relations, namely, the (1) $\textbf{EOS}$ expressed as the functional dependence of $\bf{lgp}[lg\rho,lgT]$, and (2) the relation between $lg\rho$ and $lgT$. 

The former relation of $\bf{lgp}[lg\rho,lgT]$ is a generic material-specific EOS. 

The latter relation is a one-to-one correspondences between $lg\rho$ and $lgT$. It is based on our assumption of the thermal structure in the planetary interior. For isothermal assumption, $lgT$ is constant, which essentially turns $lgp$ to be a single variable function only dependent on $lg\rho$. For isentropic assumption, the tangential slope of $\frac{dlgT}{dlg\rho}$ is a dimensionless quantity called Gr\"{u}neisen parameter $\gamma$: a material property typically on the order of unity along an isentrope. In most cases, the isentropic assumption is preferred for planet deep interior, for there is often energy source in the planetary interior that drives an interior convection and maintains the temperature profile close to adiabatic (isentropic). However, there are some exceptions, such as the existence of thermal boundaries at the upper and/or lower boundary of each homogeneous convective layer. Within the thermal boundary, the heat transport mechanism switches to conduction which renders a steeper temperature gradient. 

In this manner, the boundary condition for integrating the differential equations is set as (1) prescribing a starting central density $\rho_0$ at the center, where $m=0$ and also $r=0$; (2) prescribing a surface density $\rho_s$, where the outward integration stops. 

This approach avoids iterative solving for any aimed quantities, but will produce a final total mass and radius once the integration is completed. It will simultaneously solve for the density, temperature, and pressure profiles along the radial direction of planet interior, thus providing a full picture of the 1-D planet model. 

Then, by continuously changing the central density $\rho_0$, one generates a set of mass-radius points, from which, one can generate a smooth mass-radius curve corresponding to the specific EOS. 

\subsection{Gr\"{u}neisen Parameter}

Gr\"{u}neisen parameter, denoted as lower-case $\gamma$, is an important dimensionless parameter that relates to both the slopes of isentropes and melting phase-transitions in a T-$\rho$ phase diagram. We reserve the capital-letter $\Gamma$ for a different but related dimensionless quantity upcoming in a later discussion. 

It is defined as the isochoric derivative of Pressure $P$ respect to specific internal energy $u$, multiplied by the specific volume $v \equiv 1/\rho$: 

\begin{equation}
    \gamma \equiv v \cdot \underbrace{\Bigg( \frac{\partial P}{\partial u} \Bigg)_{v}}_{\text{differentiated along an isochor}} = \frac{1}{\rho} \cdot \Bigg( \frac{\partial P}{\partial u} \Bigg)_{\rho}
\end{equation}

This is especially convenient, such as in Mie-Gr\"{u}neisen EOS, if Pressure $P$ is expressed by independent variable $v$ and $u$:

\begin{equation}
    P = P(v,u)
\end{equation}

so that the full differential of $P$ can be expressed as: 

\begin{align}
    dP &= \Bigg(\frac{\partial P}{\partial v} \Bigg)_{u} \cdot dv + \Bigg( \frac{\partial P}{\partial u} \Bigg)_{v} \cdot du \\\nonumber
       &= \Bigg(\frac{\partial P}{\partial v} \Bigg)_{u} \cdot dv + \Bigg( \frac{\gamma}{v} \Bigg) \cdot du \\\nonumber
\end{align}

On the other hand, as we more often encounter, the specific internal energy is expressed in its natural variables $v$, and $s$ (specific entropy, or entropy per unit mass):
\begin{equation}
    u = u(v,s)
\end{equation}

and according to the first law of thermodynamics, the full differential of $u$ equals: 

\begin{equation}\label{Eq:InternalEnergyDifferential}
    du = -P \cdot dv + T \cdot ds
\end{equation}

Through Maxwell relations, it can be shown that $\gamma$ relates to the slope of an isentrope in T-$\rho$ space as: 
\begin{equation}\label{Eq:GruneisenParameter1}
    \gamma = \underbrace{\bigg(\frac{\partial \ln{T}}{\partial \ln{\rho}} \bigg)_s}_{\text{differentiated along a reversible adiabat}} = \frac{\rho}{T} \cdot \bigg(\frac{\partial T}{\partial \rho} \bigg)_s
\end{equation}

It characterizes the fractional temperature increase ($\delta T/T$) along an isentrope (constant $s$-path) due to fractional compression ($\delta \rho/\rho$). 

We define another dimensionless parameter capital-$\Gamma$ as the : 

\begin{equation}
    \Gamma \equiv \underbrace{\Bigg( \frac{\partial \ln{P}}{\partial \ln{\rho}} \Bigg)_{s}}_{\text{differentiated along a reversible adiabat}}
\end{equation}

Their ratio $\gamma/\Gamma$ gives yet another dimensionless parameter: 

\begin{equation}
    \frac{\gamma}{\Gamma} = \underbrace{\Bigg( \frac{\partial \ln{T}}{\partial \ln{P}} \Bigg)_{s}}_{\text{differentiated along a reversible adiabat}} = \bigg(\frac{\partial \ln{T}}{\partial \ln{\rho}} \bigg)_s 
    \Bigg/ 
    \Bigg( \frac{\partial \ln{P}}{\partial \ln{\rho}} \Bigg)_s
\end{equation}

To compare $\gamma$ with $\Gamma$ and $\gamma/\Gamma$ and emphasize their differences and relations:

\begin{itemize}
    \item $\gamma \equiv \bigg( \frac{\partial \ln{T}}{\partial \ln{\rho}} \bigg)_{s}$ is the logarithmic-slope of $T-\rho$ along an isentrope. We call it the \emph{Gr\"{u}neisen parameter} or the \emph{first} adiabatic index. 
    \item $\Gamma \equiv \bigg( \frac{\partial \ln{P}}{\partial \ln{\rho}} \bigg)_{s}$ is the logarithmic-slope of $P-\rho$ along an isentrope. We call it the \emph{second} adiabatic index. 
    \item $\gamma/\Gamma = \bigg( \frac{\partial \ln{T}}{\partial \ln{P}} \bigg)_{s}$ is the logarithmic-slope of $T-P$ along an isentrope. We call it the \emph{third} adiabatic index. 
\end{itemize}

In principle, both $\Gamma$ and $\gamma$ and $\gamma/\Gamma$ vary along an isentrope, but rather slowly. Therefore, in certain approximation, they can be treated as constants, and this shall bring huge advantage in considering qualitative models, such as a polytropic model.

\subsection{Isentrope (Reversible Adiabat)}

Along an isentrope, since $ds = 0$, Eq.~\ref{Eq:InternalEnergyDifferential} simplifies to $du = -P \cdot dv$. Then, 

\begin{equation}
    P = -\underbrace{\bigg(\frac{\partial u}{\partial v} \bigg)_s}_{\text{differentiated along a reversible adiabat}} = -\bigg(\frac{\partial u}{\partial (1/\rho)} \bigg)_s
\end{equation}

where the subscript $s$ means the derivative is taken under constant $s$.

However, as often the case in literature, the functional dependence of specific energy $u$ on specific entropy $s$ is not explicitly given. Rather, specific entropy $s$, or specific internal energy $u$ is given as a function of density $\rho$ and temperature T. Then, isentropic path can be calculated solely by knowing the functional dependence of $P=P(T,\rho)$ and $u=u(T,\rho)$ via: 

\begin{itemize}
    \item \emph{integrative scheme}, see~\cite{Nettelmann2012JUPITERH-REOS.2}, Appendix A: method for calculating entropy. 
        \begin{align}\label{Eq:entropyIntegrative}
            s(T,\rho) &= \frac{u(T,\rho)}{T} - \int_{T_0,\rho_0}^{T,\rho} d\bigg(\frac{f(T',\rho')}{T'} \bigg) +s_0 \\\nonumber
                      &= \frac{u(T,\rho)}{T} - \Bigg(\int_{\rho_0}^{\rho} d\rho'\cdot \frac{1}{\rho'^2} \cdot \frac{P(T_0,\rho')}{T_0} - \int_{T_0}^{T} dT' \cdot \frac{u(T',\rho)}{T'^2} \Bigg) +s_0\\\nonumber
        \end{align}
        where $f=f(T,\rho)$ is the specific Helmholtz free energy. 
    \item \emph{differential scheme}, see~\cite{Becker2013IsentropicInteriors}, Equation 2 therein: 
        \begin{equation}\label{Eq:entropyDifferential}
            \gamma = \bigg(\frac{\partial \ln{T}}{\partial \ln{\rho}} \bigg)_s 
                   =\bigg(\frac{\partial lgT}{\partial lg\rho} \bigg)_s  
                   = \Bigg( \frac{P}{\rho \cdot u} \Bigg) \cdot \frac{\bigg(\frac{\partial lgP}{\partial lgT} \bigg)_{\rho}}{\bigg(\frac{\partial lgu}{\partial lgT} \bigg)_{\rho}} 
                   = \frac{1}{\rho} \cdot \frac{\bigg(\frac{\partial P}{\partial T} \bigg)_{\rho}}{\bigg(\frac{\partial u}{\partial T} \bigg)_{\rho}} 
        \end{equation}
        where the subscript $\rho$ means the derivative is taken under constant density.
        Note that the quantity $\bigg( \frac{P}{\rho \cdot u} \bigg)$ is dimensionless. $\rho$, when multiplies to $u$, converts it from energy per unit mass, into energy per unit volume, which is the same unit as force per unit area, that is, pressure. 
\end{itemize}

Here we concentrate on the \emph{differential scheme}. 
The full differential increment in $P=P(T,\rho)$ or $u=u(T,\rho)$ is separated by the effects of increment in Temperature $T$ or increment in density $\rho=\frac{1}{v}$ as: 

\begin{align}
    dP &= \Bigg(\frac{\partial P}{\partial T} \Bigg)_{\rho} \cdot dT + \Bigg( \frac{\partial P}{\partial \rho} \Bigg)_{T} \cdot d\rho \\\nonumber
    du  &= \Bigg(\frac{\partial u}{\partial T} \Bigg)_{\rho} \cdot dT + \Bigg( \frac{\partial u}{\partial \rho} \Bigg)_{T} \cdot d\rho \\\nonumber
\end{align}

When we fix the specific volume $v$ or density $\rho$, that is moving along an \emph{isochor}, as in the definition of Gr\"{u}neisen Parameter $\gamma$, then, $dv=0$, or equivalently, $d\rho=0$, and we have: 

\begin{align}
    \delta P\Bigr\rvert_{v} &= \delta P\Bigr\rvert_{\rho} = \Bigg(\frac{\partial P}{\partial T} \Bigg)_{v} \cdot dT = \Bigg(\frac{\partial P}{\partial T} \Bigg)_{\rho} \cdot dT \\\nonumber
    \delta u\Bigr\rvert_{v} &= \delta u\Bigr\rvert_{\rho} = \Bigg(\frac{\partial u}{\partial T} \Bigg)_{v} \cdot dT =  \Bigg(\frac{\partial u}{\partial T} \Bigg)_{\rho} \cdot dT  \\\nonumber
\end{align}

Then Gr\"{u}neisen Parameter $\gamma$ can be re-written as: 

\begin{equation}\label{Eq:GruneisenParameter2}
    \gamma = v \cdot  \frac{\bigg(\partial P/\partial T \bigg)_{v}}{\bigg(\partial u/\partial T \bigg)_{v}}  = \frac{1}{\rho} \cdot \frac{\bigg(\partial P/\partial T \bigg)_{\rho}}{\bigg(\partial u/\partial T \bigg)_{\rho}}
\end{equation}

It should be noted that pressure $P$ increases faster along an isentrope compared to an isotherm for the same amount of increase in density $\rho$: 

\begin{equation}\label{Eq:PressureIncrement}
    \bigg( \frac{\partial P}{\partial \rho} \bigg)_{s} = \underbrace{\bigg( \frac{\partial P}{\partial \rho} \bigg)_{T}}_{\text{isothermal part}} + \bigg( \frac{\partial P}{\partial T} \bigg)_{\rho} \cdot \underbrace{\bigg( \frac{\partial T}{\partial \rho} \bigg)_{s}}_{\text{due to $\gamma$}}
\end{equation}

multiply density $\rho$ across both sides of Eq.~\ref{Eq:PressureIncrement}, we have: 

\begin{equation}\label{Eq:PressureIncrement2}
    \underbrace{\bigg( \frac{\partial P}{\partial \ln{\rho}} \bigg)_{s}}_{K_s,\text{~isentropic bulk modulus}} = \underbrace{\bigg( \frac{\partial P}{\partial \ln{\rho}} \bigg)_{T}}_{K_T,\text{~isothermal bulk modulus}} + \rho \cdot \bigg( \frac{\partial P}{\partial T} \bigg)_{\rho} \cdot \underbrace{\bigg( \frac{\partial T}{\partial \rho} \bigg)_{s}}_{\text{due to $\gamma$}}
\end{equation}

Then, plugging Eq.~\ref{Eq:GruneisenParameter1} and Eq.~\ref{Eq:GruneisenParameter2} for Gr\"{u}neisen Parameter $\gamma$, we have:

\begin{equation}\label{Eq:PressureIncrement3}
    K_s = K_T + \underbrace{\frac{T}{\rho} \cdot \frac{\Bigg[\bigg( \frac{\partial P}{\partial T} \bigg)_{\rho} \Bigg]^2}{\bigg( \frac{\partial u}{\partial T} \bigg)_{\rho}}}_{\text{always positive}}
\end{equation}

Therefore, $K_s > K_T$ and it is more difficult to compress matter along an isentrope, compared to an isotherm.

%%%%%%%%%%%%%%%%%%%%%%%%%%%%%%%%%%%%%%%%%%%%%%%%%%%%%%%%%%%%%%%%%%%%%%%%%%%%%
%%%Mantle-Core as coupled oscillator
%%%Reynold number and fractal geometry as dissipation propagating down through many-orders-of-magnitude in dimensions
%%%Planet as a boiling pot

\subsection{Melting}

\emph{Melting}, or fusion, refers to the phase transition from solid to liquid. The same process in the opposite direction from liquid to solid is called \emph{Solidification}: 

\begin{equation}
       \text{Solid} \xtofrom[\text{Solidification}]{\text{Melting}} \text{Liquid}
\end{equation}

\begin{figure}[h!]
\centering
\includegraphics[scale=0.2]{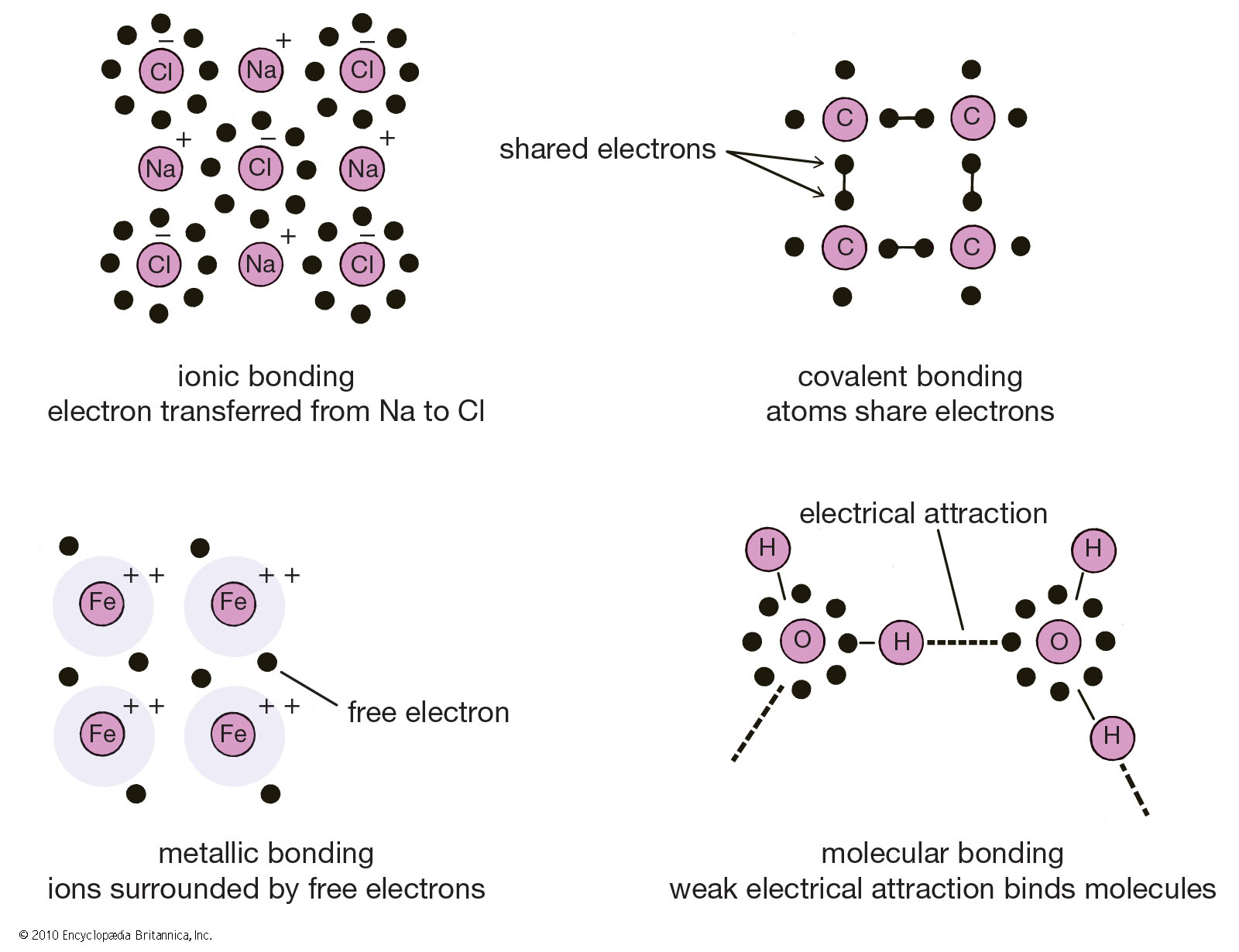}
\caption{Different types of bonding in crystals. When \emph{melting} occurs, some bondings are disrupted or \emph{dislocated}, to allow flow. Copyright 2010 by Encyclopædia Britannica, Inc.~\citep{EncyclopediaBritannica_crystalbonding}. }
\label{fig:types-bonding-crystals}
\end{figure}

On the microscopic scale, \emph{melting} is associated with the breaking of certain bonds in a crystal (Fig.~\ref{fig:types-bonding-crystals}), and is sometimes described by the \emph{dislocation} theory~\citep{Stacey:1977}. 

For a single-component system, \emph{melting} is often described by a \emph{melting curve}, in the $P$-$T$ phase diagram or $\rho$-$T$ phase diagram. 

When melting occurs, not only the Temperature $T$ is fixed at $T_M$, but also the Pressure $P$ is fixed at $P_M$. Therefore, the melting process is simultaneously a \emph{reversible} \emph{isothermal} and \emph{reversible} \emph{isobaric} process. However, there is typically a volume (density) change associated, and heat absorption (or release) involved. 

The differential relation between $P_M$ and $T_M$ if we move along the \emph{melting curve} is expressed by the Clapyron-Clausius relation as: 

\begin{equation}\label{Eq:ClapyronClausius}
   \underbrace{\Bigg( \frac{\partial P_M}{\partial T_M} \Bigg)_{M}}_{\text{along melting}} = \frac{\Delta s}{\Delta v} = \frac{\Delta h}{T_M \cdot \Delta v}
\end{equation}

At high pressure, we are mostly concerned with the \emph{volume-driven} fluid-to-solid phase transition due to compression, not the \emph{entropy-driven} fluid-to-solid phase transition due to cooling. 

With sufficient amount of compression, a fluid transitions into a solid accompanied by a finite (specific) volume change ($\Delta v$). 
Unlike the gas-liquid phase boundary that vanishes above the Critical Point~\textbf{C.P.}, fluid-solid phase boundary exists up to ultra-high pressure and temperature. The (specific) volume change ($\Delta v$) means discrete density $\rho$ jump along the \emph{melting curve}. Therefore, viewed in the $T$-$\rho$ diagram, the \emph{melting curve} is actually represented by two separate curves, one representing the density of solid $\rho_{\text{solid}}$ along the \emph{melting curve}, and the other representing the density of liquid $\rho_{\text{liquid}}$ along the \emph{melting curve}. The area in between these two curves are forbidden. 

\citep{Stacey:2019} argues that the slope of \emph{melting curve} in the $\rho$-$T$ phase diagram at high-pressure is related to Gr\"{u}neisen Parameter $\gamma$ as: 

\begin{comment}
\begin{equation}\label{Eq:meltingcurveslope1}
    \underbrace{\bigg(\frac{\partial \ln{T}}{\partial \ln{\rho}} \bigg)_{M}}_{\text{along melting}} = \frac{\delta\ln{T_M}}{\underbrace{\delta \ln{\rho_{\text{S,M}}}}_{\text{solid phase along melting}}} \approx 2 \cdot \gamma 
\end{equation}
\end{comment}

\begin{equation}\label{Eq:meltingcurveslope1}
    \underbrace{\bigg(\frac{\partial \ln{T}}{\partial \ln{\rho}} \bigg)_{M}}_{\text{along melting}} \approx 2 \cdot \gamma 
\end{equation}

where the subscript $M$ means the path of differentiation is taken along the \emph{melting curve}, that is, maintaining the phase equilibrium.

In order to understand the melting process in more physical detail, let's introduce the (specific) enthalpy $\Delta h$ of melting (sometimes also known as enthalpy of fusion), that is, the total energy required to melt a unit mass of matter at a given pressure $P_M$:

\begin{equation}\label{Eq:EnthalpyofFusion1}
    \Delta h \equiv h_{\text{liquid}} - h_{\text{solid}} 
\end{equation}

From the perspective of specific internal energy $u$ and specific volume $v$, it can be written as: 
\begin{align}\label{Eq:EnthalpyofFusion2}
    \Delta h &= \bigg( u+P_M \cdot v \bigg)_{\text{liquid}} - \bigg( u+P_M \cdot v \bigg)_{\text{solid}} \\\nonumber
             &= \bigg( u_{\text{liquid}} + u_{\text{solid}} \bigg) + P_M \cdot \bigg( u_{\text{liquid}} - u_{\text{solid}} \bigg) \\\nonumber
             &= \Delta u + P_M \cdot \Delta v \\\nonumber
\end{align}

$\Delta u$ generally contains two parts: the increase in the kinetic energy of atomic motion $\Delta u_K$, and the increase in the potential energy of the atomic interactions among neighbors $\Delta u_P$. 

so that, 

\begin{equation}
    \Delta u = \Delta u_K + \Delta u_P
\end{equation}

\citep{Stacey:2019} attributes  melting to the free proliferation of \emph{dislocations}. Because melting involves no temperature change $T=T_M$, then, $\Delta u_K =0$, and, 

\begin{equation}
    \Delta u = \Delta u_P
\end{equation}

From the perspective of specific entropy $s$, it can be written as: 

\begin{align}\label{Eq:EnthalpyofFusion3}
    \Delta h &= \bigg( T_M \cdot s \bigg)_{\text{liquid}} + \bigg( T_M \cdot s \bigg)_{\text{solid}} \\\nonumber
             &= T_M \cdot \bigg( s_{\text{liquid}} - s_{\text{solid}} \bigg) \\\nonumber
             &= T_M \cdot \Delta s \\\nonumber
\end{align}

\citep{Stacey:2019} argues that at high pressure, both solid and liquid have close-packing atomic arrangements. The difference between them is well-characterized by the \emph{dislocation} theory. So the (specific) entropy increase of melting is: 

\begin{equation}
    \Delta s = \Delta u_P /T_M
\end{equation}

The interpretation of this formula is that the (specific) entropy increase $\Delta s$ at melting is primarily contributed by the increase of spatial degree of freedom. The thermal degree of freedom stays roughly the same. Since for vibrations above a certain frequency (\textbf{Frenkel frequency}), molecules in a liquid behave like those in a solid, and can support shear phonons~\citep{Bolmatov_FrenkelFrequency}. Thus, the heat capacities are similar between the solid and liquid phases along the \emph{melting curve}, provided that the higher frequencies of the thermal vibrations dominate. 

\begin{comment}
$\Delta s$ can also be separated into two parts~\citep{Stacey:2019}: 

\begin{equation}
    \Delta s = \underbrace{}_{\text{owing to dislocations}} + \underbrace{\gamma \cdot c_v \cdot \Delta v}_{owing to expansion}
\end{equation}
\end{comment}

However, for the melting of \textbf{superionic} H$_2$O ice, because its protons are already mobilized and can flow freely like in liquid, $\Delta s$ is less than what is expected for melting of a complete solid phase. Melting there only regards the oxygen atoms in a crystal lattice. Therefore, there is an additional factor of 3 when calculating the slope of \emph{melting curve} of the \textbf{superionic} H$_2$O ice in the $\rho$-$T$ phase diagram at high-pressure: 

\begin{equation}\label{Eq:meltingcurveslope2}
    \underbrace{\bigg(\frac{\partial \ln{T}}{\partial \ln{\rho}} \bigg)_{M}}_{\text{along melting of superionic ice}} \approx 6 \cdot \gamma
\end{equation}

Please note that Eq.~\ref{Eq:meltingcurveslope1} and Eq.~\ref{Eq:meltingcurveslope2} are only approximate, and should be checked against experimental data or DFT-MD (Density Functional Theory-Molecular Dynamics) calculations. 

%\subsection{Vaporization}

\subsection{Ionization}

It is commonly known that under ambient conditions (25\Celcius, 1 bar), the H$_2$O molecules in the liquid dissociates into a small amounts H$^{+}$ and OH$^{-}$ ions in a dynamical equilibrium: 

\begin{equation}
       \text{H$_2$O} \xtofrom[\text{Recombination}]{\text{Dissociation}} \text{H$^{+}$} + \text{OH$^{-}$}
\end{equation}

The degree of ionization, characterized by the mole fractions of ions, increases with both increased pressure and increased temperature~\citep{Goncharov:2005}. 

\begin{figure}[h!]
\centering
\includegraphics[scale=1.0]{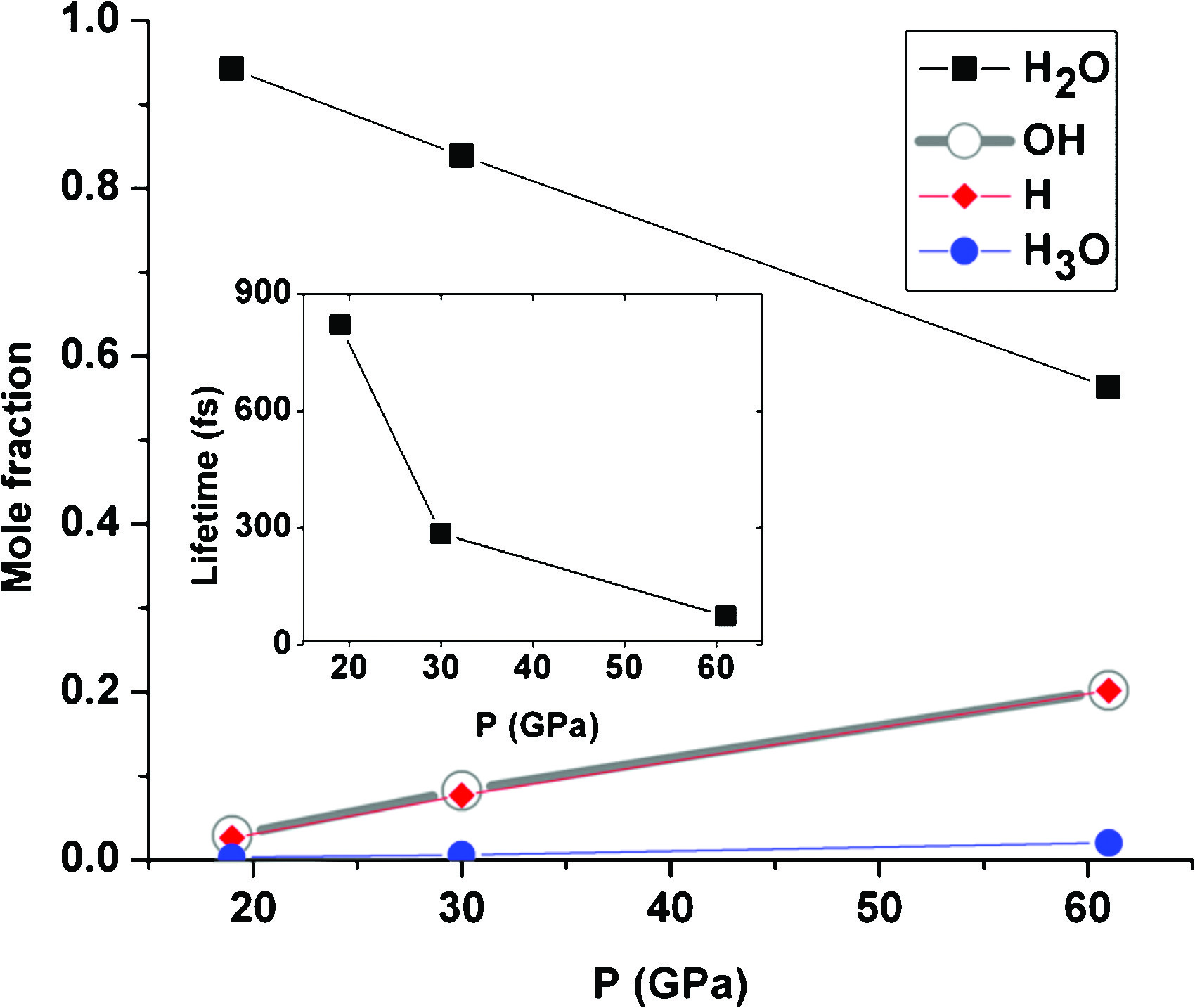}
\caption{The concentrations of chemical species as predicted by the first principles simulations in ~\citep{Goncharov:2009} are shown as a function of pressure. Inset: The lifetime of the H2O molecule is shown as a function of pressure. All other species had lifetimes between 30 and 50 fs, independent of pressure. Copyright 2009 by AIP Publishing.}
\label{fig:Goncharov_2009_Fig_5}
\end{figure}

Experiments show an enormous increase in $\Delta h$ (enthalpy of fusion) along the melting line of H$_2$O at around $\sim$40 GPa~\citep{Goncharov:2009}. They attribute this measurement to significant molecular dissociation of H$_2$O upon melting (see Fig.~\ref{fig:Goncharov_2009_Fig_5}). Thus, the fluid in this regime is termed \emph{ionic} fluid. The \emph{ionic} fluid will have higher electric conductivity ($\sigma$ in S/cm) compared to \emph{molecular} fluid, owing to the mobility of ions. 

We expect similar phenomenon of increased molecular dissociation into ions to occur in fluid Ammonia (NH$_3$) under increased pressure and temperature: 

\begin{equation}
       \text{NH$_3$} \xtofrom[\text{Recombination}]{\text{Dissociation}} \text{H$^{+}$} + \text{NH2$^{-}$}
\end{equation}

%Therefore, we come up with the simple schematic phase diagram (Fig.\ref{fig:PhaseDiagramH2OandNH3}) for both H$_2$O and NH$_3$, and their mixture, owing to their very similar chemical properties and free mixing with each other. In Fig.\ref{fig:PhaseDiagramH2OandNH3} we have ignored the complicated ice phases at lower pressures and approximate them as an ensemble of entropy-driven solid phase. 

A schematic phase diagram of water and ammonia is shown as follows: 

\begin{figure}[h!]
\centering
\hspace*{-2cm}\includegraphics[scale=0.7]{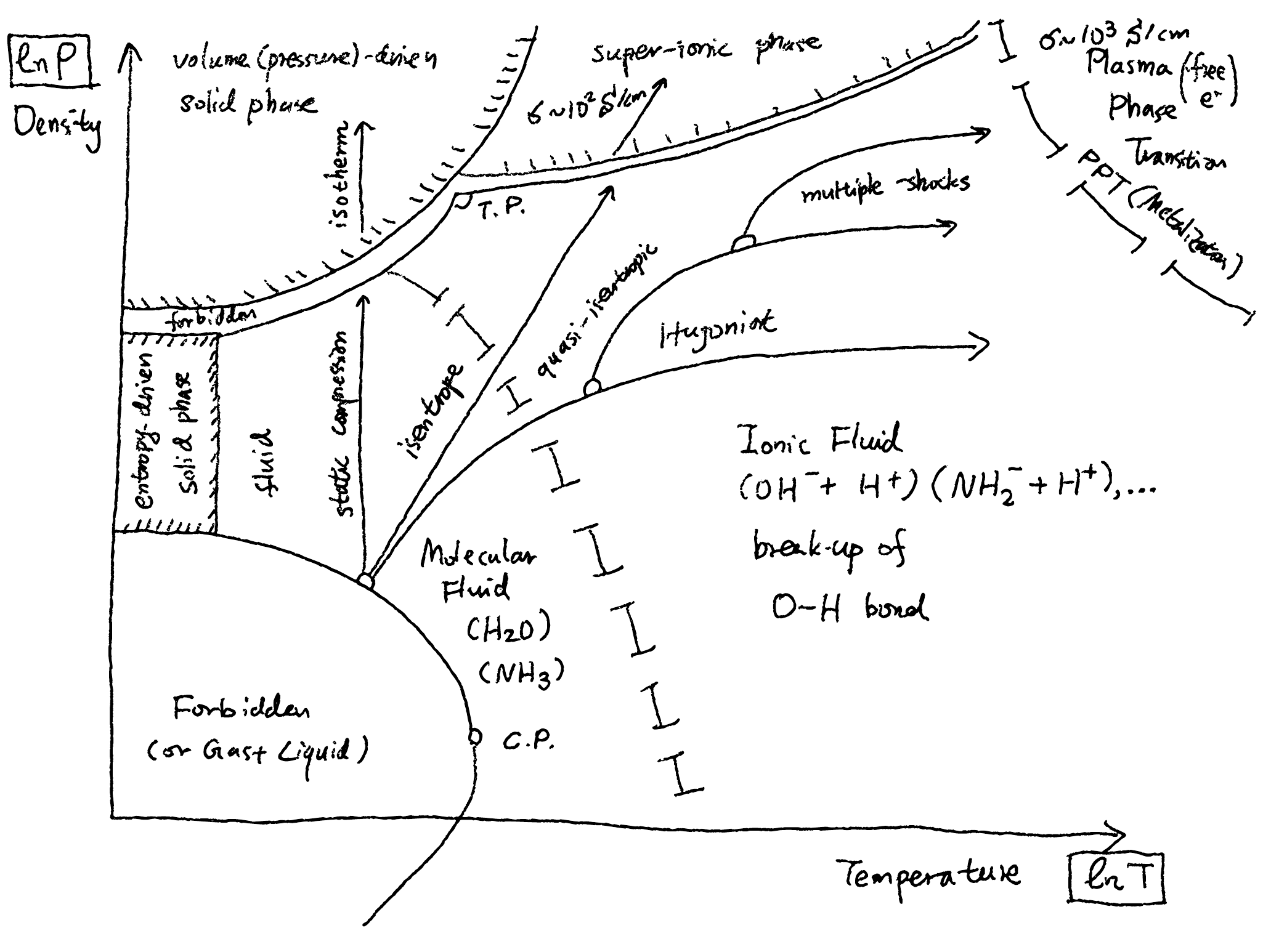}
\caption{Schematic phase diagram for water and ammonia, or their mixture. Also shown are trajectories of isotherm, isentropes, and single/multiple-shock Hugoniot. A single-shock Hugoniot will likely miss the super-ionic regime. The detailed differences among the various ice structures/phases have been ignored as they are all approaching close-packing at sufficient pressures. }
\label{fig:PhaseDiagramH2OandNH3}
\end{figure}

\subsection{Electric Conductivity}

The electric conductivity $\sigma$ (in S/cm) is defined as: 

\begin{equation}
       \sigma \equiv \frac{\overbrace{\mathbf{J}}^{\text{current density}}}{\underbrace{\mathbf{E}}_{\text{electric field magnitude}}}
\end{equation}

It can be caused by: 
\begin{itemize}
    \item de-localized (mobile) ions
    \item de-localized (mobile) protons (H$^{+}$)
    \item de-localized (mobile) electrons (e$^{-}$)
\end{itemize}

The range of $\sigma$ spans over many orders-of-magnitude:

\begin{figure}[h!]
\centering
\includegraphics[scale=0.7]{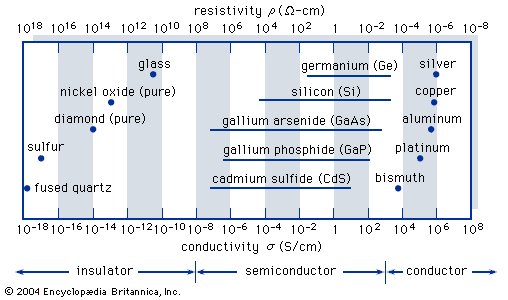}
\caption{Range of electric conductivity ($\sigma$ in S/cm). Copyright 2004 by Encyclopædia Britannica, Inc.~\citep{EncyclopediaBritannica_electricalconductor}}
\label{fig:conductivities}
\end{figure}

Here we concentrate on the electric conductivity resulting from electrons (e$^{-}$). 

Quote from the \emph{Encyclopædia Britannica}~\citep{EncyclopediaBritannica_electricity}~\citep{EncyclopediaBritannica_electricalconductor}~\citep{EncyclopediaBritannica_bandtheory}: 

\begin{displayquote}
Materials are classified as conductors, insulators, or semiconductors according to their electric conductivity. The classifications can be understood in atomic terms. Electrons in an atom can have only certain well-defined energies, and, depending on their energies, the electrons are said to occupy particular \emph{energy levels}. In a typical atom with many electrons, the lower energy levels are filled, each with the number of electrons allowed by a quantum mechanical rule known as the \emph{Pauli exclusion principle}. Depending on the element, the highest energy level to have electrons may or may not be completely full. If two atoms of some element are brought \emph{close enough} together so that they \emph{interact}, the two-atom system has two closely spaced levels for each level of the single atom. If 10 atoms \emph{interact}, the 10-atom system will have a cluster of 10 levels corresponding to each single level of an individual atom. In a \emph{condensed matter}, the number of atoms and hence the number of levels is extremely large; most of the higher energy levels overlap in a continuous fashion except for certain energies in which there are no levels at all. Energy regions with levels are called \emph{energy bands}, and regions that have no levels are referred to as \emph{band gaps}.
\end{displayquote}

\begin{figure}[h!]
\centering
\includegraphics[scale=0.8]{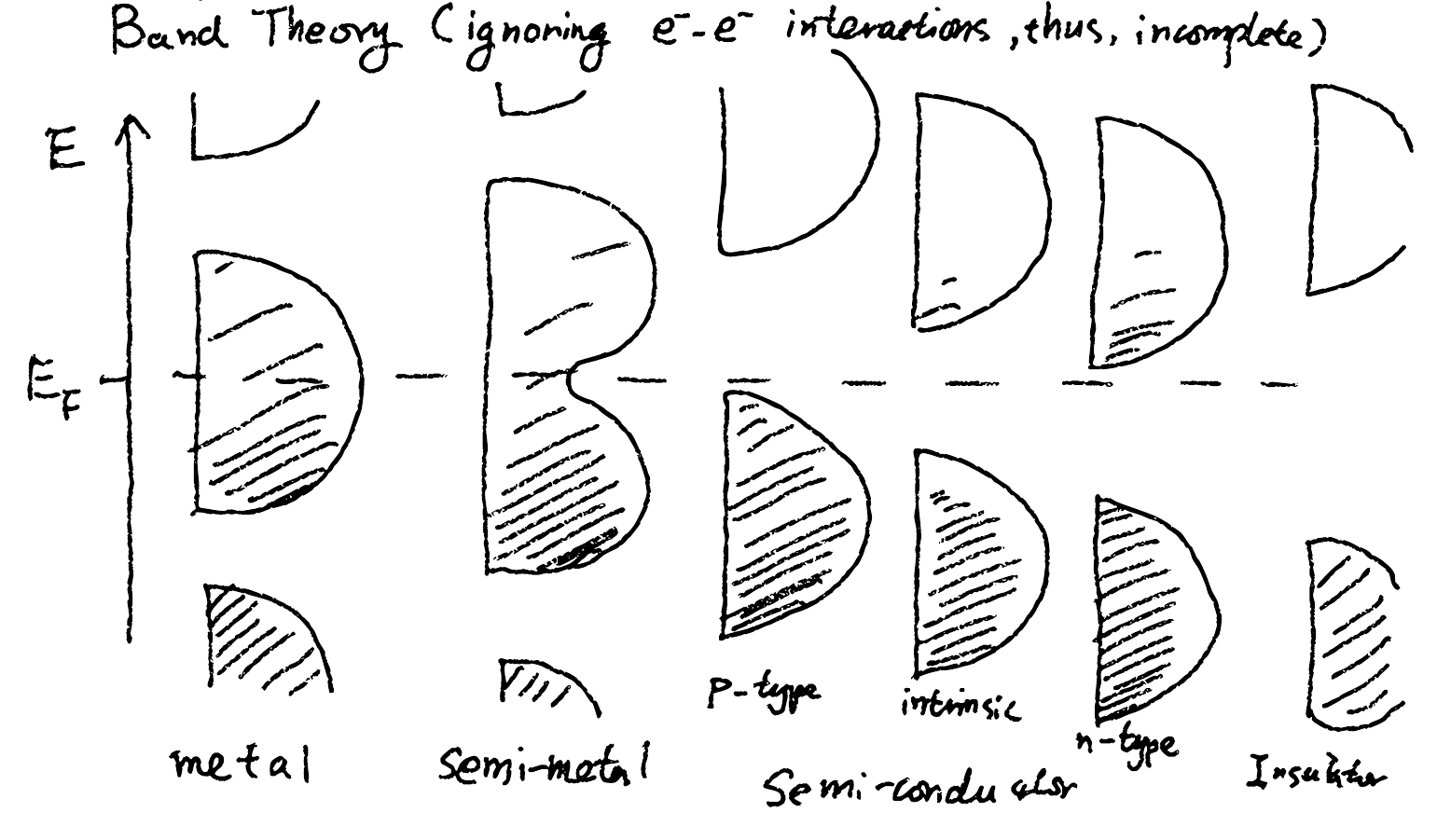}
\caption{Band Theory. The envelope of bands can be understood as a representation of the \emph{density of states}. The topology of band determines the nature of the type of electrical conductivity in the material. Here we only consider Pauli exclusion principle and have ignored e$^{-}$-e$^{-}$ interactions/coupling, thus, posing an incomplete theoretical picture. The probability of the occupancy of the energy levels fill up roughly to the \emph{Fermi energy} $E_F$, and can spill over a few $k_B T$ upwards. }
\label{fig:band_theory_20200518}
\end{figure}

From the perspective of energy, here we show a schematic drawing of the energy levels (band and gap), for different types of conductors, insulators, or semiconductors: Fig~\ref{fig:band_theory_20200518}

In general, the temperature-dependence of $\sigma$ is different for a metal and an insulator (semi-conductor): 

For metal: 

\begin{equation}
    \sigma = \sigma_0 \cdot \frac{T}{T_0} \propto T
\end{equation}

For insulator (semi-conductor):

\begin{equation}
    \sigma = \sigma_0 \cdot \exp{\Bigg( -\frac{E_{a}}{k_B T} \Bigg)}
\end{equation}

where $E_a$ is the activation energy for a given insulator (semi-conductor). The $\sigma$ of metals depends on the temperature and the ambient conditions much less because in metals, the state-of-conduction is a \emph{non-excited state}. Meanwhile, in order to make an insulator (semi-conductor) conductive, it is necessary to transform them into an \emph{excited (activated) state}~\citep{Kireev1978}. 

In the next sections, we will show that both increase in Temperature and Density (or Pressure) will increase the conductivity of a material. See also in~\citep{Yoo2020}. 

%\subsubsection{Viewpoint of Space}
%Another way to look at it is that the great pressure excites valence electrons across a mobility gap of a few eV into the conduction band of the dense fluid~\cite{Nellis2017}. 
\subsection{Metallization \\due to Density Increase}

\subsubsection{Scaling Relations}

At high enough density, that is, with significant compression, everything is expected to become \emph{metal}. The fundamental reason is simple: as atom spacing becomes small, their outermost electronic orbitals are forced to overlap, and electrons become de-localized and are shared with other atoms~\citep{Mott1982}~\citep{Hensel2015Metallization}. 

In terms of the \emph{band theory}, the topology of band (Fig~\ref{fig:band_theory_20200518}) changes when the spacings between neighboring atoms change. 

At $T = 0$ Kelvin, the transition from non-metal to metal state can be estimated by the Mott's Formula~\citep{Edwards1978}~\citep{Edwards2010Metallization}:

\begin{equation}\label{Eq:MottFormula1}
    n_c^{1/3} \cdot a_H^{*} \approx 0.25
\end{equation}

Here $n_c$ the critical atom density that corresponds to a critical distance ($d_c$) between close-packing atoms, at which a first-order, discontinuous transition from a metal to a non-metal would occur at T = 0 K. 
$n_c$ has the dimension of reciprocal-volume, thus, its product with a distance $a_H^{*}$ is dimensionless. 
$a_H^{*}$ is an \emph{equivalent} Bohr (hydrogenic) radius of the isolated (low density) atomic state (in this instance). $a_H^{*}$ can be estimated as follows~\citep{Edwards1978}: 

\begin{equation}\label{Eq:MottFormula3}
    a_H^{*} \approx \ddfrac{K_{st} \cdot \hbar^2 }{m_{e}^{*} \cdot e^2}
\end{equation}

where $K_{st}$ is the static (low-frequency) dielectric constant of the host material, and $m_{e}^{*}$ is the \emph{effective} mass of an electron in the host conduction band. 

At $T \gg 0$ Kelvin, say 1000 K, the scaling parameter $n_c^{1/3} \cdot a_H^{*}$ for transition from non-metal to metal state is relaxed to~\citep{Edwards1978}~\citep{Edwards2010Metallization}:

\begin{equation}\label{Eq:MottFormula2}
    n_c^{1/3} \cdot a_H^{*} \approx 0.38
\end{equation}

Eq.~\ref{Eq:MottFormula1} and Eq.~\ref{Eq:MottFormula2} are convenient when we apply them to understand the non-metal-to-metal transition in the T-$\rho$ phase diagram. For example, an eight-fold compression would double $d_c$ and increase both $\rho$ and $n_c$ by eight-fold. 

Experiments and calculations show that almost all \emph{elements} (not their compounds) become metal under the pressures of 200$\sim$300 GPa or above with the only exceptions of Helium, Carbon, Fluorine, Chlorine, and other noble gases or alike~\citep{Edwards2010Metallization}~\citep{Hensel2015Metallization}. For example, metallic Oxygen is obtained in lab around 100 GPa~\citep{Hensel_Metallic_Oxygen}. 

\subsubsection{Fermi-Dirac Distribution for Electrons}

The electrons in a \emph{condensed phase} can be divided into two groups; those that form the closed electron shells of the constituent atoms (the \emph{core electrons}), and the remaining electrons of higher energy (the \emph{valence electrons}). In metals the valence electrons can move more or less freely through the lattice with an \emph{effective} mass which takes into account the influence of the periodic potential resulting from the lattice ions. One usually refers to them as the \emph{conduction electrons}~\citep{Grimvall1999}.

In the \emph{Sommerfeld} model of a metal~\citep{Sommerfeld1933} in which the \emph{conduction electrons} are assumed to form a gas of fermions with energies $E$ and a density of states $N(E)$. 

The probability of a state with the energy $E$ being occupied by an electron in a state of thermodynamic equilibrium  follows the \emph{Fermi-Dirac} distribution function as: 
%= f_0 \bigg(\frac{p^2}{2m_e},T\bigg)
\begin{equation}\label{Eq:FermiDiracDistribution}
    f_0 (E,T) = \Bigg(\exp{\bigg( \frac{E-E_{F}}{k_B \cdot T} + 1\bigg)} \Bigg)^{-1}  = \Bigg(\exp{\bigg( \frac{\frac{p^2}{2 m_e}-E_{F}}{k_B \cdot T} + 1\bigg)} \Bigg)^{-1} 
\end{equation}

Here, $E_F$ is the \emph{Fermi level}, defined as the Gibbs thermodynamic potential per particle, i.e., the \emph{chemical potential} $\mu$ at $T=0$ Kelvin: $E_F \equiv \mu(0)$. $E_F$ numerically equals to the work that must be spent to add one particle (here one electron) to the system~\citep{Kireev1978}~\citep{Kittel2004}. Here we have ignored electron-electron interactions such as e$^{-}$-e$^{-}$ pairing which leads to super-conductivity. 

$f_0 (E,T)$ experiences essential changes when the energy $E$ is changed from $\sim(E_{F}-2k_B T$) to $\sim(E_{F}+2k_B T)$. 
$f_0 (E,T)$ can be integrated over the full \emph{momentum-space} to find the mean energy (per electron) of all the electrons of a degenerate system: 

\begin{equation}\label{Eq:FermiDiracDistribution2}
   \langle E \rangle = \ddfrac{\int_0^{\infty} \frac{p^2}{2m_e} \cdot f_0 \bigg(\frac{p^2}{2m_e},T\bigg) \cdot (4\pi p^2) dp}{\int_0^{\infty} f_0 \bigg(\frac{p^2}{2m_e},T\bigg) \cdot (4\pi p^2) dp}
\end{equation}

In the two limiting cases, 

\begin{equation}\label{Eq:FermiDiracDistribution3}
   \langle E \rangle \approx 
            \begin{cases}
                \frac{3}{2} k_B T & \text{~when $k_B T \gg E_F$}, \\
                \frac{3}{5} E_F & \text{~when $k_B T \ll E_F$}.
            \end{cases}
\end{equation}

As can be seen from Eq.~\ref{Eq:FermiDiracDistribution3}, the energy of degenerate e$^{-}$ gas at low temperature is independent of temperature, and thus, it does not contribute to the specific heat of the body, and this fully explains the Dulong-Petit law for solids~\citep{Kireev1978}. $\langle E \rangle=\frac{3}{5} E_F$ typically amounts to several electron-volts (eV). 
\begin{comment} %In other words, quantum theory achieved a great success in explaining the independence of specific heat in solids, at low temperatures, from electron motion, since according to classical theory the electrons should have contributed $\frac{3}{2} k_B T$ to the specific heat. 
\end{comment} 
However, at elevated temperature, when $k_B T > E_F$, the energy of the electron gas is determined by its temperature. The cross-over from quantum to classical statistics is set as follows: 

\begin{equation}
  \theta \equiv \bigg( T/T_{F} \bigg) \approx 1
\end{equation}

In other words, the dimensionless \emph{degeneracy parameter} $\theta\equiv T/T_{F}$ is on the order of unity. $T_{F}$ is the \emph{Fermi temperature} defined from \emph{Fermi energy} $E_F$ as follows: 

\begin{equation}
  T_{F} \equiv \frac{E_{F}}{k_{B}} \approx \frac{\hbar^2}{2 \cdot m_{e}^{*} \cdot k_{B}} \cdot \Bigg(3 \cdot \pi^2 \cdot n \Bigg)^{2/3}
\end{equation}

where $n$ is the number density of (conduction) electrons, that is, the number density of electrons in the conduction band, and $m_{e}^{*}$ is the \emph{effective} mass of an electron in the host's conduction band. The Fermi energy of a metal is typically of the order of 2$\sim$10 eV~\citep{Ashcroft1976}~\citep{Ashcroft1976_2}. 
%This has to do with whether shock temperatures and pressures are sufficiently large to ionize electrons from inner shells of atoms~\cite{Nellis2017}. 

The matter in this cross-over regime is sometimes referred to as the \emph{Warm Dense Matter} (WDM). It is the current interest of high-pressure shock wave experiments~\cite{Nellis2017}. This cross-over regime is characterized by atomic matter for which potential energy of electron-ion interactions is comparable to kinetic energy of electrons.

\subsection{Plasma Phase Transition (PPT) \\due to Temperature Increase}

At high enough temperature, everything will become plasma. Plasma is a gaseous or gas-like phase characterized by free electrons (e$^{-}$) due to thermal excitations. Owing to the free electrons (e$^{-}$), it has high electric conductivity ($\sigma$ in S/cm) compared to \emph{ionic} fluid. 

Plasma Phase Transition (\emph{PPT}) is related with the \emph{metallization of fluid}, as the \emph{PPT} boundary sometimes continues smoothly into the denser regime in the Temperature-Density (T-$\rho$) phase diagram which then belongs to the liquid-like phase rather than a typical gas-like phase. The following is a simple schematic phase diagram for Hydrogen in T-$\rho$ space: 

\begin{figure}[h!]
\centering
\includegraphics[scale=0.5]{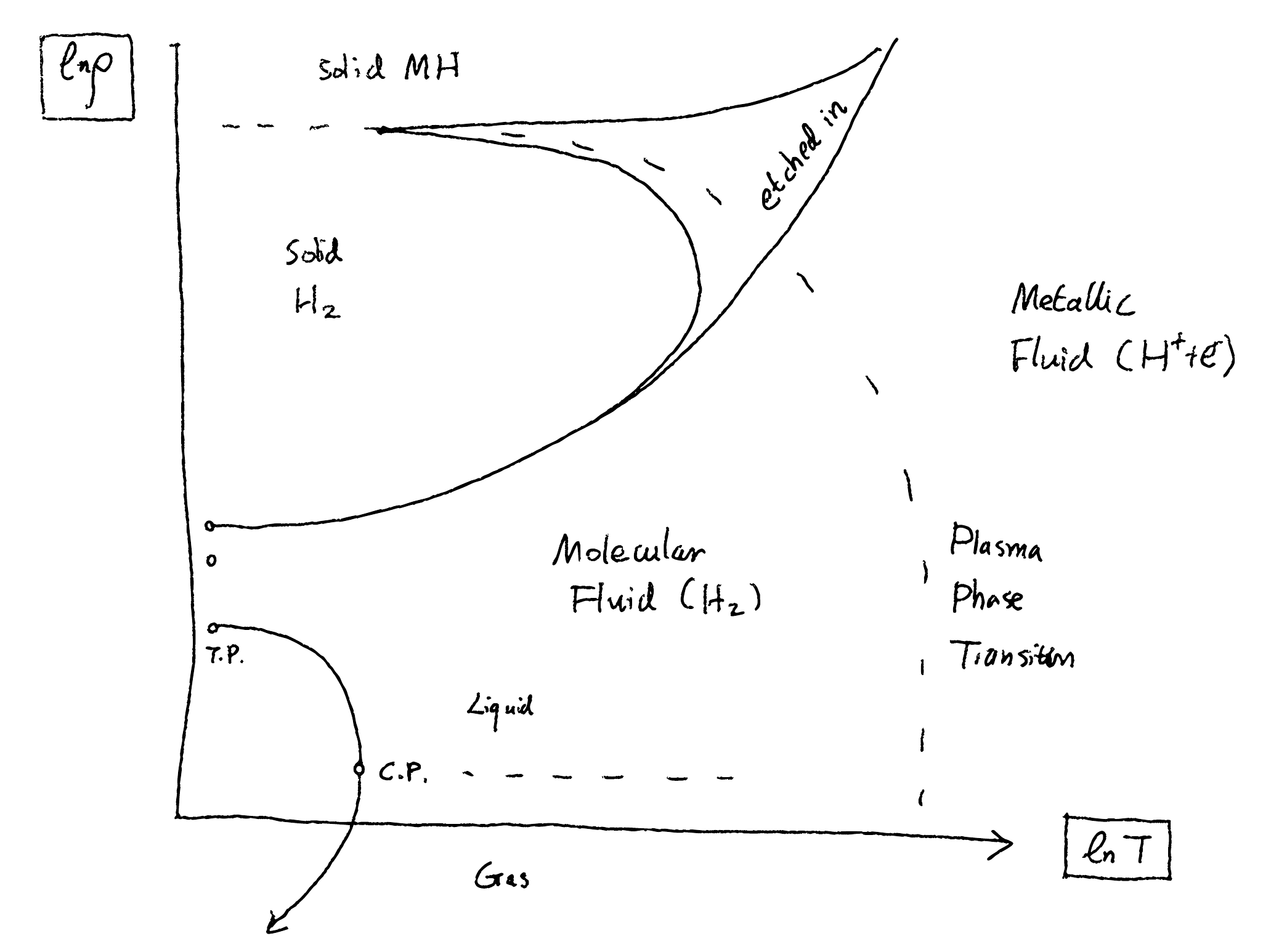}
\caption{Schematic Phase Diagram for Hydrogen.}
\label{fig:PhaseDiagramH2}
\end{figure}

As seen in Fig.~\ref{fig:PhaseDiagramH2}, the \emph{PPT} draws the boundary between \emph{fluid molecular hydrogen} (H$_2$) and \emph{fluid metallic hydrogen} (H$^{+}$ and free electrons). 

The \emph{fluid metallic hydrogen} is immiscible with \emph{fluid helium} (which is a noble gas) at low temperature, but eventually becomes miscible with \emph{fluid helium} at higher temperature. 

\begin{figure}[h!]
\centering
\includegraphics[scale=0.4]{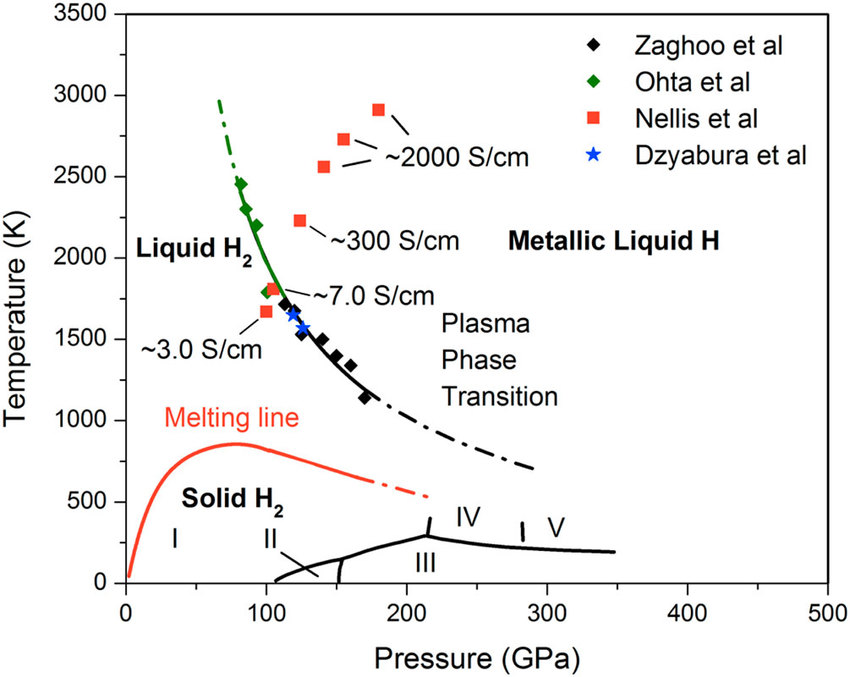}
\caption{EOS of dense hydrogen showing Pressure/Temperature squares at which electrical conductivities of fluid hydrogen were measured under dynamic compression~\cite{Nellis2017}. Copyright 2016 by American Physics Society. }
\label{fig:EOS_dense_hydrogen_Nellis}
\end{figure}

According to calculations in upcoming chapter, Fig.~\ref{fig:H2EOSMazevet2019lgTlgrho}, Fig.~\ref{fig:H2EOSMazevet2019lgrholgP}, Fig.~\ref{fig:H2EOSBecker2014lgTlgrho}, and Fig.~\ref{fig:H2EOSBecker2014lgrholgP}, \emph{PPT} of hydrogen occurs around 100 GPa, or correspondingly, at a density slightly below 1 g/cc. Jupiter's interior may still be too hot to miss this miscibility gap. Pressures in Uranus' and Neptune's envelopes are everywhere lower than \emph{PPT} at around $10^6$ bar or 100 GPa. Therefore, \emph{PPT} is most relevant to Saturn, and may explain its abnormal intrinsic luminosity and cooling history~\cite{Stevenson2017Ge131:PlanetaryEvolution}. In more detail, \emph{PPT} in \emph{fluid hydrogen} also has a temperature-dependence, in addition to its dependence on Pressure $P$, as shown in the $P$-$T$ phase diagram of dense hydrogen (Fig.~\ref{fig:EOS_dense_hydrogen_Nellis}, experiments show a rapid increase in electrical conductivity ($\sigma$ in S/cm) across \emph{PPT}~\citep{Nellis2006}). 

\subsection{Mixing and Un-Mixing}

\subsubsection{Thermodynamics of Mixing}
The fundamental reason for mixing and un-mixing is the differences in inter-molecular forces or specific molecular effects between different species, even though they are chemically non-reacting. 

Quantitatively, one can estimate the Gibbs free energy of mixing ($\Delta G_{\text{mixing}}$): 

\begin{align}\label{Eq:GibbsEnergyofMixing}
    \Delta G_{\text{mixing}} \equiv &= G_{\text{after mixing}}-G_{\text{before mixing}} \\\nonumber
                    &= \Delta H_{\text{mixing}}- T \cdot \Delta S_{\text{mixing}} \\\nonumber
                    &= \bigg( H_{\text{after mixing}} - H_{\text{before mixing}} \bigg)- T \cdot \Delta S_{\text{mixing}} \\\nonumber
\end{align}

For a given Temperature T, and the number fraction of each individual component participating in the mixing,  
\begin{itemize}
    \item If $\Delta G_{\text{mixing}} < 0$, then mixing is favored
    \item If $\Delta G_{\text{mixing}} > 0$, then un-mixing is favored
\end{itemize}

$\Delta H_{\text{mixing}}$ is the enthalpy of mixing, and $\Delta S_{\text{mixing}}$ is the entropy of mixing. 

$\Delta H_{\text{mixing}}$ can be calculated from ab initio simulations~\citep{Lorenzen2009}: 

\begin{align}\label{Eq:EnthalpyofMixing}
    \Delta H_{\text{mixing}} &= H_{\text{after mixing}} \bigg(x_1,x_2,...,x_N \bigg) - H_{\text{before mixing}} \\\nonumber
             &= H_{\text{after mixing}} \bigg(x_1,x_2,...,x_N \bigg) - \sum_{i=1}^N x_i \cdot H_i \\\nonumber
             &= \Delta U + P \bm{\cdot} \underbrace{\Delta V}_{\text{volume effect comes in here}} \\\nonumber
\end{align}

where the enthalpy of the mixture ($H_{\text{after mixing}}$) is a function of $x_1,x_2,...,x_N$, which are the number fractions of every individual components participating in the mixing, assuming no chemical reactions occurring among them, and satisfying the normalization condition: $\sum_{i=1}^N x_i =1$. 

Without a priori knowledge of the spatial distribution or orientation within the mixture, such as clustering or non-uniform distribution, $\Delta S_{\text{mixing}}$ can often be approximated by the entropy of ideal mixing: 

\begin{align}\label{Eq:EntropyofMixing}
    \Delta S_{\text{mixing}} &\approx \Delta S_{\text{ideal mixing}} \\\nonumber
             &\approx -k_B \cdot \sum_{i=1}^N x_i \cdot \ln{x_i} %\\\nonumnber
\end{align}

$\Delta S_{\text{mixing}}$ is always positive, and it is weighted by Temperature T in Eq.~\ref{Eq:GibbsEnergyofMixing}. Therefore, at high-enough temperature, mixing is always favored. 
$\Delta H_{\text{mixing}}$ primarily depends on the \emph{energy state} of the system. If the components interact with each other attractively, that is, by forming bonds and sharing electrons such as in \emph{metal}, and then the total energy of the system is reduced after mixing, and $\Delta H_{\text{mixing}}$ is small or even negative. On the other hand, if the components interact repulsively, then bringing them together cost a lot of energy and $\Delta H_{\text{mixing}}$ is positive and large. 

One can think of Temperature T as playing the role of fulcrum on a lever, and the load on one side is  $\Delta H$, and the load on the other side is $\Delta S$. We call this lever $\Delta G$: 

\begin{equation}
    \Delta G = \Delta H - \underbrace{T}_{\text{fulcrum}} \bm{\cdot} \Delta S
\end{equation}

%\balance[3]{1}{2}

\subsubsection{Relation to Metallization}
In particular, \emph{non-metal-to-metal transition} plays a key role for mixing and un-mixing of different components in planetary interior, i.e., differentiation. The binary mixture of hydrogen-helium under pressure is just one example. Generally speaking, metal (fluid) and metal (fluid) can readily mix with each other, while metal (fluid) and non-metal (fluid) cannot. So mixing/un-mixing is viewed as the same problem as \emph{metallization}. 

The story of compounds may be more complicated. For example, in silicates, the oxygen anions (having acquired e$^{-}$s and silicon cations (having donated e$^{-}$s), having assumed closed-shell structure and somewhat equivalent to noble gases, are much more stable than their elemental forms, and thus can remain non-metallic up to higher pressure (a few TPa or 1000’s GPa level). This is confirmed by both static-compression and shock-wave experiments.  

According to our earlier work on the simple scaling laws of planet interior pressure with its mass and material property~\cite{Zeng:2016c}, we find out that:
\begin{itemize}
\item For rocky planets and water worlds, their characteristic interior pressure scales approximately \emph{linear} in planet mass:     \begin{equation}
        P\sim(M/M_{\oplus}) \cdot (1-\frac{2}{3} \cdot \text{(Ice mass fraction)}) \cdot \text{~100 GPa}
    \end{equation}
\item For degenerate gas giants, their characteristic interior pressure scales approximately \emph{quadratic} in planet mass: 
    \begin{equation}
        P\sim(M/M_{\text{Jupiter}})^2 \cdot \text{~1 TPa}
    \end{equation}
\end{itemize}

For planets more massive than Earth, their core-mantle boundary of any sort may gradually become less distinct and more fuzzier with increasing mass, due to the metallization and increased miscibility of materials in their deep interior. The trends of mutual mixing of Fe-metal, silicates, ices, and all other materials dissolved in them increase. A planet core may still exist, because denser material is favored in terms of long-range gravitational energy towards the center. However, on the short-range, in terms of chemical gradation and entropy, mixing is favored more and more. 

This is the most natural explanation for Jupiter’s fuzzy core~\cite{Wahl2017JunoCoreFuzzy}, as advocated by B. Militzer and D. Stevenson. In conclusion, the planets more massive than Earth will have different interior differentiation. The exact result is also evolution-history-dependent or path-dependent. 

\subsection{Transport Properties}

Transport properties describe how a particular physical quantity is transferred through a system, for example diffusivity, viscosity, thermal conductivity, and electric conductivity (which we have just discussed). In recent years, there is a growing interest in calculating and measuring transport properties in high-pressure experiments, because they play important role in modelling and understanding the physics of planetary interiors. 

\subsubsection{Prandtl Number}
One important dimensionless parameter that characterizes the transport properties is the \emph{Prandtl number} $\mathbf{Pr}$: 

\begin{equation}\label{Eq:PrandtlNumber}
    \mathbf{Pr} \equiv \frac{\overbrace{\text{momentum}}^{\text{orderly motions}} \text{~diffusivity}}{\underbrace{\text{thermal}}_{\text{dis-orderly motions}} \text{~diffusivity}} = \frac{\overbrace{\mu}^{\text{dynamic viscosity}} / \overbrace{\rho}^{\text{density}}}{\underbrace{k}_{\text{thermal conductivity}} / (\underbrace{c_p}_{\text{specific heat capacity}} \bm{\cdot} \rho)}
\end{equation}

It characterizes the relative importance between two energy transport modes: conduction versus convection. It then determines the relative thickness of a thermal boundary layer ($\delta_{\text{thermal boundary layer}}$) versus thickness of velocity boundary layer ($\delta_{\text{velocity boundary layer}}$) in a convective cell. 

Since there is no length-scale involved in Eq.~\ref{Eq:PrandtlNumber}, $\mathbf{Pr}$ is an EOS property which only depends on the type of fluid and the state the fluid is in. For most gases over a wide Temperature-Pressure range, $\mathbf{Pr}$ is approximately constant, due to the micro-physics involved that in the gas-phase both processes are caused by the kinetic motions of gas molecules. 

\begin{itemize}
    \item $\mathbf{Pr} \ll 1$ (Liquid Metals, Earth's Core): Heat diffuses very efficiently and quickly. Thus, heat conduction is more significant compared to convection. Then, 
        \begin{equation}
            \delta_{\text{thermal boundary layer}} \ll \delta_{\text{velocity boundary layer}}
        \end{equation}
    
    \item $\mathbf{Pr} \ll 1$ (Oils, Earth's Mantle): Heat diffuses very inefficiently and slowly. Thus, convection dominated. Then, 
        \begin{equation}
            \delta_{\text{thermal boundary layer}} \gg \delta_{\text{velocity boundary layer}}
        \end{equation}
        
    \item $\mathbf{Pr} \sim 1$ (Many gases): Heat diffusion and convection are of comparative importance. Therefore, both effects need to be dealt with simultaneously, such as in aerodynamics. 
\end{itemize}

\subsubsection{Rossby Number}
Another important dimensionless parameter that characterizes the large-scale transport properties is the \emph{Rossby number} $\mathbf{Ro}$. It characterizes the \emph{large-scale} flow in the surface-layer of a planet, such as atmospheric motions and flow in ocean. 

\begin{equation}\label{Eq:RossbyNumber}
    \mathbf{Ro} \equiv \frac{\left| \text{Inertial~}\overrightarrow{Force} \right|}{\left| \text{Coriolis~}\overrightarrow{Force}  \right|} = \frac{\left| \bm{v\cdot \nabla v} \right|}{\left| \bm{\Omega \times v} \right|} \sim \frac{U^2/\overbrace{L}^{\text{characteristic length scale}}}{U \cdot \underbrace{\Omega}_{\text{angular frequency of rotation}}}
\end{equation}

On Earth, $\mathbf{Ro}$ accounts for phenomena such as jet streams, long waves in westerlies, and polar fronts, etc. On Hot Jupiters, it shall account for the flow structure in and out of the hot-spot caused by host-stellar irradiation. 

\clearpage
%%%%%%%%%%%%%%%%%%%%%%%%%%%%%%%
\section{Variational Principle and Virial Theorem}

\subsection{Basics: Hydrostatic Equilibrium Re-visited}
Owing to the simplicity of the form of specific internal energy differential along an isentrope (reversible adiabat) with constant specific entropy $s=s_0$: 
\begin{equation}
    du = -P \cdot dv
\end{equation}

We can construct a Lagrangian, which by means of \emph{variational principle}, is equivalent to the differential equation of hydro-static equilibrium and mass-conservation~\citep{Zeng2016VARIATIONALINTERIORS}. 

Assuming spherical symmetry, let us again use $m$ as the independent variable, which is the mass enclosed within shell radius $r$. 
We define a new variable $\omega$, which is the volume of the sphere contained within $r$. 
We represent the full differential with respect to mass $m$ as an overhead dot-symbol ($\dot {}$), so that: 

\begin{align}
    \omega &\equiv \bigg( \frac{3}{4\pi} \bigg)^{-1} \cdot r^3 \\\nonumber
    \dot{\omega} &\equiv \frac{d\omega}{dm} = v = \frac{1}{\rho}  \\\nonumber
    \ddot{\omega} &\equiv \frac{d\dot{\omega}}{dm} = \frac{dv}{dm}  \\\nonumber
\end{align}

The total energy $E_{\text{total}}$ of the planetary body is a combination of its total internal energy $E_{\text{internal}}$, which include both the thermal energy and the elastic energy stored due to reversible-adiabatic (isentropic) compression, and its own gravitational energy $E_{\text{grav}}$. We can write it in integral form as: 

\begin{align}
    E_{\text{total}} &= E_{\text{internal}} + E_{\text{grav}} \\\nonumber
                    &= \int_{m=0,\text{~center}}^{m=M,\text{~surface}} \Bigg(u-\frac{G\cdot m}{r} \Bigg)\cdot dm \\\nonumber
\end{align}

Now, let's introduce the \emph{Lagrangian} $\Lagr = \Lagr(m;\omega,\dot{\omega})$, with \emph{natural} variable $m$ and \emph{dependent} variables $(\omega,\dot{\omega})$, separated by the ";" sign, as: 

\begin{align}\label{Eq:Lagrangian1}
    \Lagr &= \Lagr(m;\omega,\dot{\omega}) \\\nonumber
          &= -\Bigg(u-\frac{G \cdot m}{r} \Bigg) \\\nonumber
          &= -\Bigg( u(\dot{\omega})-\frac{G \cdot m}{\bigg(\frac{3}{4\pi}\bigg)^{\frac{1}{3}} \cdot {\omega}^{\frac{1}{3}}}\Bigg) \\\nonumber
\end{align}

Then, the integral form of total planet energy $E_{\text{total}}$ becomes a \emph{functional} (function of function), which varies depending on the functional form $\omega(m)$, and its first-order derivative $\dot{\omega}(m)$: 

\begin{align}\label{Eq:Lagrangian2}
    E_{\text{total}} &= \int_{m=0,\text{~center}}^{m=M,\text{~surface}} -\Lagr(m;\omega,\dot{\omega}) \cdot dm \\\nonumber
                    &= \int_{m=0,\text{~center}}^{m=M,\text{~surface}} \Bigg( \underbrace{u(\dot{\omega})}_{\textbf{Part I}}-\underbrace{\frac{G \cdot m}{\bigg(\frac{3}{4\pi}\bigg)^{\frac{1}{3}} \cdot {\omega}^{\frac{1}{3}}}}_{\textbf{Part II}} \Bigg) \cdot dm \\\nonumber
\end{align}

Then, the problem of solving for planetary interior structure is reduced to finding the appropriate functional form of $\omega(m)$ which minimizes the total energy $E_{\text{total}}$ of the planet, provided that the specific entropy $s=s_0$ within it remains constant. 

\textbf{Part I} in Eq.~\ref{Eq:Lagrangian2} is the EOS of material. It only depends on $\dot{\omega}$. It comes from the specific internal energy $u=u(v,s_0)$ dependence on specific volume $v=\dot{\omega}$ and specific entropy $s=s_0$. So $u$ is a uni-variate function: $u=u(v)$. We then write out the derivative of $u$ with respect to $v$ and represent with the prime-symbol ($'$):  

\begin{align}
    u' &\equiv u'(v) = u'(\dot{\omega})= \frac{du}{dv} = -P \\\nonumber
    u'' &\equiv u''(v) = u''(\dot{\omega}) = -\frac{dP}{dv} = -\frac{dP}{dm} \cdot \frac{dm}{dv} = -\frac{dP}{dm} \cdot \frac{1}{\ddot{\omega}} \\\nonumber
\end{align}

From another perspective, \textbf{Part I} (EOS) is a short-range interaction so it only depends on $\dot{\omega}$, while \textbf{Part II} (gravitational interaction) is long-range so it only depends on $\omega$. 

We could then apply the one-dimensional Euler–Lagrange equation to Eq.~\ref{Eq:Lagrangian2}: 

\begin{equation}\label{Eq:Lagrangian3}
    \frac{d}{dm}\Bigg( \underbrace{\frac{\partial \Lagr}{\partial \dot{\omega}}}_{\text{act only on \textbf{Part I}}} \Bigg) = \underbrace{\frac{\partial \Lagr}{\partial \omega}}_{\text{act only on \textbf{Part II}}}
\end{equation}

Recall that:

\begin{equation}\label{Eq:Lagrangian3b}
    \frac{\partial \Lagr}{\partial \dot{\omega}} = -\frac{\partial u(\dot{\omega})}{\partial \dot{\omega}} = P
\end{equation}

Then, this gives rise to the following second-order differential equation: 

\begin{equation}\label{Eq:Lagrangian4}
    \frac{dP}{dm} = -u''(\dot{\omega}) \cdot \ddot{\omega} = -\Bigg(\frac{4\pi}{81} \Bigg)^{1/3} \cdot G \cdot \frac{m}{{\omega}^{4/3}}
\end{equation}

The negative sign ($-$) suggests the Pressure $P$ is decreasing from center outward, which is what we expected. 
After removing the negative sign ($-$) in Eq.~\ref{Eq:Lagrangian4}, we can re-cast it into the following second-order differential equation: 

\begin{equation}\label{Eq:Lagrangian5}
    u''\bigg(\dhyd{\omega}{m}\bigg) \cdot \dnhyd[2]{\omega}{m} = \Bigg(\frac{4\pi}{81} \Bigg)^{1/3} \cdot G\cdot m\cdot \omega^{-4/3}
\end{equation}

There are several important features of Eq.~\ref{Eq:Lagrangian5}: 

\begin{itemize}
    \item $\dnhyd[2]{\omega}{m}>0$, since $u''>0$. That is, $\omega(m)$ is a \emph{concave} in $m$. 
    \item $\dhyd{\omega}{m}\Big\rvert_{m=0} = \frac{1}{\rho_{\text{center}}}$. 
    \item $\dnhyd[2]{\omega}{m}\Big\rvert_{m=0} = 0$. Close to the center, $\omega(m)$ is approximately \emph{linear} in $m$. 
\end{itemize}

Eq.~\ref{Eq:Lagrangian5} can solved, to obtain $\omega(m)$, $\dot{\omega}(m)$, and $\ddot{\omega}(m)$ as a \emph{explicit} functions in $m$, by supplying the appropriate Internal-Energy-EOS: 

\begin{equation}
    u=u(v,s_0)
\end{equation}

and by supplying the appropriate \emph{boundary conditions}: 

\begin{subequations}\label{Eq:BCAll}
\begin{empheq}[left=\empheqlbrace]{align}
\label{Eq:BC1}
         &\omega(0) = 0, \\
\label{Eq:BC2}
         &\dhyd{\omega}{x}\Bigg\rvert_{m=M} = \frac{1}{\rho_{\text{surface}}}.
\end{empheq}
\end{subequations}

We can also write Eq.~\ref{Eq:Lagrangian4} in an integral form: 

\begin{equation}\label{Eq:Lagrangian6}
    P(m) = P_{\text{center}} - \int_{0}^{m} \Bigg(\frac{4\pi}{81} \Bigg)^{1/3} \cdot \frac{G\cdot m' \cdot dm'}{{[\omega(m')]}^{4/3}}
\end{equation}

Eq.~\ref{Eq:Lagrangian6} can also be understood as an Integro-differential Equation: 

\begin{equation}\label{Eq:Lagrangian7}
    P\Bigg(\bigg(\frac{d\omega}{dm} \bigg)^{-1},s_0 \Bigg) = P_{\text{center}} - \int_{0}^{m} \Bigg(\frac{4\pi}{81} \Bigg)^{1/3} \cdot \frac{G\cdot m' \cdot dm'}{{[\omega(m')]}^{4/3}}
\end{equation}

If the appropriate Pressure-Density-EOS is supplied as:

\begin{equation}\label{Eq:pEOS}
    P=P(\rho,s_0)
\end{equation}

Again, the goal is to solve for $\omega(m)$ that satisfy Eq.~\ref{Eq:Lagrangian6} and the EOS Eq.~\ref{Eq:pEOS}. 

\subsection{Virial Theorem}

Recall that Euler-Lagrange Equation gives: 

\begin{equation}\label{Eq:Lagrangian3a}
    \frac{d}{dm}\Bigg( \underbrace{\frac{\partial \Lagr}{\partial \dot{\omega}}}_{\text{act only on \textbf{Part I}}} \Bigg) = \underbrace{\frac{\partial \Lagr}{\partial \omega}}_{\text{act only on \textbf{Part II}}}
\end{equation}

Multiply Eq.~\ref{Eq:Lagrangian3a} with $\omega$ on both sides, integrate from an arbitrary starting $m_1$ to an arbitrary ending $m_2$, and substitute the independent variable in the integrant as $m'$, we have: 

\begin{equation}\label{Eq:VirialTheorem1}
\int_{m'=m_1}^{m'=m_2} \omega \cdot \frac{d}{dm'}\left( \frac{\partial \Lagr}{\partial \dot{\omega}} \right) \cdot dm' = \int_{m'=m_1}^{m'=m_2} \omega \cdot \frac{\partial \Lagr}{\partial \omega} \cdot dm'
\end{equation}

The \textbf{RHS} of Eq.~\ref{Eq:VirialTheorem1} can be written as: 

\begin{equation}\label{Eq:VirialTheorem2}
\int_{m'=m_1}^{m'=m_2} \omega \cdot \frac{\partial \Lagr}{\partial \omega} \cdot dm' = \int_{m'=m_1}^{m'=m_2} \frac{\partial \Lagr}{\partial \ln{\omega}} \cdot dm'
\end{equation}

The \textbf{LHS} of Eq.~\ref{Eq:VirialTheorem1} can be integrated by parts as: 

\begin{equation}\label{Eq:VirialTheorem3}
\int_{m'=m_1}^{m'=m_2} \omega \cdot d\left( \frac{\partial \Lagr}{\partial \dot{\omega}} \right) = \underbrace{\omega \cdot \left( \frac{\partial \Lagr}{\partial \dot{\omega}} \right) \bigg|_{m'=m_1}^{m'=m_2}}_{\textbf{Part A}} - \underbrace{\int_{m'=m_1}^{m'=m_2} d\omega \cdot \left(  \frac{\partial \Lagr}{\partial \dot{\omega}} \right)}_{\textbf{Part B}}
\end{equation}

According to Eq.~\ref{Eq:Lagrangian3b}, \textbf{Part A} of Eq.~\ref{Eq:VirialTheorem3} is just the product of Pressure and Volume evaluated at the upper and lower limits: 

\begin{equation}\label{Eq:VirialTheorem4}
\underbrace{\omega \cdot \left( \frac{\partial \Lagr}{\partial \dot{\omega}} \right) \bigg|_{m'=m_1}^{m'=m_2}}_{\textbf{Part A}} = P(m_2) \cdot \omega(m_2) - P(m_1) \cdot \omega(m_1)
\end{equation}

\textbf{Part B} of Eq.~\ref{Eq:VirialTheorem3} can be re-written as:

\begin{equation}\label{Eq:VirialTheorem5}
\underbrace{\int_{m'=m_1}^{m'=m_2} d\omega \cdot \left(  \frac{\partial \Lagr}{\partial \dot{\omega}} \right)}_{\textbf{Part B}} = \int_{m'=m_1}^{m'=m_2} dm' \cdot \frac{d\omega}{dm'} \cdot \frac{\partial \Lagr}{\partial \dot{\omega}} = \int_{m'=m_1}^{m'=m_2} dm' \cdot \dot{\omega} \cdot \frac{\partial \Lagr}{\partial \dot{\omega}} = \int_{m'=m_1}^{m'=m_2} \frac{\partial \Lagr}{\partial \ln{\dot{\omega}}} \cdot dm'
\end{equation}

Therefore, combining all of these (Eq.~\ref{Eq:VirialTheorem1}, Eq.~\ref{Eq:VirialTheorem2}, Eq.~\ref{Eq:VirialTheorem3}, Eq.~\ref{Eq:VirialTheorem4}, and Eq.~\ref{Eq:VirialTheorem5}) into a single equation, we have: 

\begin{equation}\label{Eq:VirialTheoremA}
    P(m_2) \cdot \omega (m_2) - P(m_1) \cdot \omega (m_1) = \int_{m'=0}^{m'=m} \Bigg( \underbrace{\frac{\partial}{\partial \ln{\omega}}}_{\text{act only on $\mathbf{\omega}$}} + \underbrace{\frac{\partial}{\partial \ln{\dot{\omega}}}}_{\text{act only on $\mathbf{\dot{\omega}}$}} \Bigg) \Lagr(m';\omega, \dot{\omega}) \cdot dm' 
\end{equation}

Eq.~\ref{Eq:VirialTheoremA} is the general form of \emph{virial theorem}, applicable to any $m_1$ and $m_2$. It is convenient when applied to a single layer that has a uniform composition within a multi-layered planet. At both the upper and lower boundary of this layer with adjacent layers, Pressure is continuous and Volume is continuous, and thus, their product $P(m) \cdot \omega (m)$ must be continuous. 

One special case is to take the integral from center $m'=0$ to surface $m'=M$.\\
$P(m) \cdot \omega (m)$ vanishes at the center since Volume is zero there.\\
$P(m) \cdot \omega (m)$ vanishes at the surface since Pressure is zero there.\\

Then we have: 

\begin{equation}\label{Eq:VirialTheoremB}
    0 = \int_{m'=0}^{m'=M} \Bigg( \underbrace{\frac{\partial}{\partial \ln{\omega}}}_{\text{act only on $\mathbf{\omega}$}} + \underbrace{\frac{\partial}{\partial \ln{\dot{\omega}}}}_{\text{act only on $\mathbf{\dot{\omega}}$}} \Bigg) \Lagr(m';\omega, \dot{\omega}) \cdot dm'
\end{equation}

If, however, there is certain pressure exerted onto the planet surface, due to for example the presence of an envelope, or an external field, then, 

\begin{equation}\label{Eq:VirialTheoremC}
    P_{\text{surface}} \cdot V_{\text{surface}} = \int_{m'=0}^{m'=M} \Bigg( \underbrace{\frac{\partial}{\partial \ln{\omega}}}_{\text{macroscopic}} + \underbrace{\frac{\partial}{\partial \ln{\dot{\omega}}}}_{\text{microscopic}} \Bigg) \Lagr(m';\omega, \dot{\omega}) \cdot dm'
\end{equation}

As denoted above, part of the operator gives the macroscopic interaction, that is gravitation, and the other part of the operator gives the microscopic interaction, that is elasticity. 

The \emph{virial theorem} can be viewed as a global relationship between the \emph{elastic} energy and the \emph{gravitational} energy of the bound object--\emph{planet}. %It is no more than a global manifestation of the \emph{variational principle}. 

The form of Eq.~\ref{Eq:VirialTheoremA} invokes that perhaps we should re-cast the \emph{boundary conditions} in a more symmetrical way as follows: 

\begin{subequations}\label{Eq:VirialBoundaryConditions}
\begin{empheq}[left=\empheqlbrace]{align}
    \label{Eq:VirialBoundaryCondition1}
         &P(0) \cdot \omega(0) = \omega \cdot \frac{\partial \Lagr}{\partial \dot{\omega}} \bigg|_{m'=0} =0,\\
    \label{Eq:VirialBoundaryCondition2}
         &P(M) \cdot \omega(M) = \omega \cdot \frac{\partial \Lagr}{\partial \dot{\omega}} \bigg|_{m'=M} =0.
\end{empheq}
\end{subequations}

\subsubsection{Relation to Similarity Transformation}

This result (Eq.~\ref{Eq:VirialTheoremB}) can also be viewed from the \emph{variational principle} itself, by considering a small variation of the total action $\mathcal{S}$ about the equilibrium: 

\begin{equation}\label{eq:virial05}
0 = \delta \mathcal{S} = \int_{m'=0}^{m'=M} \left( \frac{\partial \Lagr}{\partial \omega} \cdot \delta\omega + \frac{\partial \Lagr}{\partial \dot{\omega}} \cdot \delta\dot{\omega} \right) \cdot dm'
\end{equation}

Let's consider a particular kind of variation: the \emph{similarity} variation as $\delta \omega = \alpha \cdot \omega$. Here we take $\alpha$ as a small number (and a constant), then $\delta \dot{\omega} = \alpha \cdot \dot{\omega}$. 

It satisfies the first boundary condition of $\omega(0)=0$ at $m=0$ automatically. 

But it seems to violate the second boundary condition of $\dot{\omega}(\text{M})=1/{\rho_{\text{surface}}}$ at $m=M$ with this proportional variation or \emph{similarity transformation}. 

However, noticing that the pressure $P=\frac{\partial L}{\partial \dot{\omega}}$ is zero at the surface, so the effect of this variation vanishes at the surface ($m=M$) also. Therefore, by adopting this particular choice of $\delta \omega$, the same conclusion is reached:  

\begin{equation}\label{eq:virial06}
0 = \delta \mathcal{S} = \int_{m'=0}^{m'=M} \left( \frac{\partial \Lagr}{\partial \omega} \alpha \cdot \omega + \frac{\partial \Lagr}{\partial \dot{\omega}} \alpha \cdot \dot{\omega} \right) dm' = \alpha \cdot \int_{m'=0}^{m'=M} \left( \frac{\partial \Lagr}{\partial \ln\omega}+ \frac{\partial \Lagr}{\partial \ln\dot{\omega}} \right) dm'
\end{equation}

From this perspective, \emph{virial theorem} can be viewed as a special case or a direct consequence of the \emph{variational principle} (stationary action principle) itself. 

The same argument of \emph{similarity transformation} has been invoked by R.P. Feynman to derive the \emph{virial theorem} for the generalized \emph{Thomas-Fermi} model of atoms~\citep{Feynman1949}. 

\subsection{Normal Modes}

Now, if we consider the adiabatic radial motions (or pulsations) of a planetary body, we could also formulate it in the \emph{variational principle}. This time, we need to include kinetic energy of motion in the Lagrangian $\Lagr$. 

Here, for convenience, we denote partial-derivative relative to mass $\frac{\partial}{\partial m}$ as subscript $_{m}$, and partial-derivative relative to time $\frac{\partial}{\partial t}$ as subscript $_{t}$. 

Now, the volume $\omega$ not only depends on mass $m$, but also depends on time $t$, to account for motions (in the radial direction): 

\begin{equation}
    \omega = \omega \bigg( \underbrace{m, t}_{\text{now two independent (natural) variables}} \bigg)
\end{equation}

Correspondingly, the Lagrangian $\Lagr$ becomes:

\begin{equation}
    \Lagr = \Lagr \bigg( \underbrace{m, t}_{\text{now two independent (natural) variables}}; \omega, \omega_{m}, \omega_{t} \bigg)
\end{equation}

The full action \emph{functional} $S$ that we want to extremize with respect to $\omega(m,t)$ is:

\begin{equation}
    \mathcal{S}[\omega] = \iint \Lagr \bigg( m,t ; \omega, \omega_{m}, \omega_{t} \bigg) \cdot dm \cdot dt
\end{equation}

More specifically, in terms of the lower and upper limits of the integral:

\begin{equation}
    \mathcal{S}[\omega] = \int_{t=t_a}^{t=t_b} \int_{m=0}^{m=M} \Lagr \bigg( m,t ; \omega, \omega_{m}, \omega_{t} \bigg) \cdot dm \cdot dt
\end{equation}

where $t_a$ and $t_b$ are arbitrary starting and ending time moments. 

The \emph{variational principle} is understood as:

\begin{equation}
    \delta \mathcal{S}[\omega] = 0 
\end{equation}

The specific kinetic energy $\mathcal{T}$ (in the radial direction) anywhere inside the planet, that is, energy of radial motion per unit mass: 

\begin{equation}
    \mathcal{T} = \frac{1}{2} \cdot \Bigg( \frac{\partial r}{\partial t} \Bigg)^2
\end{equation}

Recall the definition of $\omega$ earlier, we have: 

\begin{equation}
    r = \bigg( \frac{3}{4\pi} \bigg)^{1/3} \cdot \omega^{1/3}
\end{equation}

so, $\mathcal{T}$ can be expressed in terms of $\omega$, and its partial-derivative with respect to time $t$, that is, $\omega_t = \frac{\partial \omega}{\partial t}$

\begin{align}
    \mathcal{T} &= \frac{1}{2} \cdot \Bigg( \frac{\partial}{\partial t} r \Bigg)^2 \\\nonumber
                &= \frac{1}{2} \cdot \bigg( \frac{3}{4\pi} \bigg)^{2/3} \cdot \Bigg( \frac{\partial}{\partial t} \bigg( \omega^{1/3} \bigg) \Bigg)^2 \\\nonumber
                &=\frac{1}{2} \cdot \bigg( \frac{3}{4\pi} \bigg)^{2/3} \cdot \Bigg( \frac{1}{3} \cdot \omega^{-2/3} \cdot \omega_t \Bigg)^2 \\\nonumber
\end{align}

The overall Lagrangian $\Lagr$ is classically defined as the difference between the kinetic energy term $\mathcal{T}$ and the potential energy term $\mathcal{U}$ (that is why initially we keep a negative sign ($-$) in front of $\mathcal{U}$, in order to be consistent with the definition here): 

\begin{align}\label{Eq:EulerLagrange1}
    \Lagr &= \mathcal{T} - \mathcal{U} \\\nonumber
      &= \frac{1}{2} \cdot \bigg( \frac{3}{4\pi} \bigg)^{2/3} \cdot \Bigg( \frac{1}{3} \cdot \omega^{-2/3} \cdot \omega_t \Bigg)^2 -\Bigg( u(\omega_m)-\frac{G \cdot m}{\bigg(\frac{3}{4\pi}\bigg)^{\frac{1}{3}} \cdot {\omega}^{\frac{1}{3}}}\Bigg) \\\nonumber
       &= \frac{1}{18} \cdot \bigg( \frac{3}{4\pi} \bigg)^{2/3} \cdot \omega^{-4/3} \cdot \omega_t^2 -\Bigg( u(\omega_m)-\frac{G \cdot m}{\bigg(\frac{3}{4\pi}\bigg)^{\frac{1}{3}} \cdot {\omega}^{\frac{1}{3}}}\Bigg) \\\nonumber
\end{align}

Now, the Euler-Lagrange Equation becomes: 

\begin{equation}\label{Eq:EulerLagrange2}
    \frac{\partial \Lagr}{\partial \omega} - \frac{\partial}{\partial t} \bigg( \frac{\partial \Lagr}{\partial \omega_t} \bigg) - \frac{\partial}{\partial m} \bigg( \frac{\partial \Lagr}{\partial \omega_m} \bigg) = 0
\end{equation}

Eq.~\ref{Eq:EulerLagrange2} will eventually lead to a \emph{wave equation}, where we can solve for its characteristic frequencies (\emph{eigenvalues}) and characteristic solutions (\emph{eigenfunctions}). 

Plugging in $\Lagr$ from Eq.~\ref{Eq:EulerLagrange1}, we have: 

\begin{align}\label{Eq:EulerLagrange3}
    \frac{\partial \Lagr}{\partial \omega} &= -\frac{2}{27} \cdot \bigg( \frac{3}{4\pi} \bigg)^{2/3} \cdot \omega^{-7/3} \cdot \omega_t^2 - \frac{1}{3} \cdot \bigg(\frac{3}{4\pi}\bigg)^{-1/3} \cdot G \cdot m \cdot {\omega}^{-4/3} \\\nonumber
    \frac{\partial \Lagr}{\partial \omega_{t}} &= \frac{1}{9} \cdot \bigg( \frac{3}{4\pi} \bigg)^{2/3} \cdot \omega^{-4/3} \cdot \omega_t \\\nonumber
    \frac{\partial}{\partial t} \bigg( \frac{\partial \Lagr}{\partial \omega_t} \bigg) &= -\frac{4}{27} \cdot \bigg( \frac{3}{4\pi} \bigg)^{2/3} \cdot \omega^{-7/3} \cdot \omega_t^2 + \frac{1}{9} \cdot \bigg( \frac{3}{4\pi} \bigg)^{2/3} \cdot \omega^{-4/3} \cdot \omega_{tt} \\\nonumber
    \frac{\partial \Lagr}{\partial \omega_{m}} &=  -\frac{\partial u(\omega_{m})}{\partial \omega_{m}} = -u'(\omega_{m}) = P (\omega_{m}) \\\nonumber
    \frac{\partial}{\partial m} \bigg( \frac{\partial \Lagr}{\partial \omega_m} \bigg) &= -u''(\omega_{m}) \cdot \omega_{mm} = \frac{\partial P}{\partial m} = \bigg( \frac{\partial P(\omega_{m})}{\partial \omega_m} \bigg) \cdot \omega_{mm} \\\nonumber
\end{align}

Since the first partial-derivative of $\omega$ relative to $m$ is the reciprocal local density $\rho(m)$ at $m$: 

\begin{equation}
    \omega_m \equiv \frac{\partial \omega}{\partial m} =\frac{1}{\rho(m)} = \frac{1}{\rho}
\end{equation}

Then, we can see how $u''$ is related with the material EOS properties, in particular, the isentropic bulk modulus $K_s$ and the isentropic sound speed $c_s$: 

\begin{align}
    u''(\omega_m) &= -\frac{\partial P(\omega_{m})}{\partial \omega_m} = -\frac{\partial P(\rho)}{\partial (1/\rho)} = \rho^2 \cdot \frac{\partial P(\rho)}{\partial \rho} = \rho \cdot \frac{\partial P(\rho)}{\partial \ln{\rho}} = \rho \cdot K_s (\rho) \nonumber \\
    &= \rho^2 \cdot \Bigg( \frac{K_s}{\rho} \Bigg) = \rho^2 \cdot \Bigg( \sqrt{\frac{K_s}{\rho}} \Bigg)^2 = \rho^2 \cdot c_s^2 = \frac{c_s^2}{\omega_m^2} >0 \label{Eq:soundspeed}
\end{align}

Here, $K_s \equiv \big( \frac{\partial P(\rho)}{\partial \ln{\rho}} \big)_{s}$ is the local isentropic bulk modulus that we have defined earlier. $c_s \equiv \sqrt{\frac{K_s}{\rho}}$ is the local isentropic \emph{sound speed} in a fluid where shear modulus is zero (a fluid does not sustain shear stress). In general, both $K_s$ and $c_s$ are functions of density and can be calculated for a given EOS. 

Furthermore, $u''$ is related with the \emph{second} adiabatic index $\Gamma$: 

\begin{equation}
    u''(\omega_m) = -\frac{\partial P(\omega_{m})}{\partial \omega_m} = \rho^2 \cdot \frac{P}{\rho} \cdot \Bigg( \frac{\partial \ln{P}}{\partial \ln{\rho}} \Bigg)_{s} = \rho \cdot P \cdot \Gamma
\end{equation}

In order to do that, let's now apply the \emph{perturbation theory}, and consider only small motions deviating from the stationary state. Suppose we can decompose $\omega$ into two parts: the time-independent stationary part $\omega_0$ corresponding to the un-perturbed state (0-state), and the time-dependent perturbation $\delta \omega$ which depends on both $m$ and $t$. 

Furthermore, we apply \emph{separation of variables} to separate the perturbation $\delta \omega$ into a time-independent amplitude $\psi(m)$, and a time-dependent oscillatory part $e^{i \sigma t}$: 
\begin{align}
    \omega (m,t) &= \omega_0 + \delta\omega \nonumber\\
                 &= \omega_0 (m) + \delta\omega (m,t) \nonumber\\
                &= \omega_0 (m) + \psi (m) \cdot \exp{(i \sigma t)} \nonumber\\
                &= \omega_0 (m) \cdot \bigg( 1 + \zeta (m) \cdot \exp{(i \sigma t)} \bigg)
    \label{Eq:Perturbation}
\end{align}

Generally speaking, both $\zeta$ and $\sigma$ can be \emph{complex} numbers, in order to accommodate phase shift in superposing different normal modes. 

$\sigma$ is the (complex) angular frequency of the oscillation. 

$\zeta$ is the (complex) fractional amplitude of the oscillation. It is dimensionless and $\left| \zeta \right| \ll 1$ for small perturbation. It vanishes at the center $\zeta(0)=0$ due to spherical symmetry. 

On the other hand, $\psi$ has the dimension of Volume. 

Therefore, we can write out all the partial-derivatives of $\omega$ in $m$ and $t$ as: 

\begin{equation}
    \omega_t = \omega_0 (m) \cdot \zeta (m) \cdot (i \sigma) \cdot \exp{(i \sigma t)}
\end{equation}
\begin{align}
    \omega_{tt} &= \omega_0 (m) \cdot \zeta (m) \cdot (i \sigma)^2 \cdot \exp{(i \sigma t)} \nonumber \\
                &= \omega_0 (m) \cdot \zeta (m) \cdot (-\sigma^2) \cdot \exp{(i \sigma t)} 
\end{align}

\begin{align}
    \omega_m &= \omega_{0,m} + \bigg(\omega_{0,m} \cdot \zeta + \omega_0 \cdot \zeta_m \bigg) \cdot \exp{(i \sigma t)} \nonumber \\
             &= \omega_{0,m} + \bigg(\omega_{0} \cdot \zeta \bigg)_{m} \cdot \exp{(i \sigma t)} 
\end{align}

\begin{align}
    \omega_{mm} &= \omega_{0,mm} + \bigg(\omega_{0,mm} \cdot \zeta + 2 \cdot \omega_{0,m} \cdot \zeta_m + \omega_0 \cdot \zeta_{mm} \bigg) \cdot \exp{(i \sigma t)} \nonumber \\
                &= \omega_{0,mm} + \bigg(\omega_{0} \cdot \zeta \bigg)_{mm} \cdot \exp{(i \sigma t)}
\end{align}

Since $\left| \zeta \right| \ll 1$, we can neglect any term of the order $\zeta^2$ or higher, thus, we don't need to be worried about the $\omega_{t}^2$ term. Moreover, we know that $\omega_0 (m)$ satisfies the stationary-state Euler-Lagrange Equation. 

By only keeping the term up to linear in $\zeta$, we have (from Eq.~\ref{Eq:EulerLagrange3}): 

\begin{align}\label{Eq:EulerLagrange4}
    \frac{\partial \Lagr}{\partial \omega} &\approx -\frac{1}{3} \cdot \bigg(\frac{3}{4\pi}\bigg)^{-1/3} \cdot G \cdot m \cdot {\omega_0}^{-4/3} \cdot \bigg(1 -\frac{4}{3} \cdot \zeta \cdot e^{i \sigma t} \bigg) \\\nonumber
    \frac{\partial}{\partial t} \bigg( \frac{\partial \Lagr}{\partial \omega_t} \bigg) &\approx \frac{1}{9} \cdot \bigg( \frac{3}{4\pi} \bigg)^{2/3} \cdot \omega_0^{-1/3} \cdot (-\sigma^2) \cdot \zeta \cdot e^{i \sigma t} \\\nonumber
    \frac{\partial}{\partial m} \bigg( \frac{\partial \Lagr}{\partial \omega_m} \bigg) &\approx  -u''\Big( \omega_{0,m} + \big(\omega_{0} \cdot \zeta \big)_{m} \cdot e^{i \sigma t} \Big) \cdot \Bigg( \omega_{0,mm} + \big(\omega_0 \cdot \zeta \big)_{mm} \cdot e^{i \sigma t} \Bigg) \\\nonumber
        &\approx - \Bigg( u''\Big( \omega_{0,m} \Big) + u'''\Big( \omega_{0,m} \Big) \cdot \big(\omega_{0} \cdot \zeta \big)_{m} \cdot e^{i \sigma t} \Big) \Bigg) \cdot \Bigg( \omega_{0,mm} + \big(\omega_0 \cdot \zeta \big)_{mm} \cdot e^{i \sigma t} \Bigg) \\\nonumber
        &\approx - \Bigg( u''\Big( \omega_{0,m} \Big) \cdot \omega_{0,mm} +  \Big[ u''\Big( \omega_{0,m} \Big) \cdot \big(\omega_{0} \cdot \zeta \big)_{mm}  + \cdot u'''\Big( \omega_{0,m} \Big) \cdot \big(\omega_{0} \cdot \zeta \big)_{m} \cdot \omega_{0,mm} \Big] \cdot e^{i \sigma t} \Bigg) \\\nonumber
        &\approx - \Bigg( u''\Big( \omega_{0,m} \Big) \cdot \omega_{0,mm} +  \Big[ u''\Big( \omega_{0,m} \Big) \cdot \big(\omega_{0} \cdot \zeta \big)_{m} \Big]_{m} \cdot e^{i \sigma t} \Bigg)
\end{align}

Now, let's re-consider the stationary-state equality from Eq.~\ref{Eq:Lagrangian4}: 

\begin{equation}\label{Eq:EulerLagrange5}
    u''(\omega_{0,m}(m)) \cdot\omega_{0,mm}(m) = \Bigg[ \frac{1}{3} \cdot \bigg(\frac{3}{4\pi} \bigg)^{-1/3} \cdot G \cdot m \Bigg] \cdot (\omega_0(m))^{-4/3}
\end{equation}

It can also be written as: 

\begin{equation}\label{Eq:EulerLagrange5b}
    u'' = \Bigg[ \frac{1}{3} \cdot \bigg(\frac{3}{4\pi} \bigg)^{-1/3} \cdot G \cdot m \Bigg] \cdot \frac{\omega_0^{-4/3}}{\omega_{0,mm}}
\end{equation}

or, 

\begin{equation}\label{Eq:EulerLagrange6}
    u''  = \Bigg[ \frac{1}{3} \cdot \bigg(\frac{3}{4\pi} \bigg)^{-1/3} \cdot G \cdot m \Bigg] \cdot \omega_0^{-4/3} \Bigg/ \Bigg(\dnhyd[2]{\omega_0}{m} \Bigg)
\end{equation}

or, 

\begin{equation}\label{Eq:EulerLagrange7}
    c_s^2 = \Bigg[ \frac{1}{3} \cdot \bigg(\frac{3}{4\pi} \bigg)^{-1/3} \cdot G \cdot m \Bigg] \cdot \omega_0^{-4/3} \cdot \Bigg(\dhyd{\omega_0}{m} \Bigg)^2 \Bigg/ \Bigg(\dnhyd[2]{\omega_0}{m}  \Bigg)
\end{equation}

By supplying an appropriate EOS, such as in the form $u = u(v) = u (1/\rho)$, or $P=P(v)=P(1/\rho)$, or $c_s = c_s(v) = c_s(1/\rho)$, the complete solution of Eq.~\ref{Eq:EulerLagrange7} will give all of the followings as \emph{explicit} functions in $m$: 

\begin{itemize}
    \item $\omega_0=\omega_0 (m)$
    \item $\omega_{0,m}(m) =\dhyd{\omega_0}{m}$
    \item $\omega_{0,mm}(m) =\dnhyd[2]{\omega_0}{m}$
    \item $u[m] = u((\omega_{0,m}(m)))$
    \item $u'[m] = u'((\omega_{0,m}(m))) = -P[m]$
    \item $u''[m] = u''((\omega_{0,m}(m))) = (c_s[m])^2/\omega_{0,m}^2$
\end{itemize}

after removing the stationary-state equality, Eq.~\ref{Eq:EulerLagrange2} becomes: 

\begin{align}
    0 &= \frac{4}{9} \cdot \bigg(\frac{3}{4\pi}\bigg)^{-1/3} \cdot G \cdot m \cdot {\omega_0}^{-4/3} \cdot \zeta \cdot e^{i \sigma t} \nonumber \\
      &+ \frac{1}{9} \cdot \bigg( \frac{3}{4\pi} \bigg)^{2/3} \cdot \omega_0^{-1/3} \cdot (\sigma^2) \cdot \zeta \cdot e^{i \sigma t} \nonumber \\
      &+ \Big[ u''\Big( \omega_{0,m} \Big) \cdot \big(\omega_{0} \cdot \zeta \big)_{m} \Big]_{m} \cdot e^{i \sigma t} \label{Eq:EulerLagrange8}
\end{align}

Removing the common factor of $e^{i \sigma t}$, that is, completely removing the time-dependence and turn the equation into an ordinary differential equation only in variable $m$, and re-arranging the terms, now we have: 

\begin{equation}\label{Eq:EulerLagrange9}
    \frac{d}{dm} \Bigg(u'' \cdot  \frac{d(\omega_0 \cdot \zeta)}{dm} \Bigg) + \Bigg[\frac{4}{9} \cdot \bigg(\frac{3}{4\pi}\bigg)^{-1/3} \cdot G \cdot m \cdot {\omega_0}^{-7/3} + (\sigma^2) \cdot \frac{1}{9} \cdot \bigg( \frac{3}{4\pi} \bigg)^{2/3} \cdot \omega_0^{-4/3} \Bigg] \cdot (\omega_0 \cdot \zeta) = 0
\end{equation}

Eq.~\ref{Eq:EulerLagrange9} is a \emph{Sturm-Liouville} form. 

Now if we divide both sides Eq.~\ref{Eq:EulerLagrange9} with $u''$, and substitute in the stationary-state identity of Eq.~\ref{Eq:EulerLagrange5b}: 

\begin{equation}\label{Eq:EulerLagrange9b}
    \frac{1}{u''} \cdot \frac{d}{dm} \Bigg(u'' \cdot  \frac{d(\omega_0 \cdot \zeta)}{dm} \Bigg) 
    + \Bigg(\frac{4}{3} \cdot \frac{\omega_{0,mm}}{\omega_0} + \frac{\sigma^2}{4\pi} \cdot \frac{\omega_{0,mm}}{G\cdot m} \Bigg) \cdot (\omega_0 \cdot \zeta) = 0
\end{equation}

or equivalently, if we factor out $\omega_{0,mm}/\omega_0$ term:

\begin{equation}\label{Eq:EulerLagrange9c}
    \frac{1}{u''} \cdot \frac{d}{dm} \Bigg(u'' \cdot  \frac{d(\omega_0 \cdot \zeta)}{dm} \Bigg) 
    + \Bigg( \frac{\omega_{0,mm}}{\omega_0} \Bigg) \cdot \Bigg( \frac{4}{3}  + \frac{\sigma^2}{4\pi} \cdot \frac{\omega_{0}}{G\cdot m} \Bigg) \cdot (\omega_0 \cdot \zeta) = 0
\end{equation}

Now, we realize that the ratio $m/\omega_0$ is just the unperturbed mean density ($\overbar{\rho}$) enclosed within $m$. The term $4\pi G \overbar{\rho}$ reminds us of the \emph{dynamical timescale}, and thus, we can define a new frequency parameter $\kappa$ as: 

\begin{equation}\label{Eq:Kappa}
   \kappa(m) \equiv \sqrt{\frac{4\pi G}{3} \cdot \frac{m}{\omega_0(m)} } = \sqrt{\frac{4\pi G}{3} \cdot \overbar{\rho}} = \sqrt{\ddfrac{G\cdot m}{r^3}}
\end{equation}

$\kappa=\kappa(m)$ decreases from center outward because $\rho$ or $\overbar{\rho}$ decreases from center outward. 
$\kappa$ is in fact the angular frequency of a circular orbit at radius $r$ around mass $m$. 

Alternatively, one can also express $\omega_0(m)$ in terms of $\kappa(m)$: 

\begin{equation}\label{Eq:Kappa1}
   \omega_0(m) = \frac{4\pi G}{3} \cdot \Bigg( \frac{m}{\kappa(m)^2} \Bigg)
\end{equation}

Therefore, Eq.~\ref{Eq:Lagrangian4} can be written as: 

\begin{equation}\label{Eq:Kappa2}
   u'' \cdot \omega_{0,mm} = -\dhyd{P}{m} = \frac{1}{4\pi} \cdot \Bigg( \frac{\kappa(m)^{8/3}}{(G \cdot m)^{1/3}} \Bigg)
\end{equation}

and, 

\begin{equation}\label{Eq:Kappa3}
  \frac{u'' \cdot \omega_{0,mm}}{\omega_0(m)}  = \frac{3}{16 \pi^2} \cdot \Bigg( \frac{\kappa(m)^{14/3}}{(G \cdot m)^{4/3}} \Bigg)
\end{equation}

We can re-write Eq.~\ref{Eq:EulerLagrange9c} as: 

\begin{equation}\label{Eq:EulerLagrange9d}
    \frac{1}{u''} \cdot \frac{d}{dm} \Bigg(u'' \cdot  \frac{d(\omega_0 \cdot \zeta)}{dm} \Bigg) 
    + \Bigg( \frac{\omega_{0,mm}}{\omega_0} \Bigg) \cdot \Bigg( \frac{4}{3} + \frac{\sigma^2}{3 \kappa^2} \Bigg) \cdot (\omega_0 \cdot \zeta) = 0
\end{equation}

We recognize that $(\omega_0 \cdot \zeta)$ is just the (complex) amplitude $\psi$ of the volume perturbation defined earlier: 

\begin{equation}\label{Eq:EulerLagrange9e}
    \frac{1}{u''} \frac{d}{dm} \Bigg(u'' \frac{d\psi}{dm} \Bigg) 
    + \Bigg( \frac{\omega_{0,mm}}{\omega_0} \Bigg) \cdot \frac{1}{3} \cdot \Bigg(4 + \frac{\sigma^2}{\kappa^2} \Bigg) \cdot \psi = 0
\end{equation}

If we multiply both sides Eq.~\ref{Eq:EulerLagrange9e} with $(u'')$ to get back to the original equation, we have:

\begin{equation}\label{Eq:EulerLagrange9e1}
\frac{d}{dm} \Bigg(u'' \frac{d\psi}{dm} \Bigg) 
    + \Bigg( \frac{u'' \cdot \omega_{0,mm}}{\omega_0} \Bigg) \cdot \frac{1}{3} \cdot \Bigg(4 + \frac{\sigma^2}{\kappa^2} \Bigg) \cdot \psi = 0
\end{equation}

which, in the language of $\kappa$, becomes: 

\begin{equation}\label{Eq:EulerLagrange9e2}
    \frac{d}{dm} \Bigg(u'' \frac{d\psi}{dm} \Bigg) 
    + \frac{1}{16 \pi^2} \cdot \Bigg( \frac{\kappa(m)^{14/3}}{(G \cdot m)^{4/3}} \Bigg) \cdot \Bigg(4 + \frac{\sigma^2}{\kappa(m)^2} \Bigg) \cdot \psi = 0
\end{equation}

The advantage of expressing everything in $\kappa$ is that $\kappa$ is positive finite and monotonically decreases (slowly) from center outward. $u''=\rho P \Gamma = \rho K_s = \rho^2 c_s^2$ is positive finite and monotonically decreases (slowly) from center outward. 

Consequently, we have from Eq.~\ref{Eq:EulerLagrange9e} and Eq.~\ref{Eq:EulerLagrange9e2}:

\begin{equation}\label{Eq:EulerLagrange9e3}
    \frac{1}{u''} \frac{d}{dm} \Bigg(u'' \frac{d\psi}{dm} \Bigg) 
    + \underbrace{\frac{1}{16 \pi^2 u''} \cdot \Bigg( \frac{\kappa(m)^{14/3}}{(G \cdot m)^{4/3}} \Bigg) \cdot \Bigg(4 + \frac{\sigma^2}{\kappa(m)^2} \Bigg)}_{\text{has dimension of 1/Mass$^2$}} \cdot \psi = 0
\end{equation}

We can define a new parameter $\lambda$. To the leading order, $\lambda \sim m^{2/3}$. Thus, we can write: $\lambda(m) = a(m) \cdot m^{2/3}$, where $a(m) \equiv 4 \pi \sqrt{u''} \cdot \Bigg( \frac{(G)^{2/3}}{\kappa(m)^{7/3}} \Bigg) \cdot \Bigg( \frac{1}{\sqrt{4 + \frac{\sigma^2}{\kappa(m)^2}}} \Bigg)$, where $a(m)$ is positive finite. Near the center of planet, where both $u''$ and $\kappa$ are approximately constants, then $a(m)\approx$const, which has the dimension of Mass:

\begin{equation}\label{Eq:lambda}
   \lambda(m) \equiv 4 \pi \sqrt{u''} \cdot \Bigg( \frac{(G \cdot m)^{2/3}}{\kappa(m)^{7/3}} \Bigg) 
   \cdot 
   \Bigg( \frac{1}{\sqrt{4 + \frac{\sigma^2}{\kappa(m)^2}}} \Bigg)
\end{equation}

Then, Eq.~\ref{Eq:EulerLagrange9e3} can be re-written as: 

\begin{equation}\label{Eq:EulerLagrange9e4}
    \frac{1}{u''} \frac{d}{dm} \Bigg(u'' \frac{d\psi}{dm} \Bigg) 
    + \frac{1}{\lambda(m)^2} \cdot \psi = 0
\end{equation}

\begin{comment}
\begin{equation}\label{Eq:lambda0}
   \lambda \sim m^{2/3}
\end{equation}

Thus, we can write: 

\begin{equation}\label{Eq:lambda1}
   \lambda(m) = a(m) \cdot m^{2/3}
\end{equation}
\end{comment}

%$a(m)$ has the has the dimension of Mass$^{1/3}$. 

One can think of $\lambda$ as the (local) wavelength in the mass dimension. 
That is, after $\sim\lambda/2$ in the $m$-coordinate, the volume perturbation $\delta \omega$ changes sign from ($-$) or ($+$) or vice versa. Recall that $\sqrt{u''}= \rho c_s$ where $\rho$ is the local density, and $c_s$ is the local \emph{effective} sound speed, which measures how fast a perturbation (either in density or pressure) signal propagates in a medium. Then, 

\begin{equation}\label{Eq:lambda3}
   \frac{\lambda}{2} = 2 \pi \rho c_s \cdot \Bigg( \frac{(G \cdot m)^{2/3}}{\kappa(m)^{7/3}} \Bigg) \cdot \Bigg(\frac{1}{\sqrt{4 + \frac{\sigma^2}{\kappa(m)^2}}} \Bigg)
\end{equation}

Remember that $\sigma$ can be a \emph{complex number}, in particular, if it is \emph{imaginary}, i.e., multiple of $i$, then it means exponential growth (multiple of $-i$) or decay (multiple of $+i$) in the system (see Eq.~\ref{Eq:Perturbation}). That suggests the possibility of emergence of structure in the system. 

However, if $\sigma$ is purely \emph{imaginary}, $\abs{\sigma}$ cannot exceed $2\kappa$, in order to make $\sqrt{4 + \frac{\sigma^2}{\kappa(m)^2}}$ well-defined. This can be interpreted as that the gravitational growth or collapse cannot happen faster than the \emph{dynamical timescale}, as we expect. 

One special situation is when $\sigma=0$, then the perturbation is stable and neither grows or decays nor oscillates with time. When this happens: 

\begin{equation}\label{Eq:lambda4}
   \frac{\lambda}{2} = \pi \rho c_s \cdot \frac{(G \cdot m)^{2/3}}{\kappa^{7/3}} 
\end{equation}

This may be interpreted as the \emph{characteristic} mass units that emerge out of the total system mass $m$ of a gravitationally-bound system with signal speed $c_s$, if the whole system is only \emph{slowly} evolving and structures emerge \emph{slowly}, compared to the \emph{dynamical timescale} of $(1/\kappa)$. It may help explain the formation of ring-like structures in proto-planetary disk, as those structures are stable over at least multiple orbits. Our derivation is general. 

If things are sticky on the small-scale, then it may help form structures on the large-scale, especially true in the molecular clouds when the initial density is small. Stickiness is manifested by the molecular forces of cohesion and can be characterized by a negative internal pressure term. This cohesion would help form structures. Van der Waal force, though not very strong, acting over a large ensemble, may become significant. That $\sqrt{u''}$ in $\lambda$ may as well be $\sqrt{-u''}$, and then the sound speed $c_s$ represents how fast the stickiness propagates. :)

Generally, we should think of $u''$ as some kind of coupling at local scale that either resist or assist collapse. The critical molar density, for the stickiness (coupling) to take effect, as we will show later, is $\sim0.01$ mol/cc (or $\sim0.001$ mol/cc at lower temperature). Thus, structures not only emerge from top down, but may also emerge from bottom up. 

%Remember that $\rho$ is the local density, while $\kappa$ is related with the global mean density and mass. 
%\emph{Schr\"{o}dinger-like

%%%%%%%%%%%%%%%%%%%%%%%%%%%%%%%%%%%%%%%%%%%%%%%%%%%%%%%%%%%%%%%%%%%%%%%%%%%%%
\subsection{Derivation of Wave Equation}
If we multiply both sides Eq.~\ref{Eq:EulerLagrange9e4} with $(u'')^2$, we have:

\begin{equation}\label{Eq:EulerLagrange9f}
    u'' \frac{d}{dm} \Bigg(u'' \frac{d\psi}{dm} \Bigg) 
    + \frac{u''^2}{\lambda(m)^2} \cdot \psi = 0
\end{equation}

Now we can define a new coordinate $x$ to replace $m$: 

\begin{equation}\label{Eq:SubstitutingVariables}
    \begin{aligned}
        dx &\equiv dm/u'',\\
        x &\equiv \int \frac{dm}{u''} = \int \frac{dm}{\rho \cdot K_s} = \int \frac{d\omega_0}{K_s}\\
    \end{aligned}
\end{equation}

Then, Eq.~\ref{Eq:EulerLagrange9f} transforms into a differential equation in $x\equiv \int dm/u''$: 

\begin{equation}\label{Eq:EulerLagrange10}
    \dnhyd[2]{\psi}{x} + \frac{1}{(\lambda / u'')^2} \cdot \psi = 0
\end{equation}

\begin{comment}
\begin{equation}\label{Eq:EulerLagrange10}
    \dnhyd[2]{\psi}{x} + \underbrace{\frac{u''}{16 \pi^2} \cdot \Bigg( \frac{\kappa(m)^{14/3}}{(G \cdot m)^{4/3}} \Bigg) \cdot \Bigg(4 + \frac{\sigma^2}{\kappa(m)^2} \Bigg)}_{\text{can be expressed as function in $x$}} \cdot \psi = 0
\end{equation}
\end{comment}

Eq.~\ref{Eq:EulerLagrange10} resembles the time-independent \emph{Schr\"{o}dinger Equation}. $(\lambda / u'')$ can be thought of as the new (local) wavelength measured in the new coordinate $x$. Generally speaking, $(\lambda / u'')$ is not constant and varies with $x$. 

The new coordinate $x$ actually has its physical meaning: $dx$ is the volume element $d\omega$ \emph{modulated} by the local bulk modulus $K_s$, in the un-perturbed state (0-state). It is the \emph{natural coordinate} viewed by a wave propagating in the interior of a planetary body. In future, we will emphasize the importance of \emph{natural coordinate}, and when using it, greatly simplifies a physical problem. In fact, the least action principle with its \emph{action} $\mathcal{S}$ and \emph{Lagrangian} $\Lagr$ can be expressed in whichever coordinate more convenient for calculation. 

$x$ has the dimension of Volume/Pressure. It ranges from 0 to $x_M$
\begin{equation}
    x_M \equiv \int_{m=0}^{m=M} \frac{dm}{u''}
\end{equation}

Then, the whole problem is reduced to solving this \emph{wave equation} (Eq.~\ref{Eq:EulerLagrange9e4} or Eq.~\ref{Eq:EulerLagrange10}) for its characteristic frequencies $\sigma$ (\emph{eigenvalues}) and corresponding characteristic solutions $\psi(m)$ or $\psi(x)$ (\emph{eigenfunctions}).

$\sigma$ is quantized, that is, $\sigma$ can only take discrete values, because of the \emph{boundary conditions}. 

For Eq.~\ref{Eq:EulerLagrange9e4}, the \emph{boundary conditions} are: 

\begin{subequations}\label{Eq:BoundaryConditions1All}
\begin{empheq}[left=\empheqlbrace]{align}
\label{Eq:BoundaryCondition1a}
         &\psi(m=0) = 0, \\
\label{Eq:BoundaryCondition1b}
         &\dhyd{\psi}{m}\Bigg\rvert_{m=M} = 0.
\end{empheq}
\end{subequations}

For Eq.~\ref{Eq:EulerLagrange10}, the \emph{boundary conditions} are: 

\begin{subequations}\label{Eq:BoundaryConditions2All}
\begin{empheq}[left=\empheqlbrace]{align}
\label{Eq:BoundaryCondition2a}
         &\psi(x=0) = 0, \\
\label{Eq:BoundaryCondition2b}
         &\dhyd{\psi}{x}\Bigg\rvert_{x=x_M} = 0.
\end{empheq}
\end{subequations}

The first \emph{boundary condition} (Eq.~\ref{Eq:BoundaryCondition1a} or Eq.~\ref{Eq:BoundaryCondition2a}) occurs because of spherical symmetry, that the amplitude $\psi$ must vanish at the center of the planet.

The second \emph{boundary condition} (Eq.~\ref{Eq:BoundaryCondition1b} or Eq.~\ref{Eq:BoundaryCondition2b}) occurs because the gradient of pressure disturbance which drives the differential volume perturbation vanishes at the surface of the planet. Although not immediately clear, this condition is equivalent to requiring that all interior disturbances be reflected at the surface (as itself moves) back into the interior; that is, no pulsation energy is lost from the interior into space because all is reflected back inward from the surface~\citep{Hansen2004}. See also in~\citep{Ledoux1958}~\citep{Tassoul1968}~\citep{Collins1978}. 

Generally speaking, Equations of the form of Eq.~\ref{Eq:EulerLagrange9e} and Eq.~\ref{Eq:EulerLagrange10} are difficult to solve analytically, because the coefficient in front of $\psi$ is not a constant but varies with $m$ or $x$, which means the wavelength varies with location. 

Even the following simple example requires the introduction of a special function called \emph{Airy function} (introduced by the British astronomer G. B. Airy): 
\begin{equation}\label{Eq:Airy}
    \dnhyd[2]{\psi}{x} + x \cdot \psi = 0
\end{equation}

Historically, many techniques to solve such equations have been developed, such as the \emph{ladder operators} or the \emph{WKB-approximation}. 

Indeed, if $\abs{\lambda / u''}\ll x_M$, that is, the wavelength is small compared to the overall dimension, then $\psi$ can be approximated as: 

\begin{equation}\label{Eq:WKB}
   \psi(x) \sim A \cdot \exp{\frac{i x}{(\lambda / u'')}} + B \cdot \exp{\frac{-i x}{(\lambda / u'')}}
\end{equation}

Or, 

\begin{equation}\label{Eq:WKB2}
   \psi(x) \sim A \cdot \exp{\frac{i \int dm/u''}{(\lambda / u'')}} + B \cdot \exp{\frac{-i \int dm/u''}{(\lambda / u'')}}
\end{equation}

This is the essence of the \emph{WKB-approximation}. 

%Nonetheless, we could still explore the behavior of the solutions to Eq.~\ref{Eq:EulerLagrange11} qualitatively. Let's non-dimensionalize Eq.~\ref{Eq:EulerLagrange11} with following introduction of new variables: 

\begin{comment}
\begin{align}
    \omega_{M} &\equiv \omega_{0} \big\rvert_{x=x_M} = \frac{4\pi}{3} \cdot R_{P}^3 \nonumber \\
     y &\equiv x/x_{M} \nonumber \\
    \psi &\equiv \psi/\omega_{M} = \Bigg(\frac{\omega_0}{\omega_M} \Bigg) \cdot \zeta
    \label{Eq:EulerLagrange13}
\end{align}
\end{comment}

\subsection{Legendre Transformation}

This product: $(\omega \cdot P)$, is important. Using it, one can define another equivalent action $\widetilde{\mathcal{S}}$ and equivalent Lagrangian $\widetilde{\mathcal{L}}$ in terms of using Pressure $P$ as the dependent variable, instead of volume $\omega$. The indepedent variable is still mass $m$, as follows: 

\begin{equation}
    \mathcal{L} + \widetilde{\mathcal{L}} \equiv \dhyd{}{m}(\omega \cdot P) = \omega \cdot \dot{P} + \dot{\omega} \cdot P
\end{equation}

in other words, 

\begin{equation}
    \widetilde{\mathcal{L}} \equiv \dhyd{}{m}(\omega \cdot P) - \mathcal{L}
\end{equation}

so that, 

\begin{align*}
    \mathcal{S}[\omega] &= \int \mathcal{L} \cdot \mathrm{d}m, & \dhyd{}{m}\mathcal{S}[\omega] &= \mathcal{L},\\
    \widetilde{\mathcal{S}}[P] &= \int \widetilde{\mathcal{L}} \cdot \mathrm{d}m, & \dhyd{}{m}\widetilde{\mathcal{S}}[P] &= \widetilde{\mathcal{L}}.
\end{align*}

Here $\widetilde{\mathcal{L}}$ is the \emph{Legendre transformation} of $\Lagr$, where the variables become $P$ and $\dot{P}$ as: $\widetilde{\mathcal{L}} = \widetilde{\mathcal{L}}(m;P,\dot{P})$ versus $\Lagr = \Lagr(m;\omega,\dot{\omega})$. 

Moreover, 

\begin{equation}
    \mathrm{d}\mathcal{L} + \mathrm{d}\widetilde{\mathcal{L}} = \omega \cdot \mathrm{d}\dot{P} + \dot{P} \cdot \mathrm{d}\omega +  \dot{\omega} \cdot \mathrm{d}P + P \cdot \mathrm{d}\dot{\omega}
\end{equation}

Recall that: 

\begin{equation}
    \mathcal{L}(m;\omega,\dot{\omega}) = -\big( \underbrace{u(\dot{\omega})}_{\text{close-range interaction due to \emph{EOS}}} - \underbrace{f(\omega) \cdot m}_{\text{far-range interaction due to \emph{gravity}}} \big)
\end{equation}

Here $f(\omega)$ encodes the geometric shape of the system. In particular, for spherical symmetry, 

\begin{equation}
    f(\omega) = G/r = G \cdot \bigg( \frac{3}{4\pi} \bigg)^{-1/3} \cdot {\omega}^{-1/3}
\end{equation}

\subsection{Hamiltonian}

Hamiltonian $\mathcal{H}$ is also related with Lagrangian $\mathcal{L}$, but with the transformation carried only half-way through. 

\begin{align}
    \mathcal{H} &\equiv \mathcal{H}(m;\omega,P) \nonumber\\
                &= \bigg( \dot{\omega} \cdot \frac{\partial \Lagr}{\partial \dot{\omega}} \bigg) - \mathcal{L}(m;\omega,\dot{\omega}) \nonumber\\
                &= \big( \dot{\omega} \cdot P \big) + u(\dot{\omega}) - f(\omega) \cdot m \nonumber\\
                &= \bigg(\big( \dot{\omega} \cdot P \big) + u(\dot{\omega}) \bigg) - f(\omega) \cdot m \nonumber\\
                &= h(P) - f(\omega) \cdot m \label{Eq:Hamiltonian1}
\end{align}

where, $h(P)$ is exactly the definition of \emph{specific enthalpy}, assuming (1) the process occurs along an isentrope, and (2) no phase transition occurs. 

The \emph{Symplectic Equations} can now be written as: 

\begin{subequations}\label{Eq:CanonicalEquations}
\begin{empheq}[left=\empheqlbrace]{align}
\label{Eq:CanonicalEquation1}
         &\dot{P} \equiv \dhyd{P}{m} = f'(\omega) \cdot m, \\
\label{Eq:CanonicalEquation2}
         &\dot{\omega} \equiv \dhyd{\omega}{m} = h'(P), \\
\label{Eq:CanonicalEquation3}
         &\frac{\partial \mathcal{H}}{\partial m} = -\frac{\partial \mathcal{L}}{\partial m} = -f(\omega).
\end{empheq}
\end{subequations}

Eq.~\ref{Eq:CanonicalEquation1} is in fact the hydrostatic equilibrium equation. Eq.~\ref{Eq:CanonicalEquation2} is in fact the EOS. Eq.~\ref{Eq:CanonicalEquation3} is the mass-evolution of \emph{Hamiltonian}---\emph{Symplectic Integral}!

\subsection{Hamilton's Principle Function}

From the perspective of \emph{Hamilton-Jacobi Theory}, the above formulation can be further distilled and summarized by introducing the \emph{Hamilton's Principle Function}. The \emph{Hamilton's Principal Function} is defined as the function of the upper limit of the action integral taken along the minimal action trajectory of the system: 

\begin{equation}\label{HamiltonPrincipleFunction}
    \mathcal{S}(m,\omega) := \int^{(m,\omega)} \mathcal{L} \cdot \mathrm{d}m
\end{equation}

This is to view to total action $\mathcal{S}$ as a function of $m$ and $\omega$ on equal footing, with the upper limit of the integral variable. In this form, $\mathcal{S}$ does not explicitly depend on either $\dot{\omega}$ or $P$. Calculating the variation of with respect to the endpoint $(m,\omega)$ gives: 

\begin{subequations}\label{HamiltonPrincipleFunction2}
\begin{empheq}[left=\empheqlbrace]{align}
\label{Eq:HamiltonPrincipleFunction1}
         &\frac{\partial \mathcal{S}}{\partial \omega} = P \\
\label{Eq:HamiltonPrincipleFunction2}
         &\frac{\partial \mathcal{S}}{\partial m} = -\mathcal{H}
\end{empheq}
\end{subequations}

Then, the \emph{Hamilton-Jacobi} Equation (HJE) is a single, first-order partial-differential equation based on the \emph{Hamilton's Principle Function} $\mathcal{S}(m,\omega)$, which is equivalent of an integral minimization problem such as the \emph{Hamilton's Principle}, and can be used to determine the \emph{Geodesics}. 

\begin{equation}\label{HJE}
    -\frac{\partial \mathcal{S}(m,\omega)}{\partial m} = \mathcal{H} \bigg(m; \omega, \frac{\partial \mathcal{S}(m,\omega)}{\partial \omega} \bigg)
\end{equation}

Plugging the particular form of Hamiltonian $\mathcal{H}$ (Eq.~\ref{Eq:Hamiltonian1}) considered here: 

\begin{equation}\label{HJE2}
    \mathcal{H} \bigg(m; \omega, \frac{\partial \mathcal{S}(m,\omega)}{\partial \omega} \bigg) = h \bigg( \frac{\partial \mathcal{S}(m,\omega)}{\partial \omega} \bigg) - f(\omega) \cdot m
\end{equation}

Then, we have:

\begin{equation}\label{HJE3}
    h \bigg( \frac{\partial \mathcal{S}(m,\omega)}{\partial \omega} \bigg) - f(\omega) \cdot m + \frac{\partial \mathcal{S}(m,\omega)}{\partial m} = 0
\end{equation}

Ultimately, we want to solve for the \emph{Hamilton's Principal Function} $\mathcal{S}(m,\omega)$ using Eq.~\ref{HJE3} given that $h$ (\emph{EOS}) and $f$ (\emph{geometry and symmetry}) are provided. 

If we express $\mathcal{S}$ in radius $r$ instead of volume $\omega$, then, HJE can be written as: 

\begin{equation}\label{HJE4}
    h \bigg(\frac{1}{4\pi r^2} \cdot \frac{\partial \mathcal{S}(m,r)}{\partial r} \bigg) - \frac{G \cdot m}{r} + \frac{\partial \mathcal{S}(m,r)}{\partial m} = 0
\end{equation}

\subsubsection{An Example}

For EOS of: 

\begin{equation}
    P = K \cdot \rho^2
\end{equation}

Then, the specific enthalpy $h(P)$, by definition, is

\begin{equation}
    h(P) = 2 \cdot \sqrt{K} \cdot \sqrt{P}
\end{equation}

Let's define a new parameter: 

\begin{equation}
    k \equiv \sqrt{\frac{2\pi G}{K}}
\end{equation}

$k$ has the dimension of 1/distance. 

Then, the \emph{Hamilton's Principal Function} $\mathcal{S}(m,r)$ for this EOS is:

\begin{equation}\label{HJE5}
    \mathcal{S}(m,r) = \bigg(-\frac{G \cdot m^2}{2} \bigg) \cdot \Bigg( \frac{k}{\tan(k\cdot r)- k\cdot r} \Bigg)
\end{equation}

%where radius $r = \bigg( \frac{3}{4\pi} \bigg)^{1/3} \cdot {\omega}^{1/3}$. 

According to Eq.~\ref{Eq:HamiltonPrincipleFunction1}, the Pressure is: 

\begin{align}
    P &= \frac{\partial \mathcal{S}(m,\omega)}{\partial \omega} \nonumber\\
                &= \frac{1}{4\pi r^2} \cdot \frac{\partial \mathcal{S}(m,r)}{\partial r} \nonumber\\
                &= \frac{1}{4\pi r^2} \cdot \frac{\partial}{\partial r} \bigg( \bigg(-\frac{G \cdot m^2}{2} \bigg) \cdot \Bigg( \frac{k}{\tan(k\cdot r)- k\cdot r} \Bigg) \bigg) \nonumber\\
                &= \frac{1}{4\pi r^2} \cdot \bigg( \frac{G \cdot m^2}{2} \cdot k^2 \bigg) \cdot \Bigg( \frac{\sin{(k\cdot r)}}{(k \cdot r)\cdot \cos{(k \cdot r)} - \sin{(k\cdot r)}} \Bigg)^2  \nonumber\\
                &= \frac{G}{8\pi} \cdot \Bigg[ \frac{m \cdot k}{r} \cdot \bigg( \frac{\tan(k\cdot r)}{\tan(k\cdot r)- k\cdot r} \bigg) \Bigg]^2 \label{Eq:HamiltonPrincipleFunction3}
\end{align}

Then, 

\begin{align}
    h &= 2 \cdot \sqrt{K} \cdot \sqrt{P} \nonumber\\
      &= 2 \cdot \frac{\sqrt{2\pi G}}{k} \cdot \sqrt{\frac{G}{8\pi}} \cdot \frac{m \cdot k}{r} \cdot \bigg( \frac{\tan(k\cdot r)}{\tan(k\cdot r)- k\cdot r} \bigg) \nonumber\\
      &= \frac{G \cdot m}{r} \cdot \bigg( \frac{\tan(k\cdot r)}{\tan(k\cdot r)- k\cdot r} \bigg)
\end{align}

Then, 

\begin{align}
    \bigg( h - \frac{G \cdot m}{r} \bigg) &= \frac{G \cdot m}{r} \cdot \bigg( \frac{k\cdot r}{\tan(k\cdot r)- k\cdot r} \bigg) \nonumber\\
        &= G \cdot m \cdot \bigg( \frac{k}{\tan(k\cdot r)- k\cdot r} \bigg)
\end{align}

On the other hand, 

\begin{equation}
    \frac{\partial \mathcal{S}(m,r)}{\partial m} = -G \cdot m \cdot \Bigg( \frac{k}{\tan(k\cdot r)- k\cdot r} \Bigg)
\end{equation}

Plugging back into Eq.~\ref{HJE4}, one can see it is satisfied. 

The essence (spirit) behind this method is \emph{Power Series}, as Cauchy used to solve \emph{Kepler Equation}: $E - e\cdot \sin{E} = M$~\citep{Needham1997}. In a sense, $\tan(k\cdot r)- k\cdot r$ can be understood as a power series expansion of radius $r$ around the origin: 

\begin{equation}
    \tan(k\cdot r) - (k\cdot r) = \frac{1}{3} (k\cdot r)^3 + \frac{2}{15} (k \cdot r)^5 + \frac{17}{315} (k \cdot r)^7 + \frac{62}{2835} (k \cdot r)^9 + ...
\end{equation}

Or alternatively, in terms of \emph{Laurent Series}: 

\begin{multline}
        \frac{1}{\tan(k\cdot r) - (k\cdot r)} = \frac{3}{(k \cdot r)^3} - \frac{6}{5 (k \cdot r)} - \frac{(k \cdot r)}{175} -\frac{2 (k \cdot r)^3}{7875} \\
        -\frac{37 (k \cdot r)^5}{3031875} -\frac{118 (k \cdot r)^7}{197071875} -\frac{5506 (k \cdot r)^9}{186232921875}+...
\end{multline}

Therefore, in this particular case, $\mathcal{S}(m,r)$ can be expressed as:

\begin{multline}
        \mathcal{S}(m,r) = \bigg(-\frac{G \cdot m^2}{2} \bigg) \cdot k \cdot \bigg( \frac{3}{(k \cdot r)^3} - \frac{6}{5 (k \cdot r)} - \frac{(k \cdot r)}{175} -\frac{2 (k \cdot r)^3}{7875} \\
        -\frac{37 (k \cdot r)^5}{3031875} -\frac{118 (k \cdot r)^7}{197071875} -\frac{5506 (k \cdot r)^9}{186232921875}+... \bigg)
\end{multline}

In particular, when $(k\cdot r) = \pi/2$, then, $\mathcal{S}(m,r) = 0$. 

In particular, when $(k\cdot r) = \pi$, then, $\mathcal{S}(m,r) = \frac{G \cdot m^2}{2} \cdot \frac{k}{\pi}$.

Since only partial derivatives of $\mathcal{S}$ are involved, there could be an arbitrary additive constant to $\mathcal{S}$.

\clearpage

%%%%%%%%%%%%%%%%%%%%%%%%%%%%%%%
\section{Hydrogen EOS}

Hydrogen EOS is important as it is relevant to the interior of many gaseous planets. It has been subject to intensive experimental investigation~\citep{SilveraDias2018MetallicHydrogen}~\citep{Nellis2006}. 

\subsection{Critical Point (C.P) of Liquid Hydrogen}

The Liquid Hydrogen (H$_2$) has its critical point ($c$) as follows~\citep{NISTWebBookHydrogen}: 

\begin{align}
\rho_{c} &\approx 0.03012 \text{~g/cc} \nonumber\\
T_{c} &= 33.20 \text{~K} = -239.95 ^{\circ}\text{C} \nonumber\\
P_{c} &= 1.3 \cdot 10^{6} \text{~Pa} = 13. \text{~bar} \nonumber\\
n_{c} &\approx 0.03012/2 \approx 0.015 \text{~mol/cc} \nonumber\\
\end{align}

Here $n_{c}$ is the molar density expressed in (mol/cc) of H$_2$ at its critical point. $T_{c}$ of H$_2$ is very low and hence the hydrogen in planet interior should be considered \emph{super-critical} molecular fluid when it is above the molar density of $0.015$ mol/cc. At even higher density or temperature, it then transitions into metallic fluid.

\subsection{Tripe Point (t1) of Liquid Hydrogen}

The Liquid Hydrogen (H$_2$) has its triple point ($t1$) co-existing with solid-H$_2$ and gas-H$_2$ as follows~\citep{NISTWebBookHydrogen}~\citep{EngineeringtoolboxHydrogen}~\citep{BNLCryogenicDataNotebook}: 

\begin{align}
\rho_{t1,\text{Liquid}} &\approx 0.07691 \text{~g/cc} \nonumber\\
\rho_{t1,\text{Solid}} &\approx 0.086 \text{~g/cc} \nonumber\\
T_{t1} &= 13.96 \text{~K} = -259.2 ^{\circ}\text{C} \nonumber\\
P_{t1} &= 7.7 \cdot 10^{3} \text{~Pa} = 0.077 \text{~bar} \nonumber\\
n_{t1,\text{Liquid}} &\approx 0.07691/2 \approx 0.0383 \text{~mol/cc} \nonumber\\
n_{t1,\text{Solid}} &\approx 0.086/2 \approx 0.043 \text{~mol/cc} \nonumber\\
\end{align}

Here $n_{t1,\text{Liquid}}$ is the molar density expressed in (mol/cc) of H$_2$-liquid at its triple point. 
Here $n_{t1,\text{Solid}}$ is the molar density expressed in (mol/cc) of H$_2$-soild at its triple point. The solid-H$_2$ was obtained by James Dewar in (1899) and is one of the lowest-density solids~\citep{Dewar1899}.

See also in Fig.~\ref{fig:PhaseDiagramH2}. It is interesting that: 

\begin{equation}
    n_{t1}/n_{c} \sim 3
\end{equation}

We will re-visit this point in a later discussion and generalize it to other molecules. 

\subsection{Mazevet 2019 EOS}
\begin{figure}[h!]
\centering
\hspace*{-2cm}\includegraphics[scale=0.8]{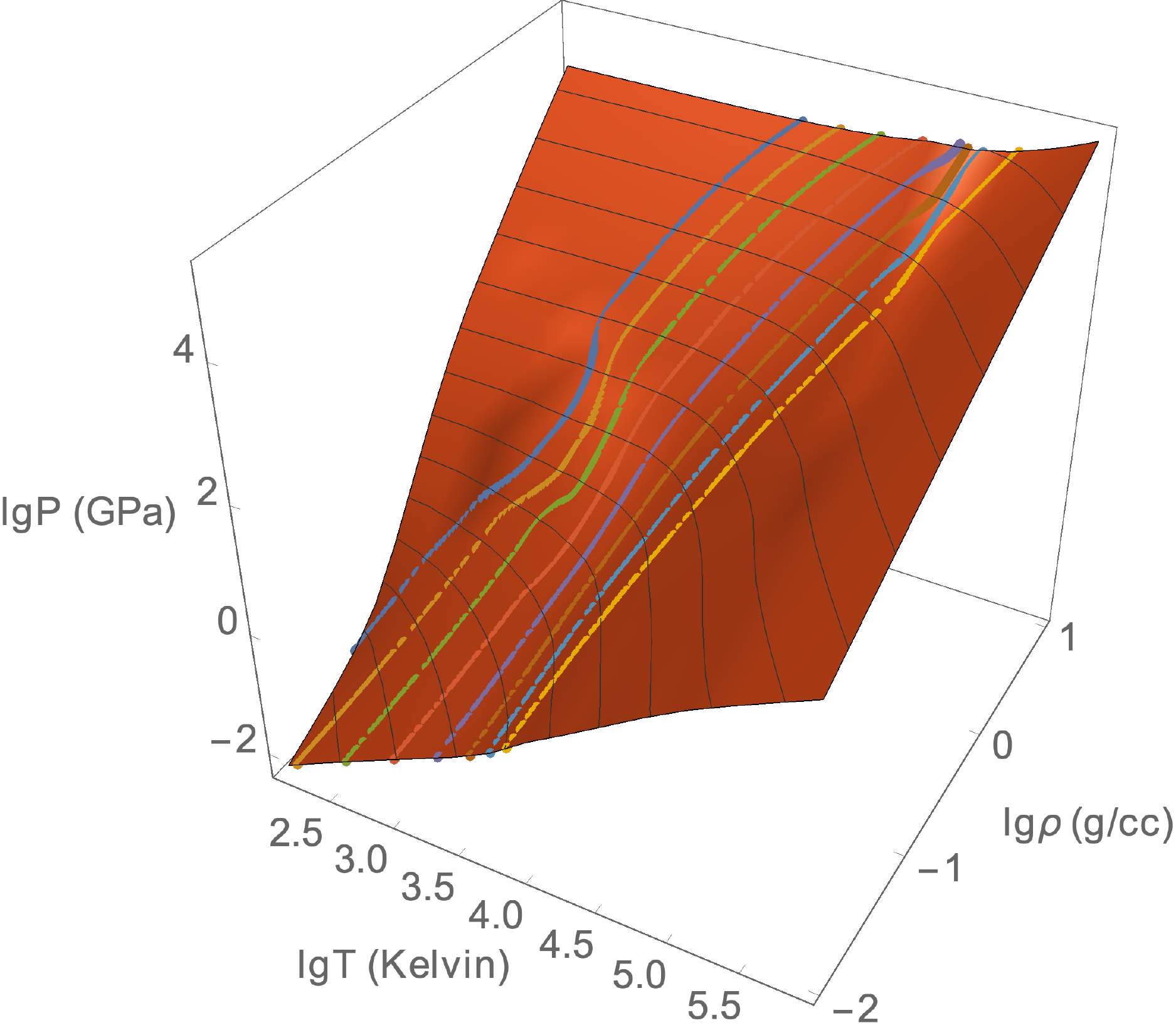}
\caption{Hydrogen EOS of Mazevet 2019 with isentropes visualized in the three dimensions of $lgT$, $lg\rho$, and $lgP$. The eight color curves corresponding to isentropes of specific entropy S (eV/1000Kelvin/atom) = {0.3, 0.4, 0.5, 0.6, 0.7, 0.8, 0.9, 1.0}. The bulge and the oscillation behaviors of the isentropes with lower specific entropies as a result of this bulge, are due to the Plasma Phase Transition (PPT). Recall that along an isentrope, $\gamma \equiv \bigg( \frac{\partial \ln{T}}{\partial \ln{\rho}} \bigg)_{s}$, and $\Gamma \equiv \bigg( \frac{\partial \ln{P}}{\partial \ln{\rho}} \bigg)_{s}$. Therefore, $\gamma$ and $\Gamma$ are just the tangential slopes of these 3-D isentropes projected onto the T-$\rho$ and P-$\rho$ plane correspondingly. }
\label{fig:H2EOSMazevet20193D}
\end{figure}

We apply this technique to Mazevet 2019 Hydrogen-EOS~\cite{Chabrier_2019}. First, we visualize this EOS in 3D the logarithmic relationship between pressure, density and temperature (Fig.\ref{fig:H2EOSMazevet20193D}). 

\begin{figure}[h!]
\centering
\hspace*{-2cm}\includegraphics[scale=0.7]{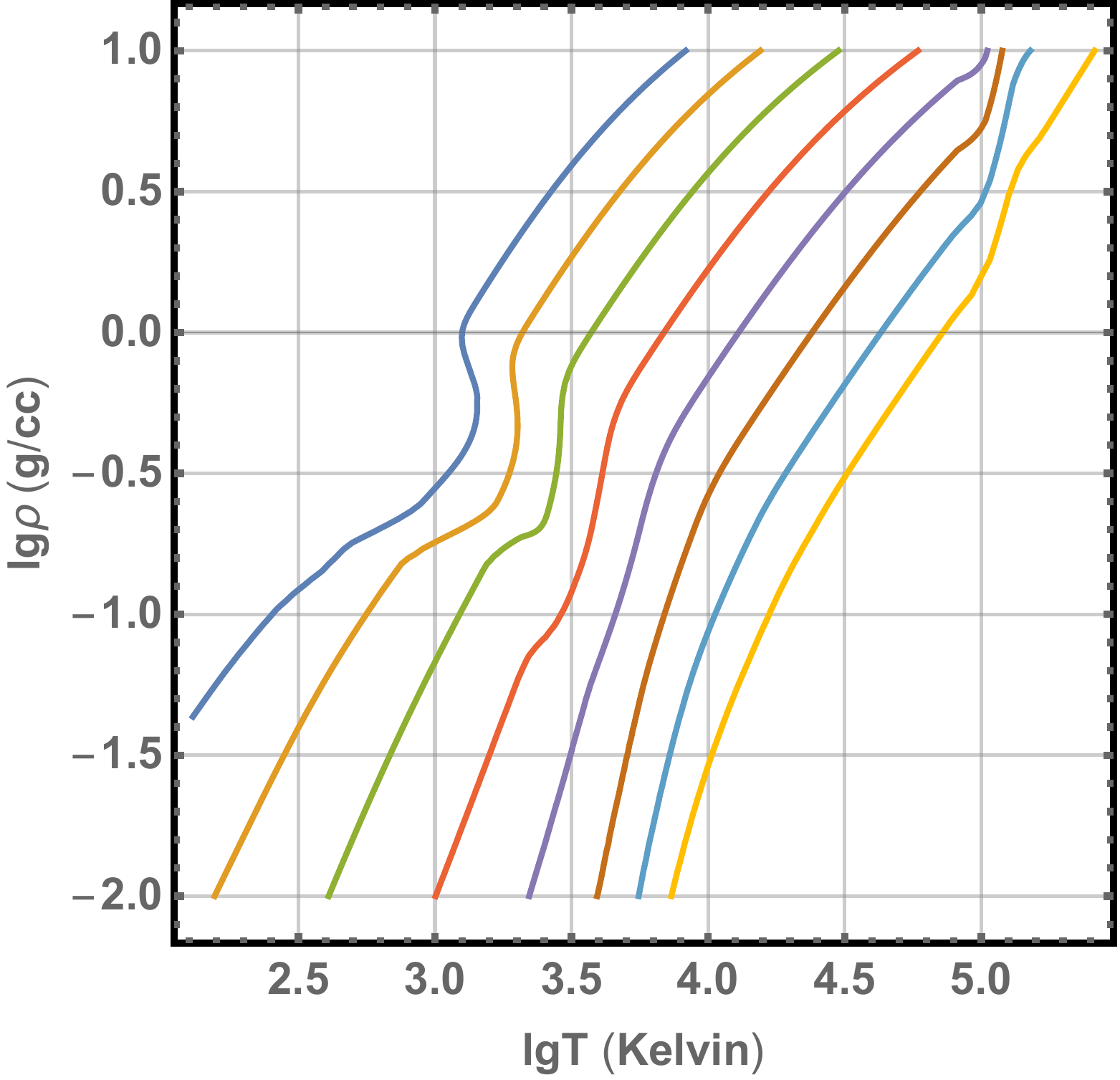}
\caption{$lgT$-$lg\rho$ relation of isentropes in the Hydrogen EOS of Mazevet 2019. The tangential slope in this plot is $\gamma \equiv \bigg( \frac{\partial \ln{T}}{\partial \ln{\rho}} \bigg)_{s}$. We call it a $\gamma$-plot. }
\label{fig:H2EOSMazevet2019lgTlgrho}
\end{figure}

\begin{figure}[h!]
\centering
\hspace*{-2cm}\includegraphics[scale=0.7]{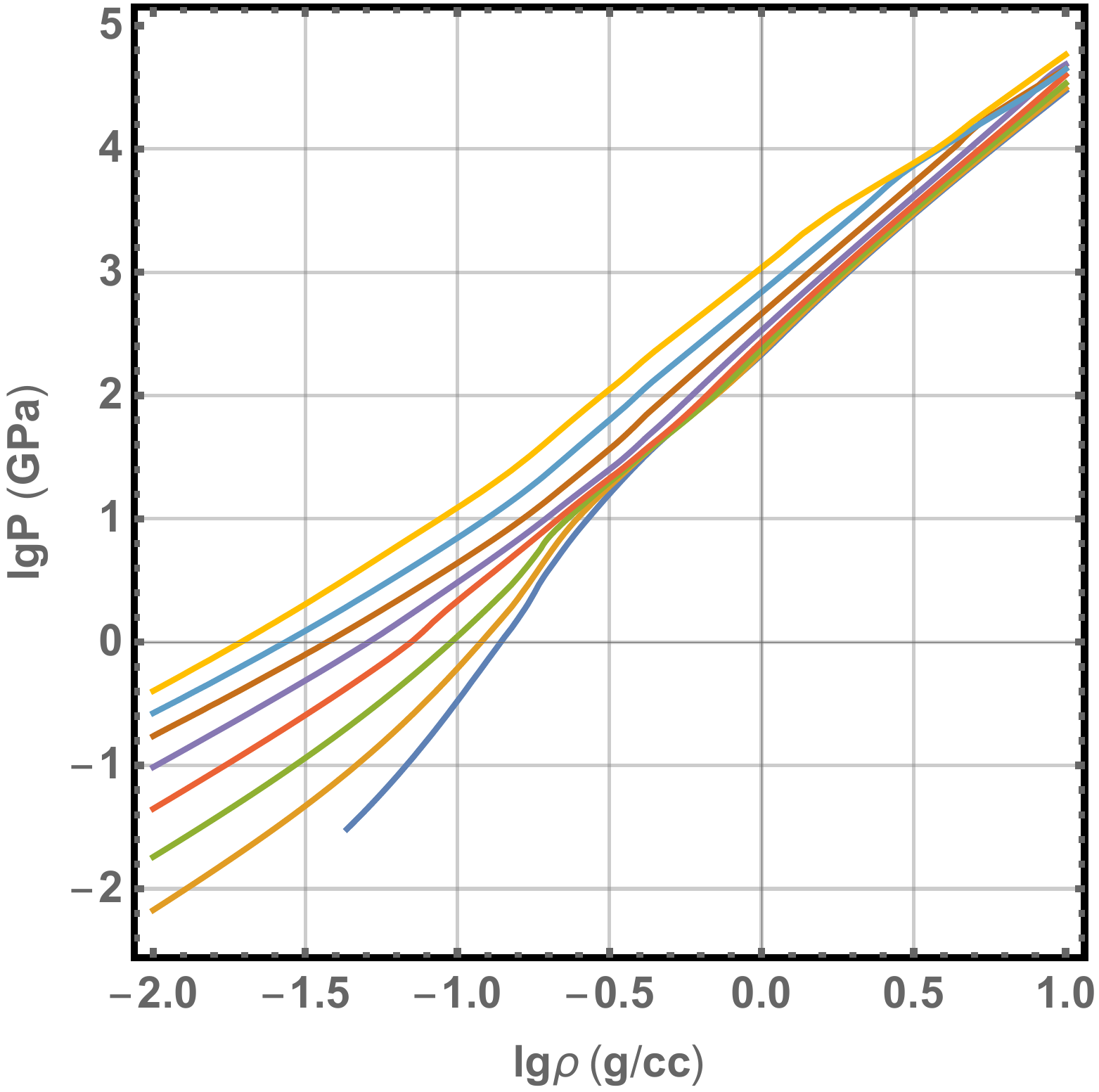}
\caption{$lg\rho$-$lgP$ relation of isentropes in the Hydrogen EOS of Mazevet 2019. The tangential slope in this plot is $\Gamma \equiv \bigg( \frac{\partial \ln{P}}{\partial \ln{\rho}} \bigg)_{s}$. We call it a $\Gamma$-plot. }
\label{fig:H2EOSMazevet2019lgrholgP}
\end{figure}

This 3D EOS can be projected onto various cardinal directions, to show the T-$\rho$ (Temperature-Density) relation (Fig.\ref{fig:H2EOSMazevet2019lgTlgrho}) or the P-$\rho$ (Pressure-Density) relation (Fig.\ref{fig:H2EOSMazevet2019lgrholgP}) along these isentropes. On the other hand, the projection onto P-T (Pressure-Temperature) plane is often encountered in illustrating a phase diagram in the literature. 

\begin{figure}[h!]
\centering
\hspace*{-3cm}\includegraphics[scale=0.3]{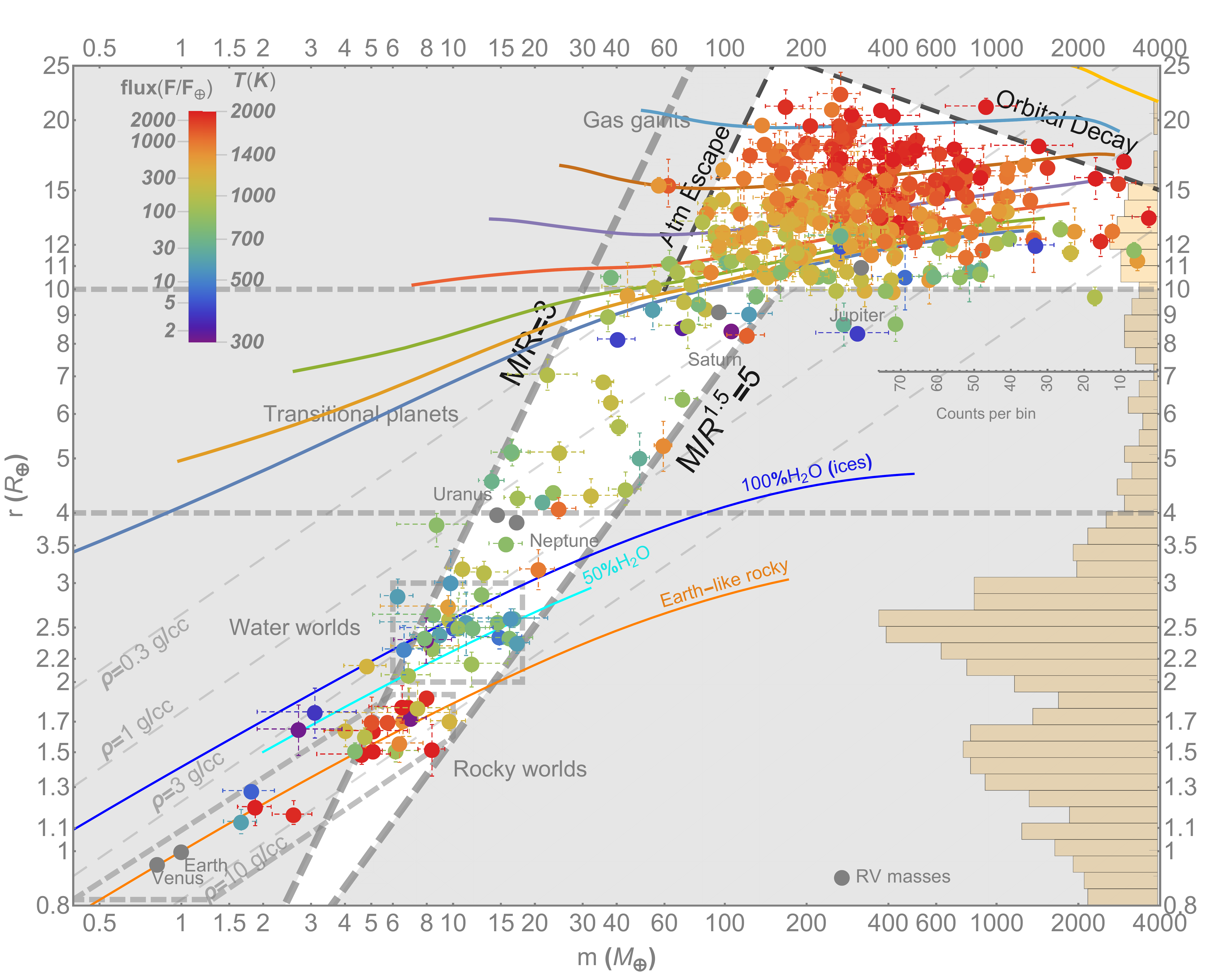}
\caption{Mass-radius curves of Hydrogen EOS of Mazevet 2019: the same color schemes of isentropes apply, but do not correspond to the temperature color scale on the upper-left. The temperature color scale is only for exoplanets plotted.  }
\label{fig:H2EOSMazevet2019MassRadius}
\end{figure}

With this understanding of EOS, one can then convert these isentropes into corresponding mass-radius curves (Fig.~\ref{fig:H2EOSMazevet2019MassRadius}), by solving the Differential Equations mentioned in the previous section (\textit{Mathematica} code is provided in a separate file). These mass-radius curves are for envelope only, not including the atmosphere. The calculation is truncated as density drops below 0.01 gram/cubic centimeters. 
There are two reasons for this truncation:
\begin{itemize}
    \item The atmosphere temperature profile is likely more isothermal due to the stellar irradiation, as opposed to isentropic that is driven by interior convection. In other words, the specific entropy of the envelope is nearly constant, and is determined by the outward energy transport in the planet interior, while that of atmosphere is influenced strongly by the host star. 
    \item The atmospheric scale height of the planet is relatively small, on the order of 0.1 Earth radius, for $H_2$ atm of 1000K. If considering 10 scale-heights, it would be a correction of 1 Earth radius. We could easily add that later. The atmosphere can sometimes be treated with modified ideal gas EOS. 
\end{itemize}

\begin{figure}[h!]
\centering
\hspace*{-3cm}\includegraphics[scale=0.3]{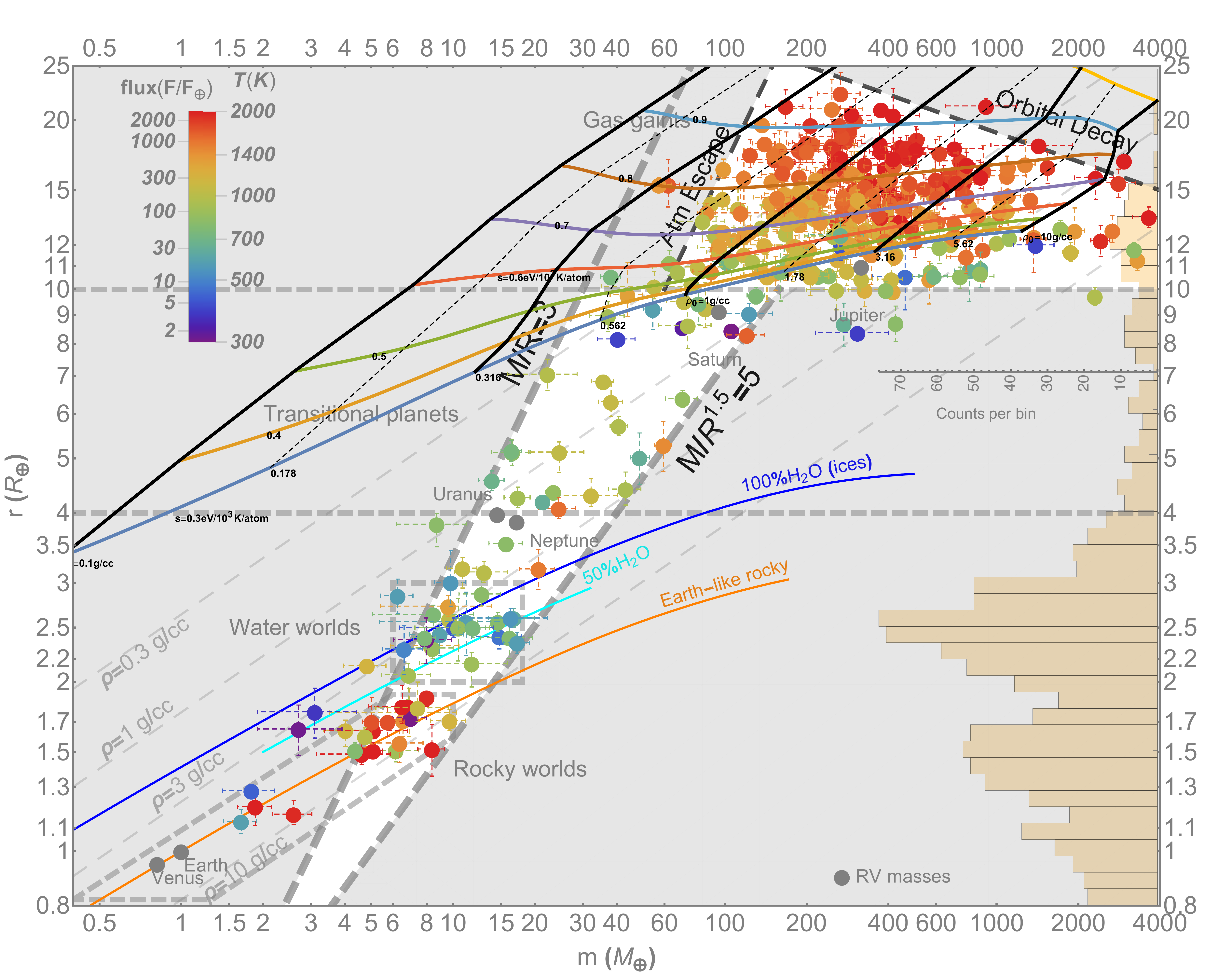}
\caption{Modified Mass-radius curves of Hydrogen EOS of Mazevet 2019}
\label{fig:H2EOSMazevet2019MassRadius2}
\end{figure}

\begin{figure}[h!]
\centering
\hspace*{-2cm}\includegraphics[scale=0.7]{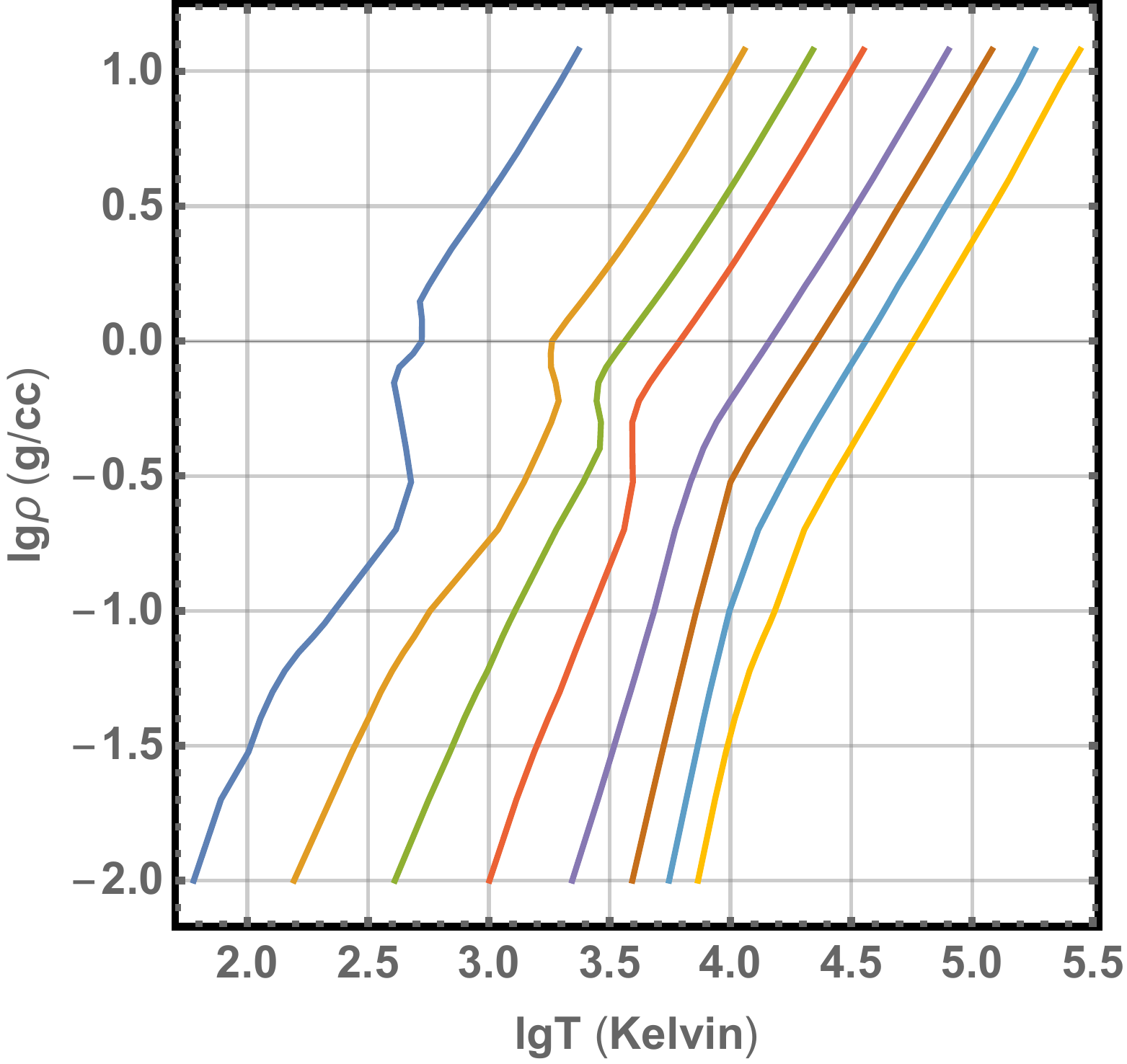}
\caption{$lgT$-$lg\rho$ relation of isentropes in the Hydrogen EOS of Becker 2014. The tangential slope in this plot is $\gamma \equiv \bigg( \frac{\partial \ln{T}}{\partial \ln{\rho}} \bigg)_{s}$. This is the $\gamma$-plot. }
\label{fig:H2EOSBecker2014lgTlgrho}
\end{figure}

\begin{figure}[h!]
\centering
\hspace*{-2cm}\includegraphics[scale=0.7]{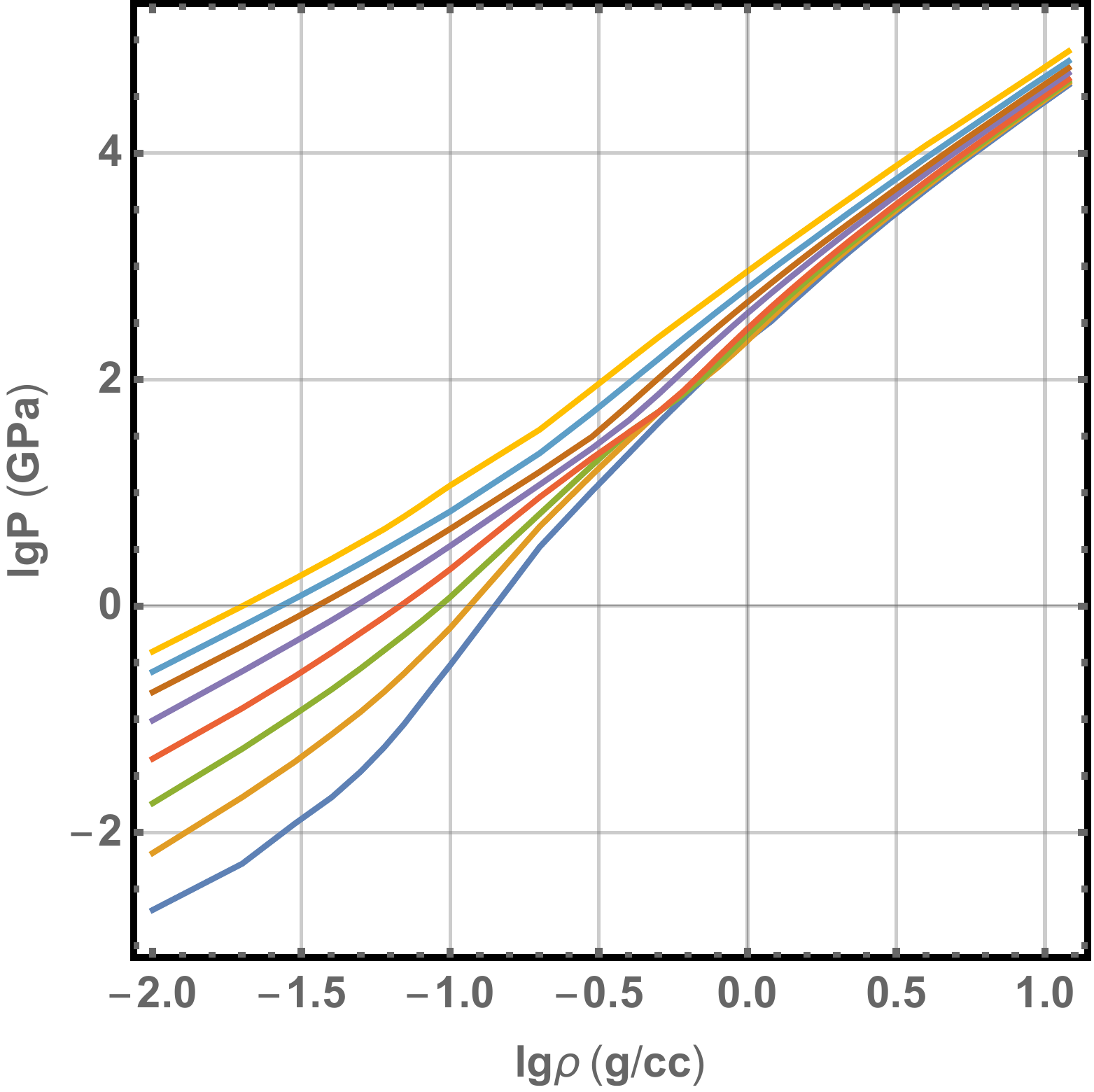}
\caption{$lg\rho$-$lgP$ relation of isentropes in the Hydrogen EOS of Becker 2014.  The tangential slope in this plot is $\Gamma \equiv \bigg( \frac{\partial \ln{P}}{\partial \ln{\rho}} \bigg)_{s}$. This is the $\Gamma$-plot. }
\label{fig:H2EOSBecker2014lgrholgP}
\end{figure}

Furthermore, one can add contours to illustrate the underlying physical meanings of these mass-radius curves. For example, one can add two sets of contours, one set for specific entropy S (eV/1000K/atom), and the other set for central density of planet in $\rho_0$ (g/cc). When a planet cools but not losing mass, it moves vertically (because mass is conserved) on the mass-radius diagram. From this point, one can get a sense of how its interior contracts (with central density and bulk density increasing) over cooling. (See Fig.~\ref{fig:H2EOSMazevet2019MassRadius2})

\subsection{Becker 2014 EOS}

Now, let's perform a detailed comparison between Mazevet 2019 EOS with that of Becker 2014 EOS from the German Rostock group (\cite{Becker2014AbDwarfs}, referred to as H-REOS.3, where 3 stands for version 3). 

%The Becker 2014 EOS neglect irradiation from host star when calculating the Brown Dwarf (BD) and Giant Planet (GP) radii. They use the radiative-convective atmospheric model of \citep{} and \citep{}. 

For Becker 2014 EOS, again we construct the temperature-density relation (Fig.\ref{fig:H2EOSMazevet2019lgTlgrho}) or the density-pressure relation (Fig.\ref{fig:H2EOSMazevet2019lgrholgP}) along eight isentropes of different specific entropies. 

\begin{figure}[h!]
\centering
\hspace*{-3cm}\includegraphics[scale=0.3]{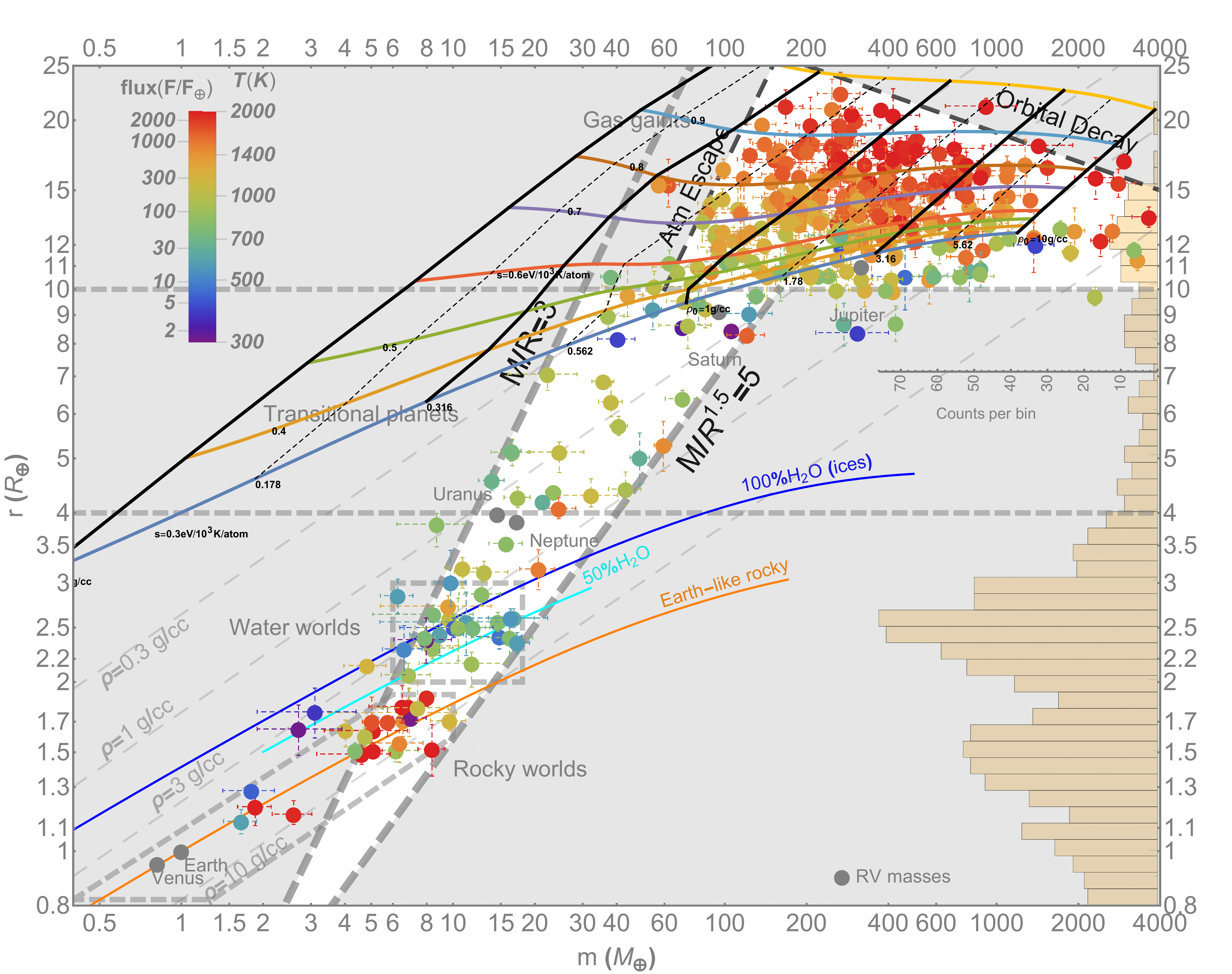}
\caption{Mass-radius curves of the Hydrogen EOS of Becker 2014}
\label{fig:H2EOSBecker2014MassRadius}
\end{figure}

Carrying on the same procedure of solving differential equations, we obtain the mass-radius curves corresponding to Becker 2014 Hydrogen EOS (Fig.~\ref{fig:H2EOSBecker2014MassRadius}). 

\begin{figure}[h!]
\centering
\hspace*{-3cm}\includegraphics[scale=0.3]{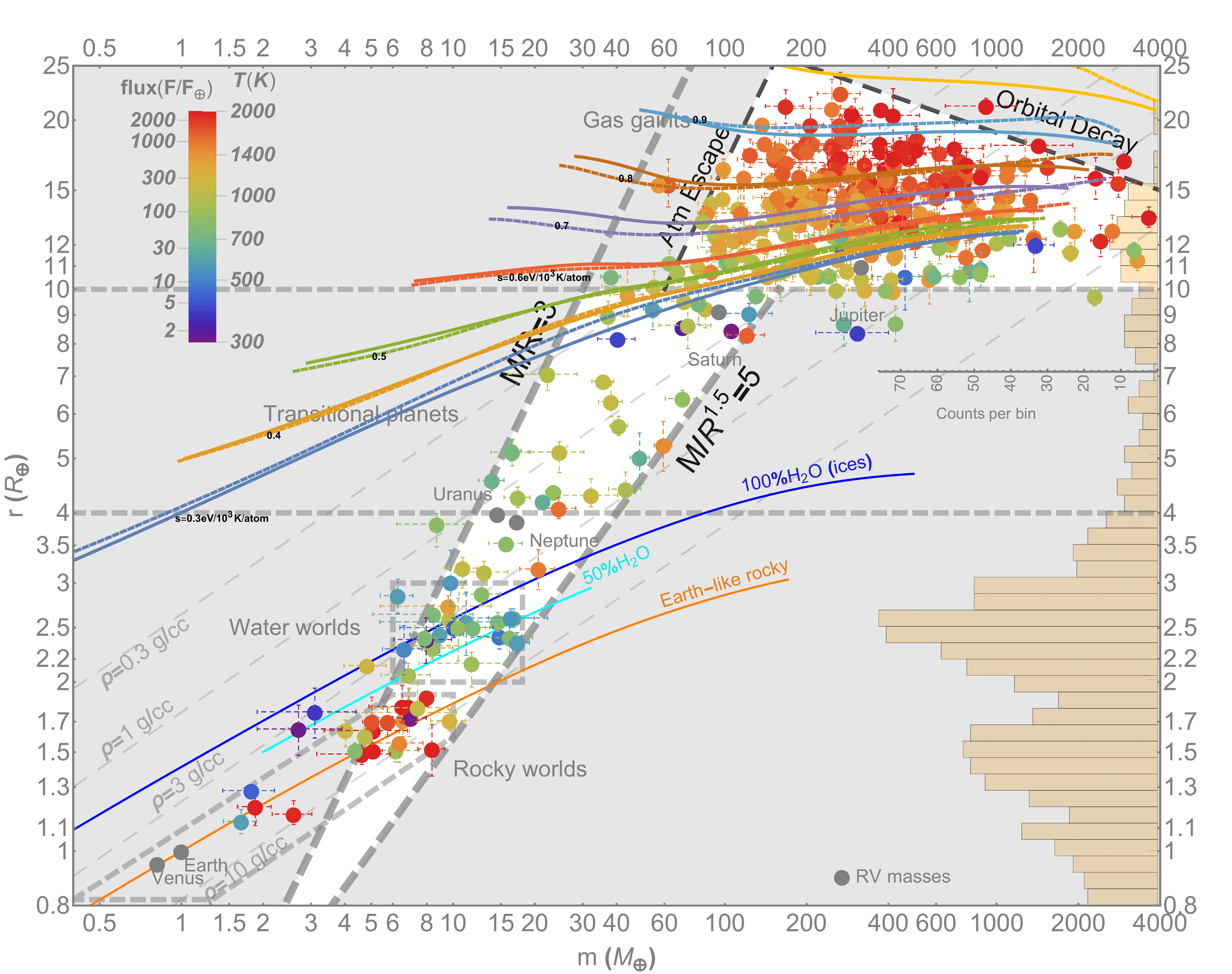}
\caption{Mass-radius curves comparison: Mazevet 2019 EOS (dashed) versus Becker 2014 EOS.}
\label{fig:H2EOSBecker2014vsMazevet2019MassRadius}
\end{figure}

Then, we make a comparison between the Mazevet 2019 EOS and Becker 2014 EOS (Fig.~\ref{fig:H2EOSBecker2014vsMazevet2019MassRadius}). 

\subsection{Discussion}

The major difference between Mazevet 2019 EOS and Becker 2014 EOS are their treatment above $10^5$ Kelvin, resulting in the biggest difference among the mass-radius curves of the highest entropy. Specifically, Becker 2014 EOS does not have a jump in thermodynamic properties such as entropy at $10^5$ Kelvin, or at density of 10 g/cc, while Mazevet 2019 EOS does. These jumps in Mazevet 2019 EOS are artificial, because the authors stitch together different EOSs calculated from different methods, applicable to different $lgT$-$lg\rho$ regimes (From Figure 1 of \cite{Chabrier_2019}, $10^5$ Kelvin is boundary between SCvH and H-EOS, and 10 g/cc is the boundary between QMD and CP98.). 

\begin{figure}[h!]
\centering
\hspace*{-2cm}\includegraphics[scale=0.7]{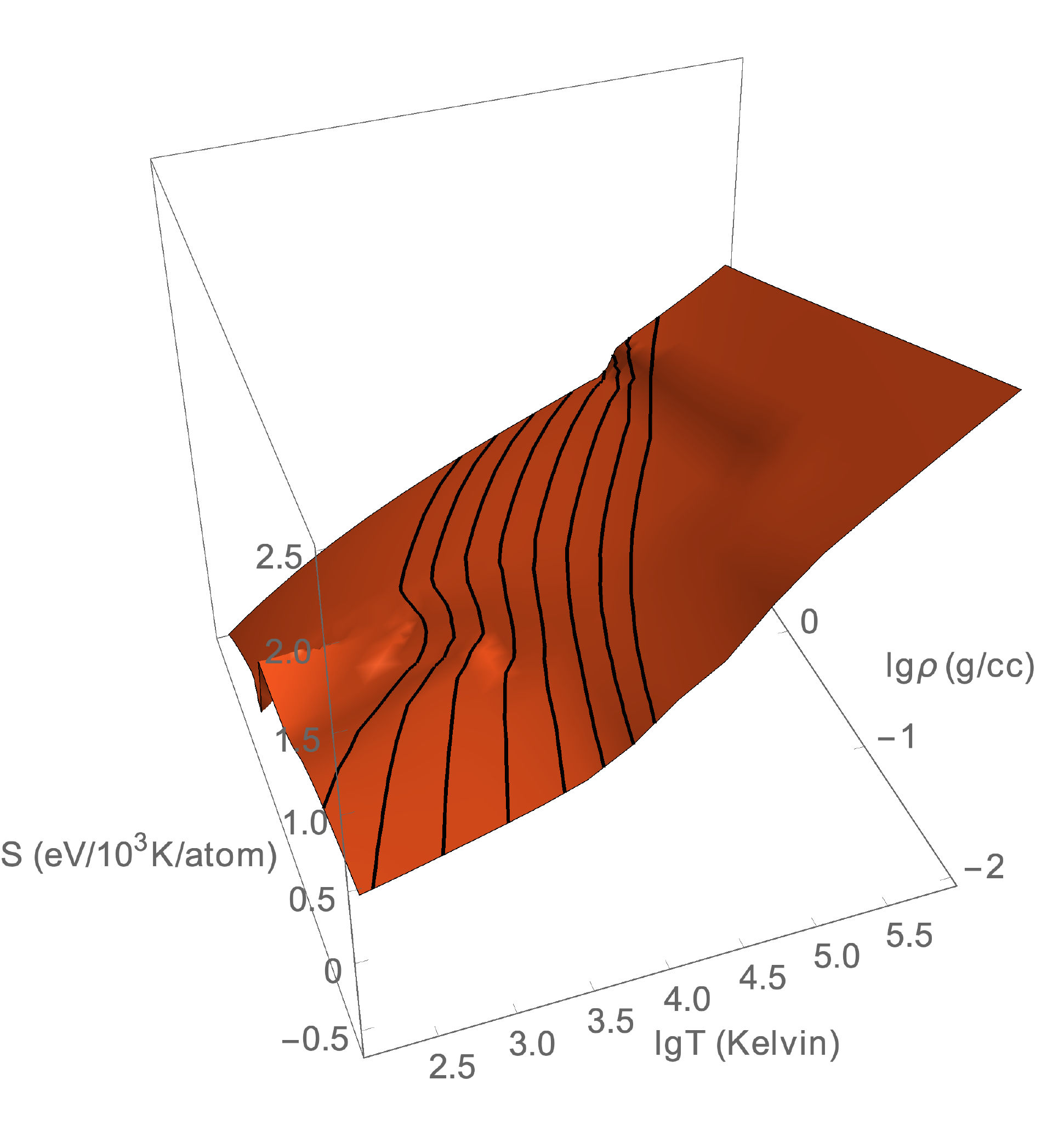}
\caption{Hydrogen EOS of Mazevet 2019 with isentropes viewed in three dimensions of $lgT$, $lg\rho$, and S (specific entropy in unit of eV/1000Kelvin/atom). The eight color curves corresponding to isentropes of S = {0.3, 0.4, 0.5, 0.6, 0.7, 0.8, 0.9, 1.0} . The bulge and the oscillation behaviors of the isentropes with lower specific entropies as a result of this bulge, are due to the Plasma Phase Transition (PPT). }
\label{fig:H2EOSMazevet20193D2}
\end{figure}

This is better visualized in 3D the relationship between specific entropy, logarithmic density and logarithmic temperature (Fig.\ref{fig:H2EOSMazevet20193D2}). 

\begin{figure}[h!]
\centering
\hspace*{-2cm}\includegraphics[scale=0.3]{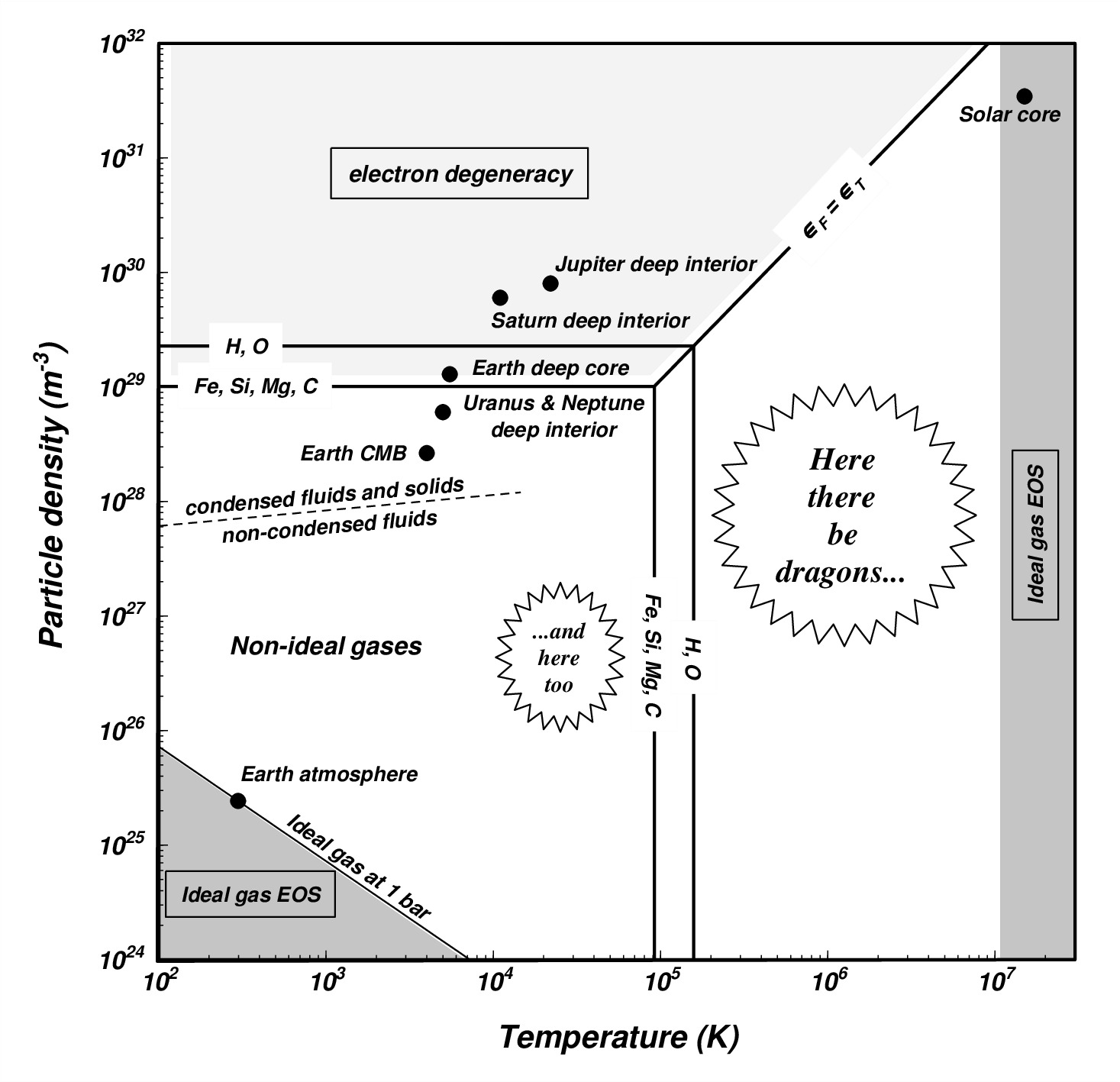}
\caption{Generic EOS diagram.  Copyright 2011 by Alberto Douce~\cite{Douce2011}}
\label{fig:Douce2011}
\end{figure}

The three hottest isentropes (the three right-most curves) go across a jump at $10^5$ Kelvin, into the so called "\textit{Dragon}" regime in a generic EOS diagram (Fig.\ref{fig:Douce2011}). That is crossing the boundary of Fermi-energy equals Thermal energy of electrons. This "\textit{Dragon}" regime is complicated because it is both partially degenerate and partially ionized, so the EOS there is not very well understood. It would require fully quantum numerical calculations such as Path-Integral Monte Carlo (PIMC) or Path-Integral Molecular Dynamics (PIMD). 

%%%To gain a deeper physical understanding into the planet interior when applying the Mazevet 2019 Hydrogen EOS, we calculate the interior profiles for 3 cases, each starting with the same central density of 10g/cc, but with different specific entropies, corresponding to the cold ($S=0.3eV/10^3K/atom$, Fig.~\ref{fig:planet_cold}), hot ($S=0.6eV/10^3K/atom$, Fig.~\ref{fig:planet_hot}), and very hot case ($S=0.9eV/10^3K/atom$, Fig.~\ref{fig:planet_very_hot}). The result is as follows: 

\begin{comment}
\begin{figure}
\centering
    \begin{subfigure}[t]{0.45\textwidth}
        \centering
        \includegraphics[width=\linewidth]{planet_cold.pdf}
        \caption{$S=0.3eV/10^3K/atom$}\label{fig:planet_cold}
    \end{subfigure} 
    \hfill
    \begin{subfigure}[t]{0.45\textwidth}
        \centering
        \includegraphics[width=\linewidth]{planet_hot.pdf}
        \caption{$S=0.6eV/10^3K/atom$}\label{fig:planet_hot}
    \end{subfigure} 
    %
   % \medskip
    \vspace{0.01cm}
    \begin{subfigure}[t]{\textwidth}
        \centering
        \vspace{0pt}% set the real top as the top
        \includegraphics[width=\linewidth]{planet_very_hot.pdf}
        \caption{$S=0.9eV/10^3K/atom$}\label{fig:planet_very_hot}
    \end{subfigure}
    \caption{Interior profiles of three cases with the same $\rho_0$ but different specific entropy. In (\subref{fig:planet_very_hot}) you can see a kink....}
\end{figure}
\end{comment}

According to Dave Stevenson~\citep{Stevenson2017Ge131:PlanetaryEvolution}, the interiors of many gas giants are simple. To a very good first-order approximation, they can be characterized by ONLY two parameters: 
\begin{itemize}
    \item central density $\rho_0$
    \item either the planet radius $r$, or its interior specific entropy $s$ (constant throughout its interior)
\end{itemize}

\begin{figure}[h!]
\centering
\hspace*{-2cm}\includegraphics[scale=0.85]{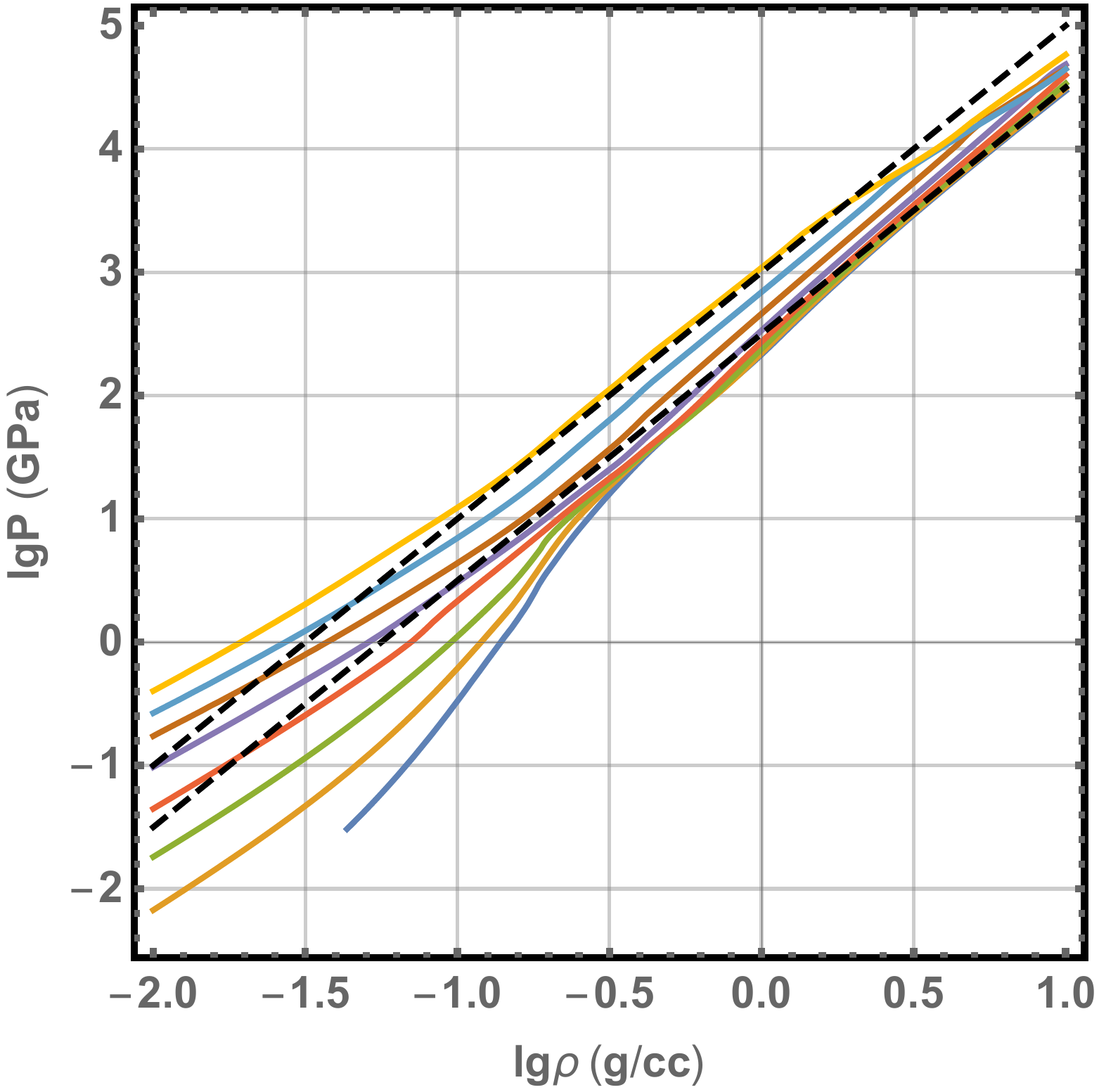}
\caption{black dashed lines correspond to $P \sim \rho^2$ in lg-lg plot. The tangential slope in this plot is $\Gamma \equiv \bigg( \frac{\partial \ln{P}}{\partial \ln{\rho}} \bigg)_{s}$. This is again the $\Gamma$-plot of Hydrogen. This plot suggests that $\Gamma \approx 2$ for fluid metallic hydrogen over many orders-of-magnitude in parameter-space. }
\label{fig:H2_eos_mazevet_2019_isentropes_2}
\end{figure}

The reason can be seen in the mass-radius curve grid itself: 
If one looks at the mass-radius diagram corresponding to Mazevet 2019 Hydrogen EOS (Fig.~\ref{fig:H2EOSMazevet2019MassRadius2}) or Becker 2014 Hydrogen EOS (Fig.~\ref{fig:H2EOSBecker2014MassRadius}), the constant central density contours are almost parallel to the constant bulk density contours (the gray dashed lines in the background marked by 0.3, 1, 3, 10 g/cc). This implies that the interior structures of these planets are homologous to one another. The density everywhere inside a hydrogen planet almost scales linearly with its central density. In most of the parameter space, the isentropes in the pressure-density (P-$\rho$) parameter-space can be very well approximated by the parabolic power law relation of $P \sim \rho^2$ (See Fig.~\ref{fig:H2_eos_mazevet_2019_isentropes_2}). 

\begin{figure}
\centering
    \begin{subfigure}[t]{0.45\textwidth}
        \centering
        \hspace*{-3.5cm}\includegraphics[scale=0.3]{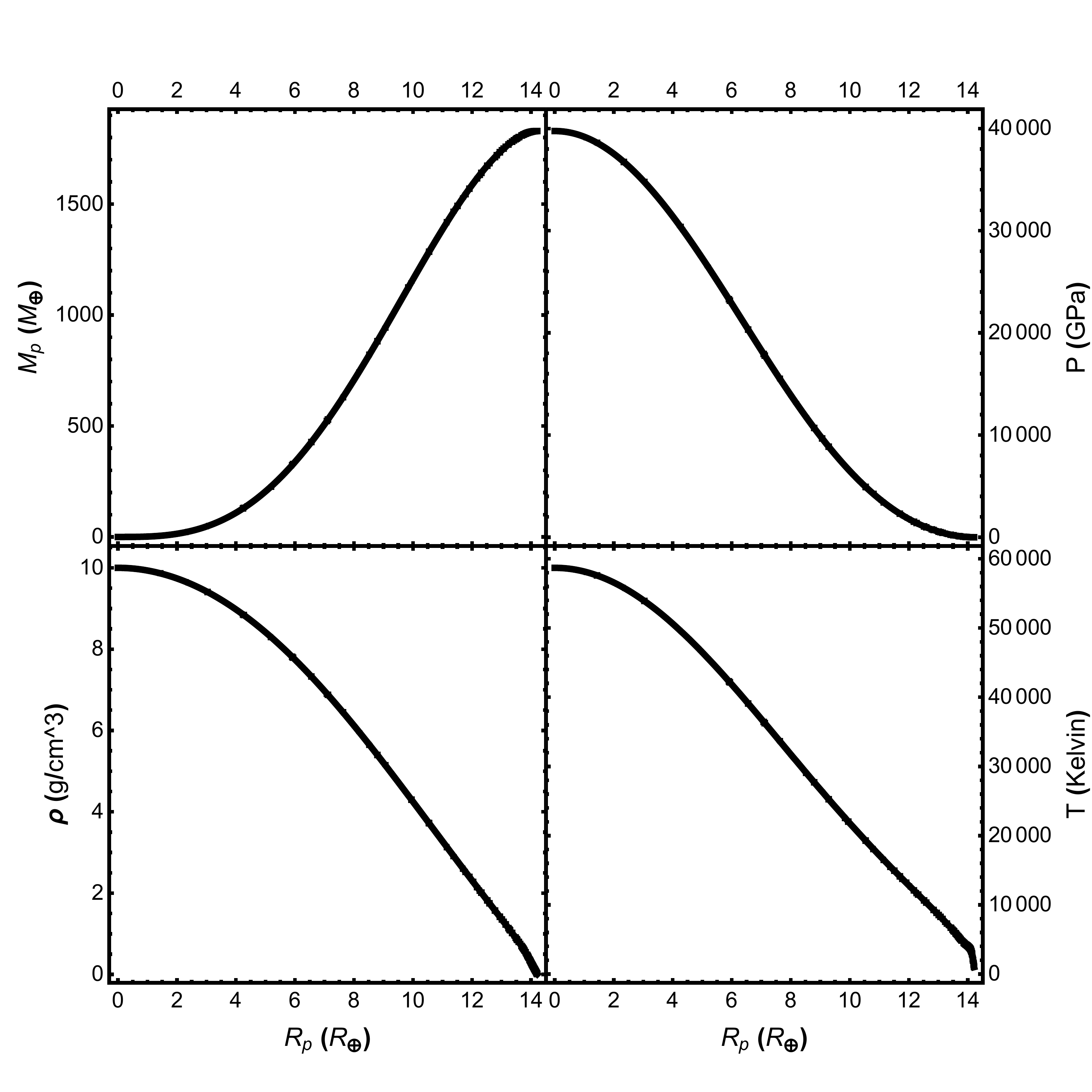}
        \caption{an example of a hot super-Jupiter (1800 M$_{\oplus}$): note that its interior density is almost linear in depth z = (R-r) except at the center, and its interior pressure profile is quadratic in depth, and its temperature profile is almost linear in depth as well. This is all due to $P \sim \rho^2$ being approximately true for metallic fluid hydrogen, and can thus be generalized to all other hot jupiters.  }\label{fig:hot_super_Jupiter}
    \end{subfigure} 
    \hfill
    \begin{subfigure}[t]{0.45\textwidth}
        \centering
        \hspace*{-0.5cm}\includegraphics[scale=0.3]{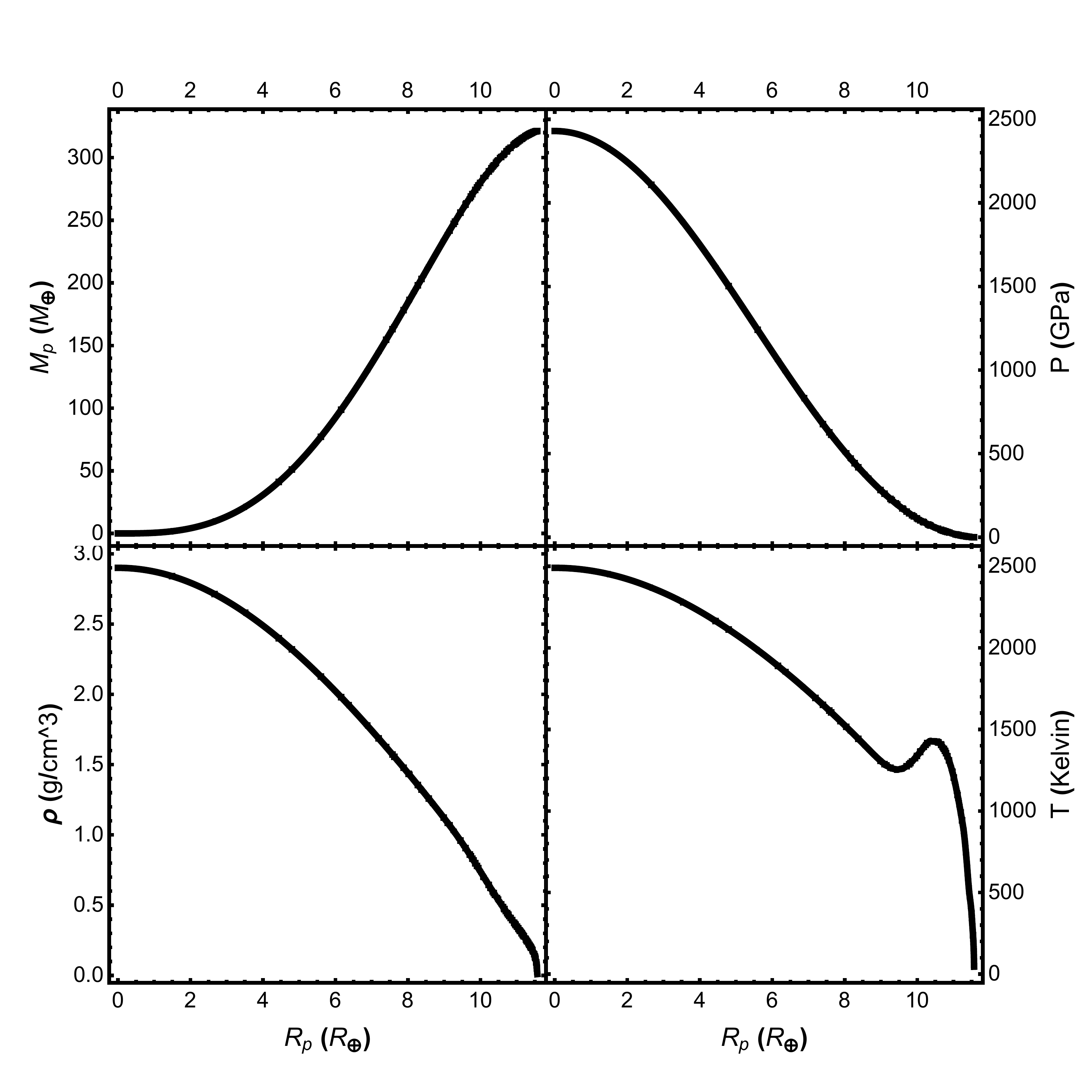}
        \caption{Jupiter (317 M$_{\oplus}$): Its temperature profile shows an oscillation behavior similar to a van der Waals EOS isotherm crossing a phase boundary. Here it is due to the metallization of Hydrogen, also called the PPT--Plasma Phase Transition. In reality, the actual temperature profile should be replaced by something similar to a Maxwell construction.  }\label{fig:Jupiter}
    \end{subfigure} 
\end{figure}

This can be better understood by the following two examples: Fig.~\ref{fig:hot_super_Jupiter}, Fig.~\ref{fig:Jupiter}. The oscillation behavior is general for any \emph{cubic} or higher-order polynomial EOS at phase transition, and can be solved by Maxwell construction. 

\clearpage
%%%%%%%%%%%%%%%%%%%%%%%%%%%%%%%
\section{H$_2$O EOS}

Recent observations have suggested the possible existence of water-rich exoplanets~\citep{PNAS:Zeng2019,Zeng2021}. Thus, the study of H$_2$O EOS and its application to the interior structure modelling of these exoplanets is essential. 

\subsection{Fluid H$_2$O: \\(Gas+Liquid+Supercritical)}

The International Association for the Properties of Water and Steam (IAPWS)~\citep{IAPWS:2016}~\citep{IAPWS:2011} constantly integrates new experimental data and constantly publishes the EOS of H$_2$O in Liquid and Vapor Phases, in tabular forms and also in analytical formulae through interpolation. 

\subsubsection{Critical Point \textbf{C.P.} of H$_2$O: \\Principle of Corresponding States}

It should be noted that J. van der Waals' idea of corresponding states was anticipated by D. Mendeleev (1870), who pointed to the usefulness of comparing volumes not at the boiling points, but at temperatures when the cohesion of the liquids is close to zero~\citep{Karapetyants1978}. This idea is based on the \emph{principle of corresponding states}~\cite{Karapetyants1978}. According to this principle, the same or similar \textbf{common} functional relationship: 

\begin{equation}
    f \Bigg( \frac{\rho}{\rho_{c}}, \frac{T_c}{T}, \frac{P}{P_{c}} \Bigg) = 0
\end{equation}

exists for similar substances. In other words, instead of expressing the properties of a fluid (liquid phase + gas phase) through $\rho$, T, and P, we now use dimensionless units for this purpose -- the reduced parameters: $\delta \equiv \frac{\rho}{\rho_c}$, $\tau \equiv \frac{T_c}{T}$, $\pi \equiv \frac{P}{P_c}$, then we can obtain a reduced \emph{"universal"} EOS $f(\delta,\tau,\pi)=0$. \emph{The goal is one arrow for a thousand birds. }

Many important corollaries follow from this principle. For example, if any two fluids, say H$_2$O and NH$_3$, have identical values of $\pi$ and $\tau$, then they should occupy an approximately identical reduced density $\delta$. However, an EOS characterized by only three constants ($\rho_{c}$,$T_{c}$,$P_{c}$) cannot be infinitely accurate. Therefore, both the principle of corresponding states and the conclusions following from it are only approximate. The errors in calculating various properties typically do not exceed 10\%, especially for homologous substances or substances with close boiling points, such as H$_2$O and NH$_3$. Fundamentally, it results from the similarities in the microscopic structures between H$_2$O and NH$_3$, that both are polar molecules with hydrogen bonds. Moreover, the bond strengths of the covalent bond O-H and N-H are similar, $\sim 100$ kcal/(mol$\cdot$bond), or equivalently, $\sim$4.5 eV/bond. 

The analytical formulae of H$_2$O are thus anchored at its Critical Point \textbf{C.P.}. Here underscript $c$ to denote various parameters at critical point~\citep{NISTWebBookWater}: 

\begin{align}
\rho_{c} &= 0.322 \text{~g/cc} \nonumber\\
T_{c} &= 647.096 \text{~K} = 373.946 ^{\circ}\text{C} \nonumber\\
P_{c} &= 2.2064 \cdot 10^{7} \text{~Pa}= 220.64 \text{~bar} \nonumber\\
n_{c} &= 0.322/18 = 0.018 \text{~mol/cc} \nonumber\\
\end{align}

Here $n_c$ is the molar density in (mol/cc) of water at its critical point. 

Reduced density $\delta$ (dimensionless), 
\begin{equation}
    \delta=\frac{\rho}{\rho_c}
\end{equation}

and \emph{inverse} reduced temperature $\tau$ (dimensionless), 
\begin{equation}
    \tau=\frac{T_c}{T}
\end{equation}

are used as independent variables, to express other thermodynamic variables and functions. 
The reason to use the \emph{inverse} reduced temperature $\tau$ is because generally speaking, the higher the temperature (the smaller the $\tau$), the closer the substance behaves as ideal gas, and thus, the series expansion based on ideal gas EOS is better expressed in $\tau$ rather than $\bigg(\frac{1}{\tau} \bigg)$. 

\subsubsection{Series Expansion based on Ideal-Gas EOS}

All other thermodynamic quantities can be derived from the the specific Helmholtz free energy $f$. (We follow our derivation here as in IAPWS)~\citep{IAPWS:2016}~\citep{IAPWS:2011} 

Fundamentally, the specific Helmholtz free energy $f$ is: 

\begin{equation}
    f = u - T \cdot s
\end{equation}

It can be cast into a dimensionless form as: 

\begin{equation}
    \psi = \frac{f}{R \cdot T} = \psi^{o}(\delta,\tau) + \psi^{r}(\delta,\tau)
\end{equation}

$\psi^{o}$ is the ideal-gas part,
$\psi^{r}$ is the residual part. 

$\psi^{o}$ and $\psi^{r}$ is each expressed as analytic formula in terms of $\delta$ and $\tau$, with coefficients determined by best-fit to experimental data. 

\begin{equation}
    \psi^{o}[\delta,\tau] = \ln{\delta} + n^{o}_1 + n^{o}_2 \cdot \tau + n^{o}_3 \cdot \ln{\tau} + \sum_{i=4}^{8} n^{o}_i \cdot \ln{[1-\exp{(-\gamma^{o}_i \cdot \tau)}]}
\end{equation}

where $n^{o}_i$ and $\gamma^{o}_i$ are dimensionless coefficients. 

Likewise,

\begin{align}
    \psi^{r}[\delta,\tau] = &\sum_{i=1}^{7} n_i \cdot \delta^{d_i} \cdot \tau^{t_i} + \sum_{i=8}^{51} n_i \cdot \delta^{d_i} \cdot \tau^{t_i} \cdot \exp{(-\delta^{c_i})} + \nonumber\\
    &\sum_{i=52}^{54} n_i \cdot \delta^{d_i} \cdot \tau^{t_i} \cdot \exp{(-\alpha_i \cdot (\delta - \epsilon_i)^2-\beta_i \cdot (\tau-\gamma_i)^2)} + \nonumber\\
    &\sum_{i=55}^{56} n_i \cdot \Delta^{b_i} \cdot \delta \cdot \psi
\end{align}

with all kinds of coefficients introduced to fine tune the formula to better match the experimental data: 

\begin{align}
\Delta &= \theta^{2} + B_{i} \cdot [(\delta-1)^2]^{\alpha_i} \nonumber\\
\theta &= (1-\tau)+A_{i} \cdot [(\delta-1)^2]^{\frac{1}{2 \cdot \beta_i}} \nonumber\\
\psi &=\exp{(-C_{i} \cdot (\delta-1)^2-D_{i} \cdot (\tau-1)^2)} \nonumber\\
\end{align}

A specific gas constant for this mass-based formulation is adopted: 

\begin{equation}
    R = 0.46151805~\text{kilojoules} \cdot \text{kg}^{-1} \cdot \text{K}^{-1}
\end{equation}

Much of the complexity is introduced to describe the behavior of H$_2$O around its Critical Point (\textbf{C.P.}), where the pressure changes tremendously over a very small density and temperature range. Beyond \textbf{C.P.}, the distinction between the liquid and vapor phases vanish, and water is supercritical, existing as small but liquid-like hydrogen-bonded clusters dispersed within a gas-like phase~\citep{ChaplinSuperCritical}. 

The pressure $P$ can be calculated from the specific Helmholtz free energy $f$ through fundamental thermodynamic relation: 

\begin{equation}
    P = \rho^2 \cdot \bigg( \frac{\partial f}{\partial \rho} \bigg)_{T}
\end{equation}

and be expressed in dimensionless variables as: 

\begin{equation}
    \frac{P(\delta,\tau)}{\rho \cdot R \cdot T} = 1 + \delta \cdot \bigg( \frac{\partial \psi^{r}}{\partial \delta} \bigg)_{\tau}
\end{equation}

Likewise, all other thermodynamic functions and variables can be derived from the specific Helmholtz free energy $f$ through appropriate partial-derivatives. 

Internal energy (specific) $u$: 

\begin{equation}
    \frac{u(\delta,\tau)}{R \cdot T} = \tau \cdot \Bigg( \bigg( \frac{\partial \psi^{o}}{\partial \tau} \bigg)_{\delta} + \bigg( \frac{\partial \psi^{r}}{\partial \tau} \bigg)_{\delta} \Bigg)
\end{equation}

Entropy (specific) $s$: 

\begin{equation}
    \frac{s(\delta,\tau)}{R} = \tau \cdot \Bigg( \bigg( \frac{\partial \psi^{o}}{\partial \tau} \bigg)_{\delta} + \bigg( \frac{\partial \psi^{r}}{\partial \tau} \bigg)_{\delta} \Bigg) - \psi^{o} -\psi^{r}
\end{equation}

Enthalpy (specific) $h$:

\begin{equation}
    \frac{h(\delta,\tau)}{R \cdot T} = 1 + \tau \cdot \Bigg( \bigg( \frac{\partial \psi^{o}}{\partial \tau} \bigg)_{\delta} + \bigg( \frac{\partial \psi^{r}}{\partial \tau} \bigg)_{\delta} \Bigg) + \delta \cdot \bigg( \frac{\partial \psi^{r}}{\partial \delta} \bigg)_{\tau}
\end{equation}

\subsection{Triple Points (T.P.)}

Triple Points (\textbf{T.P.}) are important anchoring points to make the $\rho$-T phase diagram. We use underscript (t) plus a number to denote different \textbf{T.P.}. 

\subsubsection{Gas-Liquid-Ice Ih T.P. (t1)}
Furthermore, specific internal energy $u$ and specific entropy $s$ of saturated liquid at the Gas-Liquid-Ice Ih triple point have been set to zero to determine the constants of integration. The Gas-Liquid-Ice Ih triple point ($t1$) has the following properties~\citep{NISTWebBookWater}: 

\begin{align}
\rho_{t1} &= 0.99979 \approx 1 \text{~g/cc for liquid} \nonumber\\
T_{t1} &= 273.16 \text{~K} = 0.01 ^{\circ}\text{C} \nonumber\\
P_{t1} &=611.655 \text{~Pa} \approx 0.006 \text{~bar} \nonumber\\
n_{t1} &\approx 1/18 \approx 0.0555 \text{~mol/cc for liquid} \nonumber\\
\end{align}

Since the entire formulation is based on Temperature $T$ and density $\rho$ as independent variables. We ought to construct a $T$-$\rho$ Phase Diagram to visualize the behavior of water. 

The range of IAPWS formulation is the entire stable fluid (including liquid, vapor, and supercritical fluid beyond \textbf{C.P.}). It is experimentally verified for Temperature $T$ in between 273 K and 1273 K, and Pressure $P$ up to $\sim$1 GPa. Tests have shown that this IAPWS analytical formulation of the H$_2$O EOS can be safely extrapolated for density and enthalpy of undissociated H$_2$O up to $\sim$5000 K and $\sim$100 GPa at least. 

\begin{figure}[h!]
\centering
\hspace*{-2cm}\includegraphics[scale=0.4]{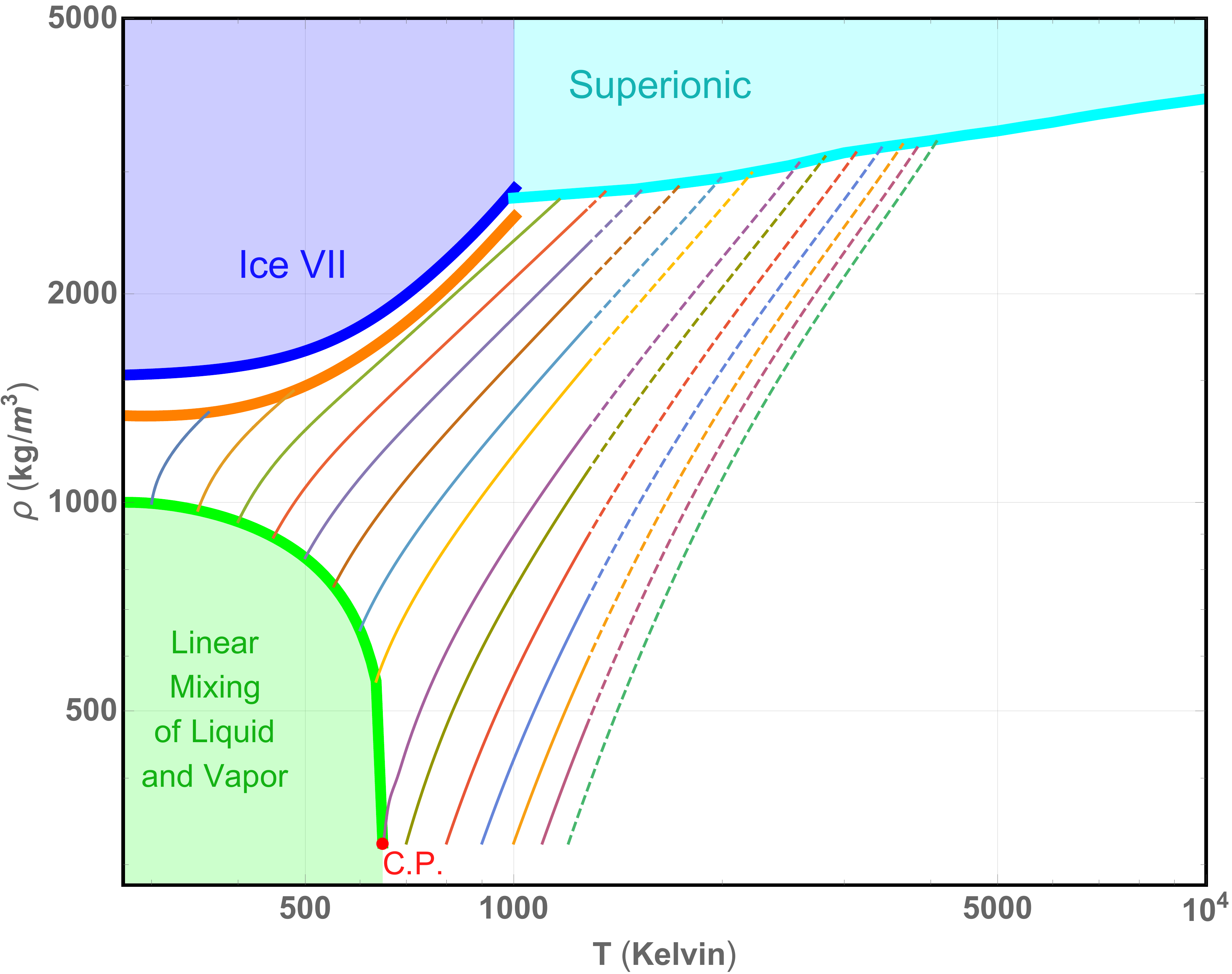}
\caption{Isentropes of Fluid-H$_2$O in T-$\rho$ space. \textbf{C.P.} stands for the Critical Point between Liquid-Gas-Supercritical Fluid. Dashed curves are extrapolations from experimentally validated regimes. The tangential slope in this plot is $\gamma \equiv \bigg( \frac{\partial \ln{T}}{\partial \ln{\rho}} \bigg)_{s}$. This is the $\gamma$-plot of water. }
\label{fig:isentropes_IAPWS_plus_ab_initio_2019_3}
\end{figure}

The isentropes (the adiabatic curves which most likely resemble the $T$-$\rho$ profiles in the interiors of H$_2$O-layer in a planetary body) are calculated and shown here (Fig.~\ref{fig:isentropes_IAPWS_plus_ab_initio_2019_3}). All the isentropes start from the Liquid-Vapor phase boundary (the solid green curve), except the one above \textbf{C.P.} that are calculated from the isochor of $\rho$=0.322 g/cc. The dashed part of the curves are extrapolation from the experimentally verified regime, all the way up to the Fluid-Solid phase boundary. Here, the Fluid-Solid phase boundary is the boundary of Fluid-Ice VII and Fluid-Superionic Ice. 

\subsubsection{Fluid-Ice VII-Ice XVIII T.P. (t2)}

The superionic ice is also called Ice XVIII (Ice-eighteen)~\citep{ChaplinSuperIonic}. It is now experimentally verified~\citep{Millot2019}. 
The triple point ($t2$) of Fluid-Ice VII-Superionic Ice is another important anchor point for plotting the phase diagram, with approximate temperature of $\sim$1000 K and pressure of $\sim$50 GPa. 

\begin{align}
\rho_{t2} &\sim 2.6 \text{~g/cc for fluid} \nonumber\\
T_{t2} &\sim 1000 \text{~K} \nonumber\\
P_{t2} &\sim 50 \text{~GPa} \approx 5\cdot10^5 \text{~bar} \nonumber\\
n_{t2} &\approx 2.6/18 \approx 0.144 \text{~mol/cc for liquid} \nonumber\\
\end{align}

\subsubsection{Ice VI-Ice VII-Liquid T.P. (t3)}

\begin{align}
T_{t3} &= 355 \text{~K} \nonumber\\
P_{t3} &= 2.216 \text{~GPa} = 2.216 \cdot 10^4 \text{~bar} \nonumber\\
\rho_{\text{VI}} &= 1.31 \text{~g/cc for Ice VI} \nonumber\\
\rho_{\text{VII}} &= 1.567 \text{~g/cc for Ice VII} \nonumber\\
\rho_{\text{Liq}} &= 1.35 \text{~g/cc for liquid} \nonumber\\
n_{\text{VI}} &\approx 1.31/18 \approx 0.0728 \text{~mol/cc for Ice VI} \nonumber\\
n_{\text{VII}} &\approx 1.567/18 \approx 0.0871 \text{~mol/cc for Ice VII} \nonumber\\
n_{\text{Liq}} &\approx 1.35/18 \approx 0.075 \text{~mol/cc for liquid} \nonumber\\
\end{align}

where $n$ is the molar density of each phase in (mol/cc) at this triple point.

\subsection{Solid: \\Various Forms of Ices}

Each form of ice requires its own EOS, because of the difference in crystal structure and compressibility among the different ice forms. There exist solid-solid phase transitions among the different ice forms. 

\subsubsection{General EOS of Ices}

Ices can assume different crystal structures, depending on the temperature and density (or pressure) conditions. For a crystal, its microscopic structure dictates its macroscopic properties. The Ice Ih (Ice-one-hexagonal) is the common form of ice found in ambient conditions on Earth surface. Its low density is due to loose packing, which is a result of the orientation of hydrogen bonding. However, in the deep interior of planet, subject to tremendous amount of pressure and heat, the ice crystals change its packing to become more symmetrical and denser. One common form of such a high-pressure ice phase is Ice VII (Ice-seven). It is believed to exist in the interiors of icy satellites such as Europa and Ganymede. Recent theoretical and experimental works suggest the de-localization or mobilization of protons in ice crystals above a certain pressure threshold. The free protons (hydrogen ions) within the lattices of oxygen ions can flow and give rise to moderate amount of electric conductivity. This new phase is thus termed the Superionic ice phase~\citep{ChaplinSuperIonic}. Although the protons are mobile and fluidic, it is still considered a solid phase, because the oxygen ions are fixed in the crystal lattice. 

The pressure-EOS of any ice, and in general, for any solids, can be constructed as follows: 

\begin{equation}
    P_{\text{total}} = P_{\text{elastic}} (\rho) + P_{\text{thermal}} (\rho,T) + P_{\text{electronic}} (\rho,T)
\end{equation}
\begin{itemize}
    \item $P_{\text{elastic}}$ is the pressure resulting from cold elastic compression, and is a function of density $\rho$ only. 
    \item $P_{\text{thermal}}$ is the lattice vibrational contribution, and is a function of both density $\rho$ and temperature T. 
    \item $P_{\text{electronic}}$ comes from the heat capacity of electrons calculated from finite-temperature Fermi-Dirac distribution, and is a function of both material density $\rho$ and temperature T. In this context, we ignore it for now. 
\end{itemize}

\subsubsection{Born-Mie EOS}

A general form of $P_{\text{elastic}}$ called Born-Mie EOS is constructed from a modified power law, with one attractive term and one repulsive term: 

\begin{equation}\label{Eq:BornMieEOS}
    P_{\text{Born-Mie}} (\rho) = \frac{3 \cdot  K_{0}}{m-n} \cdot \Bigg[ \bigg(\frac{\rho}{\rho_{0}} \bigg)^{\frac{m+3}{3}} - \bigg(\frac{\rho}{\rho_{0}} \bigg)^{\frac{n+3}{3}} \Bigg]
\end{equation}

where m and n are constants. 

Note that the pressure increment $dP$ in Born-Mie EOS can be written as: 

\begin{equation}\label{Eq:BornMieEOS2}
    dP_{\text{Born-Mie}} (\rho) = \frac{3 \cdot  K_{0}}{m-n} \cdot \Bigg[ \frac{m+3}{3} \cdot \bigg(\frac{\rho}{\rho_{0}} \bigg)^{\frac{m}{3}} - \frac{n+3}{3} \cdot \bigg(\frac{\rho}{\rho_{0}} \bigg)^{\frac{n}{3}} \Bigg] \cdot d\bigg(\frac{\rho}{\rho_0} \bigg)
\end{equation}

Thus, the bulk modulus $K$ at any density can be written as:

\begin{equation}\label{Eq:BornMieEOS3}
    K_{\text{Born-Mie}} = \frac{dP}{d\ln{(\rho/\rho_0)}} = \frac{3 \cdot  K_{0}}{m-n} \cdot \Bigg[ \frac{m+3}{3} \cdot \bigg(\frac{\rho}{\rho_{0}} \bigg)^{\frac{m+3}{3}} - \frac{n+3}{3} \cdot \bigg(\frac{\rho}{\rho_{0}} \bigg)^{\frac{n+3}{3}} \Bigg]
\end{equation}

The bulk modulus increment $dK$ in Born-Mie EOS can be written as: 

\begin{equation}\label{Eq:BornMieEOS4}
    dK = \frac{3 \cdot  K_{0}}{m-n} \cdot \Bigg[ \bigg( \frac{m+3}{3} \bigg)^2 \cdot \bigg(\frac{\rho}{\rho_{0}} \bigg)^{\frac{m}{3}} - \bigg( \frac{n+3}{3} \bigg)^2 \cdot \bigg(\frac{\rho}{\rho_{0}} \bigg)^{\frac{n}{3}} \Bigg] \cdot d\bigg(\frac{\rho}{\rho_0} \bigg)
\end{equation}

Thus, the derivative of $K$ with respect to $P$, can be written as: 

\begin{equation}\label{Eq:BornMieEOS5}
    K' \equiv \frac{dK}{dP} = \ddfrac{\bigg( \frac{m+3}{3} \bigg)^2 \cdot \bigg(\frac{\rho}{\rho_{0}} \bigg)^{\frac{m}{3}} - \bigg( \frac{n+3}{3} \bigg)^2 \cdot \bigg(\frac{\rho}{\rho_{0}} \bigg)^{\frac{n}{3}}}{\frac{m+3}{3} \cdot \bigg(\frac{\rho}{\rho_{0}} \bigg)^{\frac{m}{3}} - \frac{n+3}{3} \cdot \bigg(\frac{\rho}{\rho_{0}} \bigg)^{\frac{n}{3}}}
\end{equation}

The asymptotic behavior of $K' (\equiv dK/dP)$ (dimensionless) is as follows: 

\begin{subequations}\label{Eq:BornMieEOSdKdP}
\begin{empheq}[left=\empheqlbrace]{align}
    \label{Eq:BornMieEOSdKdP1}
         &K'_{0} \equiv \frac{dK}{dP} \bigg|_{\rho=\rho_0} =\frac{(m+3)+(n+3)}{3},\\
    \label{Eq:BornMieEOSdKdP2}
         &K'_{\infty} \equiv \frac{dK}{dP} \bigg|_{\rho \to +\infty} =\frac{m+3}{3}.
\end{empheq}
\end{subequations}

Therefore, $K' (\equiv dK/dP)$ is a decreasing function in $\rho$ or $P$. In principle, both $K'_{0}$ and $K'_{\infty}$ can be determined by experiments. 

In some cases, it is convenient to express $m$ and $n$ in terms of $K'_{0}$ and $K'_{\infty}$:

\begin{subequations}\label{Eq:BornMieEOSdKdP2}
\begin{empheq}[left=\empheqlbrace]{align}
    \label{Eq:BornMieEOSdKdP21}
         &n = 3 \cdot (K'_{0} - K'_{\infty}) - 3,\\
    \label{Eq:BornMieEOSdKdP22}
         &m = 3 \cdot K'_{\infty} - 3.
\end{empheq}
\end{subequations}

Then, Born-Mie EOS can be re-written as: 

\begin{equation}\label{Eq:BornMieEOS6}
    P_{\text{Born-Mie}} (\rho) = \frac{K_{0}}{(2 \cdot K'_{\infty} - K'_{0})} \cdot \Bigg[ \bigg(\frac{\rho}{\rho_{0}} \bigg)^{K'_{\infty}} - \bigg(\frac{\rho}{\rho_{0}} \bigg)^{(K'_{0} - K'_{\infty})} \Bigg]
\end{equation}

Thus, it is parametrized by three parameters ($K_{0}$, $K'_{0}$, and $K'_{\infty}$) to completely determine its functional form. 

Sometimes, the number of required parameters can be reduced to two, if one takes $K'_{\infty} \approx 2$, and Eq.~\ref{Eq:BornMieEOS6} can be re-written as~\citep{Holzapfel2018}:

\begin{equation}\label{Eq:BornMieEOS7}
    P_{\text{Born-Mie}} (\rho) = \frac{K_{0}}{(4 - K'_{0})} \cdot \Bigg[ \bigg(\frac{\rho}{\rho_{0}} \bigg)^{2} - \bigg(\frac{\rho}{\rho_{0}} \bigg)^{(K'_{0} - 2)} \Bigg]
\end{equation}

Eq.~\ref{Eq:BornMieEOS7} only needs two parameters ($K_{0}$, $K'_{0}$). 

%%%BM2 EOS%%%%%%%%%%%%%%%%%%%%%%%%%%%
\subsubsection{Birch-Murnagham 2nd-order (BM2) EOS}
For our purpose, a more specific analytic formula called the Birch-Murnagham 2nd-order (BM2) EOS is adopted by setting m=4 and n=2 , to approximate $P_{\text{elastic}}$, due to its compliance with experimental high-pressure data of mineral compression, and good asymptotic behavior during extrapolation. BM2 EOS has been experimentally verified for a wide range of minerals under high pressure. 

\begin{equation}\label{Eq:BM2EOS}
    P_{\text{BM2}} (\rho) = \frac{3}{2} \cdot K_{0} \cdot \Bigg( \bigg(\frac{\rho}{\rho_{0}} \bigg)^{7/3} - \bigg(\frac{\rho}{\rho_{0}} \bigg)^{5/3} \Bigg)
\end{equation}

\begin{itemize}
\item $\rho_{0}$ is a reference density at ambient conditions
\item $K_{0}$ is a reference isothermal bulk modulus at ambient conditions
\end{itemize}
Both $\rho_{0}$ and $K_{0}$ are inputs in BM2 EOS for a particular crystal phase. If this particular phase does not exist at ambient conditions (ordinary temperature and pressure), and then appropriate nominal extrapolated values are adopted, such as the case for Ice VII. 

Note that the pressure increment $dP$ in BM2 EOS can be written as: 

\begin{equation}\label{Eq:BM2EOS2}
    dP = \frac{3}{2} \cdot K_{0} \cdot \Bigg(\frac{7}{3} \cdot \bigg(\frac{\rho}{\rho_{0}} \bigg)^{4/3} - \frac{5}{3} \cdot \bigg(\frac{\rho}{\rho_{0}} \bigg)^{2/3} \Bigg) \cdot d\bigg(\frac{\rho}{\rho_0} \bigg)
\end{equation}

Thus, the bulk modulus $K$ at any density can be written as:

\begin{equation}\label{Eq:BM2EOS3}
    K = \frac{dP}{d\ln{(\rho/\rho_0)}} = \frac{3}{2} \cdot K_{0} \cdot \Bigg(\frac{7}{3} \cdot \bigg(\frac{\rho}{\rho_{0}} \bigg)^{7/3} - \frac{5}{3} \cdot \bigg(\frac{\rho}{\rho_{0}} \bigg)^{5/3} \Bigg)
\end{equation}

The bulk modulus increment $dK$ in BM2 EOS can be written as: 

\begin{equation}\label{Eq:BM2EOS4}
    dK = \frac{3}{2} \cdot K_{0} \cdot \Bigg(\bigg(\frac{7}{3}\bigg)^2 \cdot \bigg(\frac{\rho}{\rho_{0}} \bigg)^{7/3} - \bigg(\frac{5}{3}\bigg)^2 \cdot \bigg(\frac{\rho}{\rho_{0}} \bigg)^{5/3} \Bigg) \cdot d\bigg(\frac{\rho}{\rho_0} \bigg)
\end{equation}

Thus, the derivative of $K$ with respect to $P$, can be written as: 

\begin{equation}\label{Eq:BM2EOS5}
    K' \equiv \frac{dK}{dP} = \ddfrac{\bigg(\frac{7}{3}\bigg)^2 \cdot \bigg(\frac{\rho}{\rho_{0}} \bigg)^{7/3} - \bigg(\frac{5}{3}\bigg)^2 \cdot \bigg(\frac{\rho}{\rho_{0}} \bigg)^{5/3}}{\bigg(\frac{7}{3}\bigg) \cdot \bigg(\frac{\rho}{\rho_{0}} \bigg)^{7/3} - \bigg(\frac{5}{3}\bigg) \cdot \bigg(\frac{\rho}{\rho_{0}} \bigg)^{5/3}}
\end{equation}

The asymptotic behavior of $K' (\equiv dK/dP)$ (dimensionless) is as follows: 

\begin{subequations}\label{Eq:BM2dKdP}
\begin{empheq}[left=\empheqlbrace]{align}
    \label{Eq:BM2dKdP1}
         &K'_{0} \equiv \frac{dK}{dP} \bigg|_{\rho=\rho_0} =4,\\
    \label{Eq:BM2dKdP2}
         &K'_{\infty} \equiv \frac{dK}{dP} \bigg|_{\rho \to +\infty} =7/3\approx2.33.
\end{empheq}
\end{subequations}

The behavior of $K' (\equiv dK/dP)$ is illustrated in Fig.~\ref{fig:BM2dKdP}: 

\begin{figure}[h!]
\centering
\includegraphics[scale=0.45]{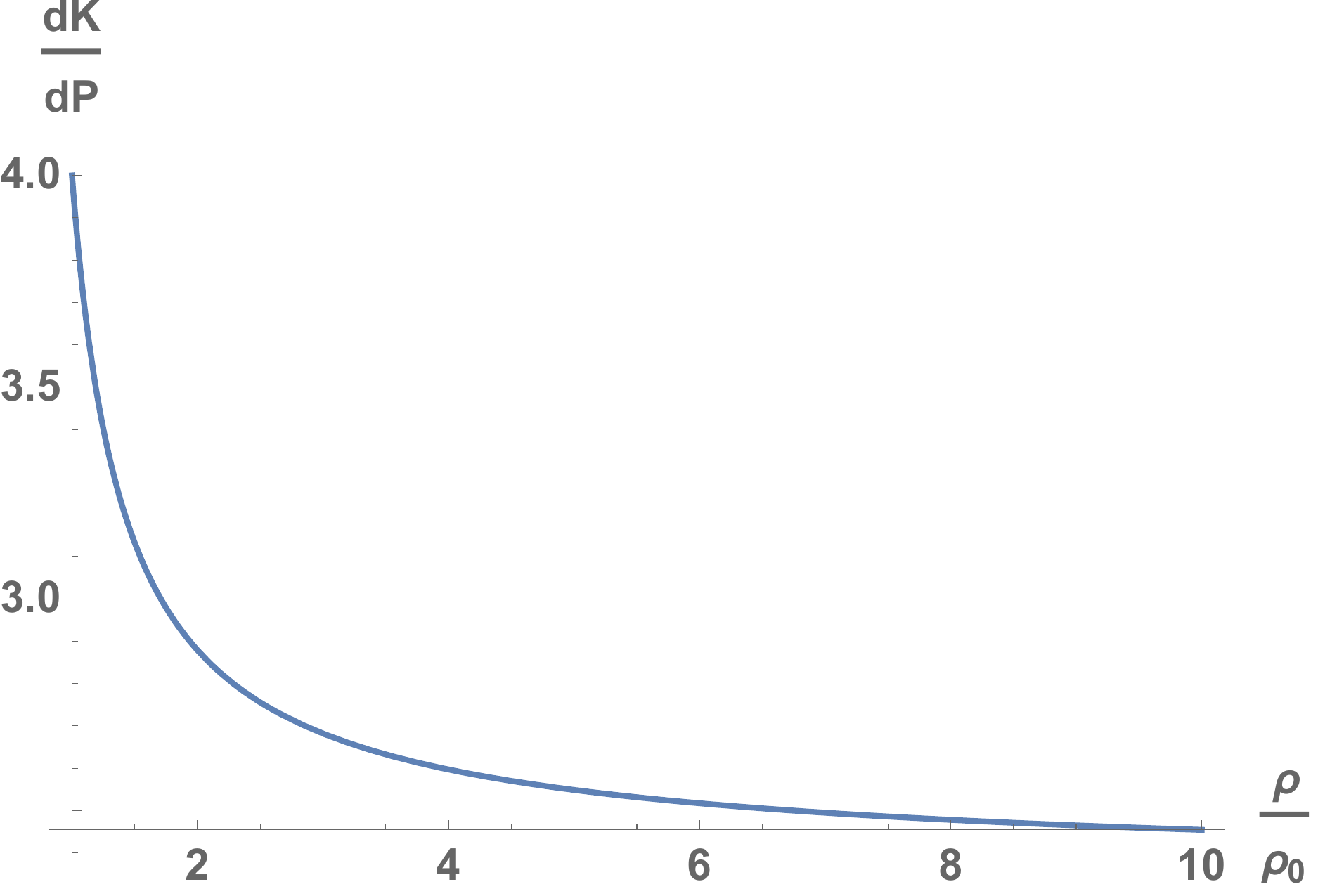}
\caption{$K' (\equiv dK/dP)$ versus $\rho/\rho_0$ }
\label{fig:BM2dKdP}
\end{figure}

The initial value of $K'_{0} = 4$ is close to many experimentally-determined values for silicates through static compression. 

\subsubsection{Thermal Pressure: \\Debye Model}
On the other hand, we approximate $P_{\text{thermal}}$ with the Debye Model~\cite{Douce2011} for $T > \theta_D$, which is generally true for planetary interiors deeper than a few kilometers: 

\begin{equation}
    P_{\text{Debye}} (\rho,T) = \gamma \cdot \frac{E_{\text{vib}}}{V} \approx (\alpha \cdot K_T)_D \cdot (T-0.23 \cdot \theta_D)
\end{equation}

\begin{itemize}
\item $\theta_D$ is the Debye temperature
\item $\alpha$ is the thermal expansion coefficient, in unit of \emph{inverse} temperature (K$^{-1}$)
\item $K_T$ is the isothermal bulk modulus, in unit of pressure (GPa)
\end{itemize}
where the underscript $D$ of $(\alpha \cdot K_T)_D$ means that the product $(\alpha \cdot K_T)$ is evaluated at the Debye temperature $\theta_D$. For $T > \theta_D$, $(\alpha \cdot K_T)$ stays approximately constant, because of one important thermodynamic identity: 

\begin{equation}\label{Eq:identity}
    \alpha \cdot K_T = \gamma \cdot \rho \cdot C_v
\end{equation}

\begin{itemize}
\item $\gamma$ is the Gr\"{u}neisen Parameter
\item $\rho$ is density
\item $C_v$ is the isochoric heat capacity
\end{itemize}
Above $\theta_D$, $C_v$ approaches a constant of $3 \cdot R$ per mole of atoms, known as the Dulong-Petit limit. This limit is because above the Debye temperature $\theta_D$, each atom in the crystal lattice realizes its full degree of freedom and becomes equivalent to an independent 3-D harmonic oscillator. 
Therefore, the above expression of $P_{\text{thermal}}$ can be further simplified into the following form: 

\begin{equation}
    P_{\text{thermal}} (\rho,T) = a \cdot (T-b)
\end{equation}
where $a = \alpha \cdot K_T$ and $b = 0.23 \cdot \theta_D$ are constants to be determined, either by theoretical approaches or by fit to experimental data. 

In summary, we write the analytic form of pressure in a solid phase as: 

\begin{equation}\label{Eq:pressure}
    P (\rho,T) = \frac{3}{2} \cdot K_{0} \cdot \Bigg[ \bigg(\frac{\rho}{\rho_{0}} \bigg)^{7/3} - \bigg(\frac{\rho}{\rho_{0}} \bigg)^{5/3} \Bigg] + a \cdot (T-b)
\end{equation}

From this relation of Eq.~\ref{Eq:pressure}, one can easily calculate the density-pressure ($\rho$-P) relation for any isotherm (with constant temperature). Because of the nature of atomic vibration in a crystal lattice, this equation tells us that regardless of the density or pressure, the thermal pressure contribution as described by Debye model is only a function of temperature. This allows the separation of variables and simplifies our calculations. 

\subsection{Ice-XVIII: Superionic Ice}

This sub-section deals with calculating the thermodynamic parameters of superionic ice, from the standpoint of BM2 EOS for cold compression, and Debye model for thermal pressure, as introduced above. 

\subsubsection{Calculating Isentropes of Superionic Ice}

The fit of the Eq.~\ref{Eq:pressure} to the experimental and ab initio simulation data of superionic ice~\cite{Millot2018ExperimentalCompression} give the following optimal fit values for the superionic phase: 

\begin{align}
K_{0} &= 110 \text{~GPa} \nonumber\\
\rho_{0} &= 2.1 \text{~g/cc} \nonumber\\
a &=0.0117 \text{~GPa/Kelvin}=\frac{1}{85.5}\cdot \frac{\text{GPa}}{\text{K}} \nonumber\\
b &=300 \text{~Kelvin} \nonumber\\
\theta_{D} &=\frac{b}{0.23} \approx 1300 \text{~Kelvin} \nonumber\\
c_v &=\frac{9R}{\mu} = 4.157 \cdot \frac{\text{J/K}}{g}
\end{align}

Since

\begin{equation}\label{Eq:isentropeDiff}
   \gamma \cdot \rho = \Bigg( \frac{d\ln{T}}{d\ln{\rho}} \Bigg)_s \cdot \rho = -\Bigg[ \frac{d\ln{T}}{d\ln{(1/\rho)}} \Bigg]_s = \rho_{\text{const}}
\end{equation}

From Eq.~\ref{Eq:identity}, 

\begin{equation}
    \rho_{\text{const}} = \frac{\alpha \cdot K_T}{c_v} = \frac{a}{c_v} = 2.8145 \text{~g/cc}
\end{equation}

Then, Eq.~\ref{Eq:isentropeDiff} can be integrated to find the functional relationship between $T$ and $\rho$ along an isentrope: 

\begin{equation}
    T = T_{0} \cdot \exp{\Bigg(-\frac{\rho_{\text{const}}}{\rho} \Bigg)}
\end{equation}

where $T_{0}$ is constant of integration and is determined by the initial condition. The initial condition refers to the specific entropy $s$ this particular isentrope corresponds to. 

Then, the total pressure along an isentrope is expressed in the following analytic form: 

\begin{equation}\label{Eq:pressureisentrope}
    P_{\text{isentrope}} (\rho) = \frac{3}{2} \cdot K_{0} \cdot \Bigg[ \bigg(\frac{\rho}{\rho_{0}} \bigg)^{7/3} - \bigg(\frac{\rho}{\rho_{0}} \bigg)^{5/3} \Bigg] + a \cdot (T_{0} \cdot e^{-\frac{\rho_{\text{const}}}{\rho}}-b)
\end{equation}

One has to be cautious as this relation of Eq.~\ref{Eq:pressureisentrope} only applies to within a single solid phase. If the isentrope encounters a phase boundary, such as a Fluid-Solid phase transition, or a Solid-Solid phase transition, then there will be a discontinuity in the specific entropy, and a discontinuity in the density (or equivalently specific volume) as well. Then this calculation needs to be carried out separately for each phase. 

\begin{figure}[h!]
\centering
\hspace*{-3.5cm}\includegraphics[scale=0.45]{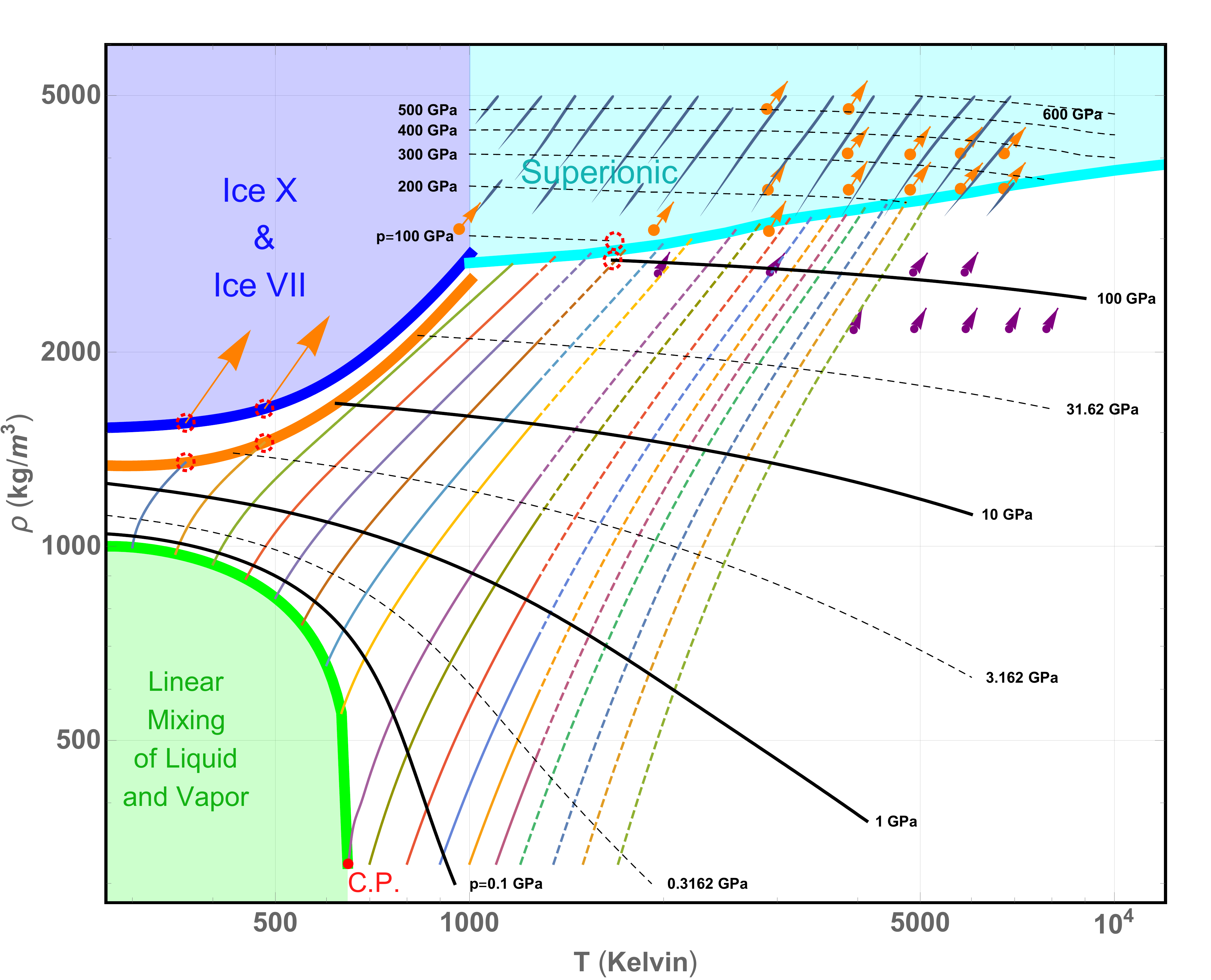}
\caption{Isentropes of fluid-H$_2$O and superionic-H$_2$O. The arrows (orange for superionic and purple for high-pressure fluid phase) show the local slope of an isentrope, that is, the Gr\"{u}neisen parameter $\gamma$ which is calculated from applying Eq.~\ref{Eq:GruneisenParameter1} to the ab initio simulation data of~\cite{Millot2018ExperimentalCompression}. Pressure contours are shown and labelled with its value in GPa. The density discontinuity between the fluid and superionic phases is not shown in this plot, but will be elaborated in more detail in the next section. }
\label{fig:isentropes_IAPWS_plus_ab_initio_2019_7}
\end{figure}

The isentropes of the superionic ice (Ice-eighteen) can be calculated from the ab initio simulation data from~\cite{Millot2018ExperimentalCompression}, as shown in Fig.~\ref{fig:isentropes_IAPWS_plus_ab_initio_2019_7}. The local slope of isentrope, i.e., the Gr\"{u}neisen parameter $\gamma$, is calculated from Eq.~\ref{Eq:GruneisenParameter1}. 

\begin{figure}[h!]
\centering
\hspace*{-0.5cm}\includegraphics[scale=0.45]{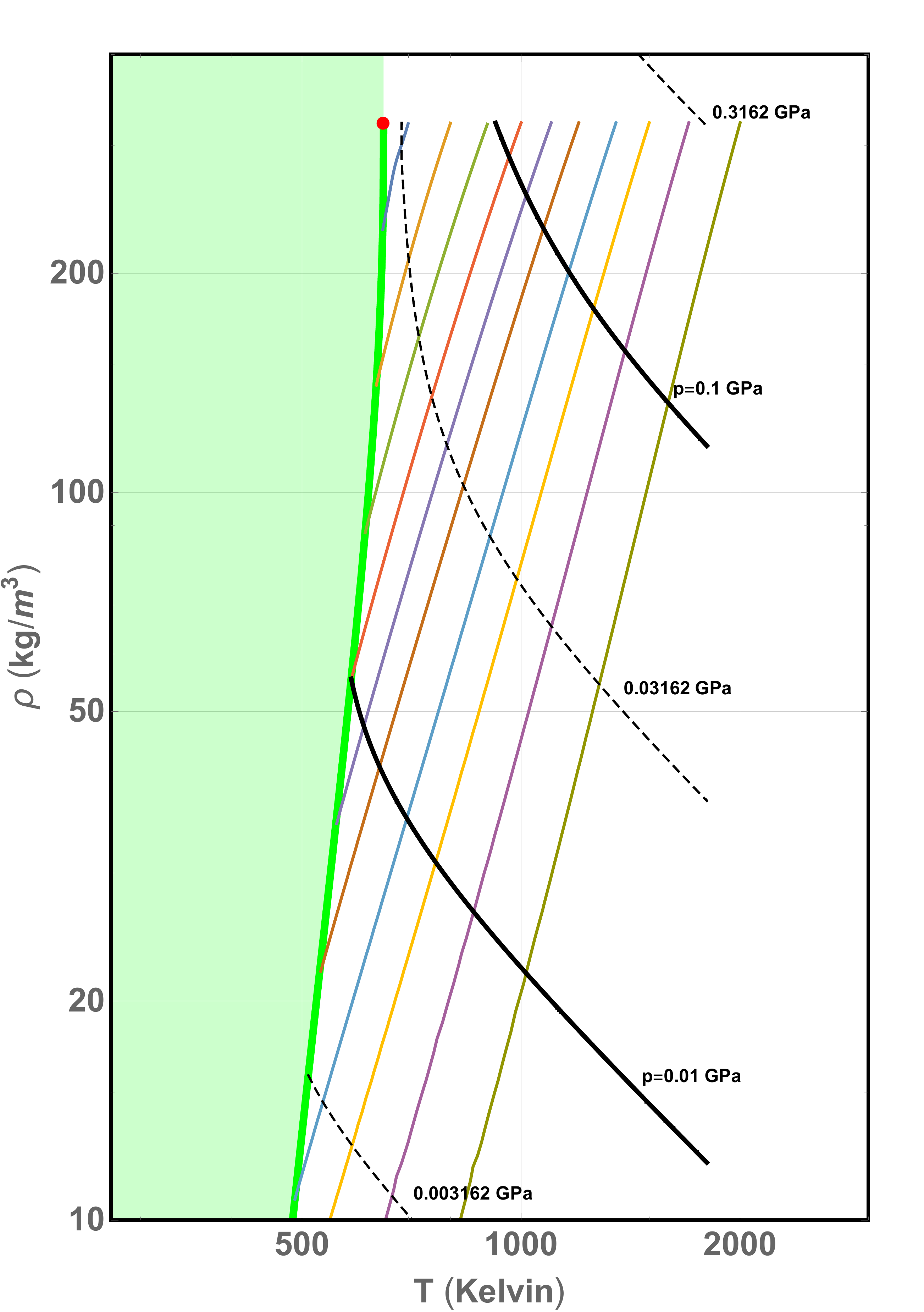}
\caption{Isentropes (colored) and Isobars (black) of H$_2$O below and near its critical point. }
\label{fig:isentropes_IAPWS_plus_ab_initio_2019_8}
\end{figure}

It is also quite interesting to see how the isentropes behave below the density of critical point (Fig.~\ref{fig:isentropes_IAPWS_plus_ab_initio_2019_8}). This regime people often refer to as \emph{Real Gas}, as it is still considered a gas phase, while the inter-molecular interaction is quite strong and cannot be neglected. One such type of EOS that we will introduce later on is called the \emph{Berthelot EOS}. 

Sometimes, it is useful to view these isentropes in a linear-plot of density $\rho$ rather logarithmic. Thus, we come up with the following Fig.~\ref{fig:h2o_eos_log_linear_plot_2019}: 

\begin{figure}[h!]
\centering
\hspace*{-3.5cm}\includegraphics[scale=0.45]{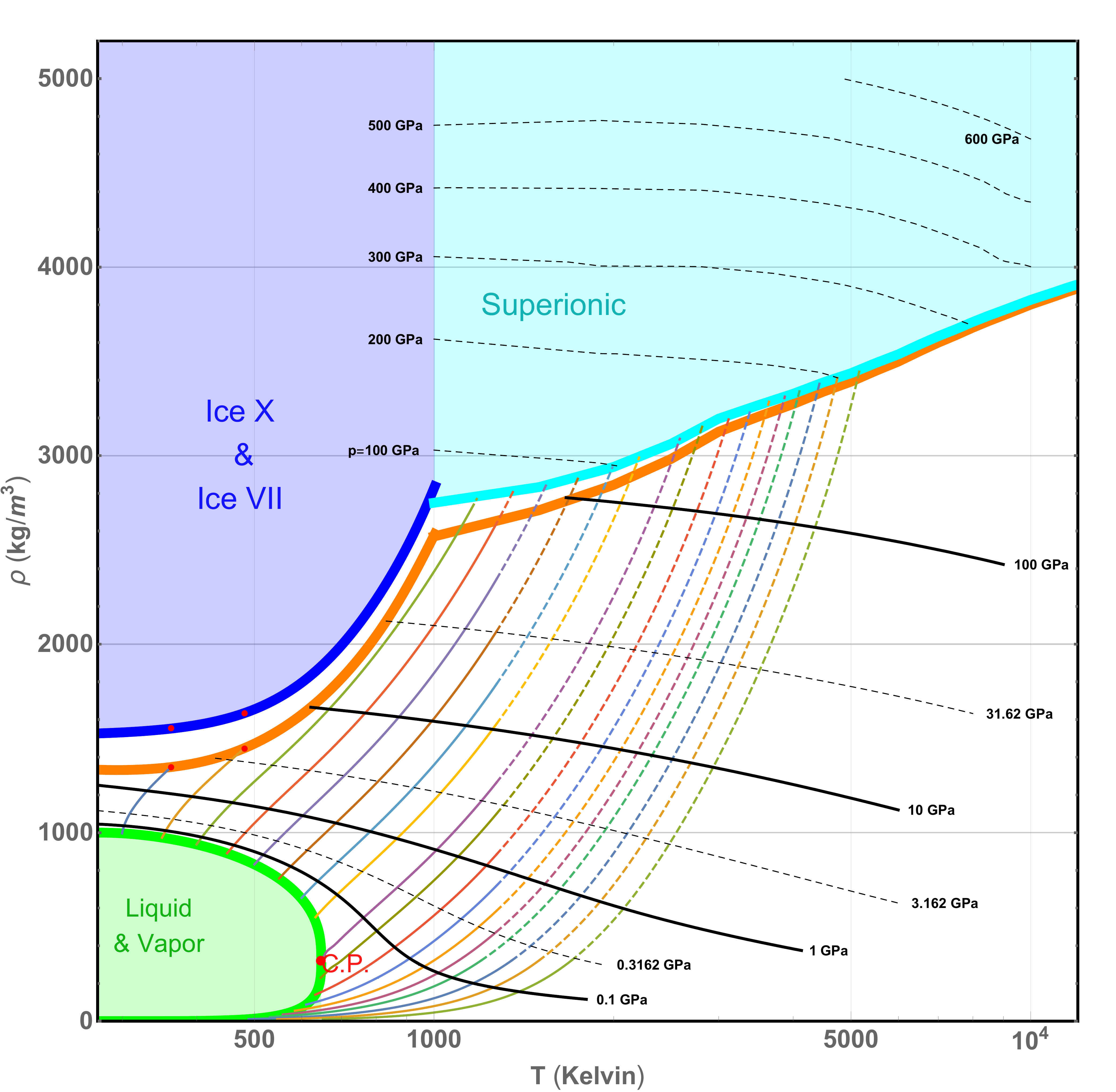}
\caption{Isentropes of H$_2$O in linear-$\rho$ and $\ln{T}$ plot. }
\label{fig:h2o_eos_log_linear_plot_2019}
\end{figure}

\subsubsection{Calculating Melting Curve of Superionic Ice}

The \emph{melting curve} here defines the phase boundary between the liquid and the superionic ice (also known as \textbf{Ice-eighteen}~\cite{ChaplinSuperIonic}). It has a shallower slope compared to the isentropes in the same regime in the temperature-density ($T$-$\rho$) phase diagram. It can be calculated according to Eq.~\ref{Eq:meltingcurveslope2}. 

\begin{figure}[h!]
\centering
\hspace*{-3.5cm}\includegraphics[scale=0.6]{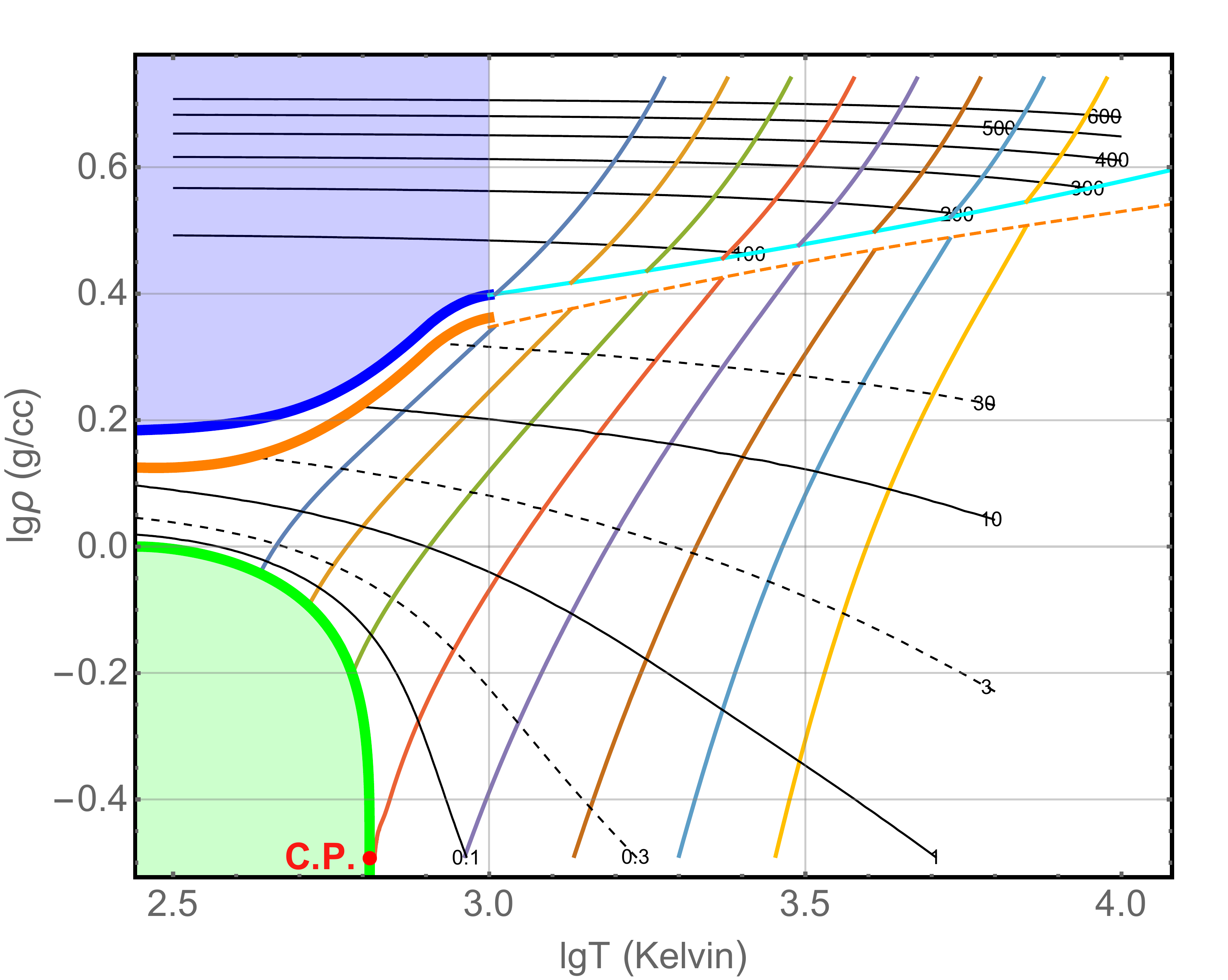}
\caption{Eight isentropes of fluid-H$_2$O and superionic-H$_2$O, plus the density discontinuity at the phase boundary, are chosen here for the purpose of calculating mass-radius curves of pure-H$_2$O planets. Pressure contours are shown in black. Their smooth behavior suggests a possible approximation by power-law for future simplification of calculations. That is $\gamma \approx$ constant.}
\label{fig:H2O_phase_diagram_Mesh_20190428}
\end{figure}

With this knowledge in mind, we refine the High-Pressure Phase Diagram of H$_2$O with the density jump at the phase transition between the fluid and superionic. See Fig.~\ref{fig:H2O_phase_diagram_Mesh_20190428}. 

\begin{figure}[h!]
\centering
\hspace*{-2cm}\includegraphics[scale=0.57]{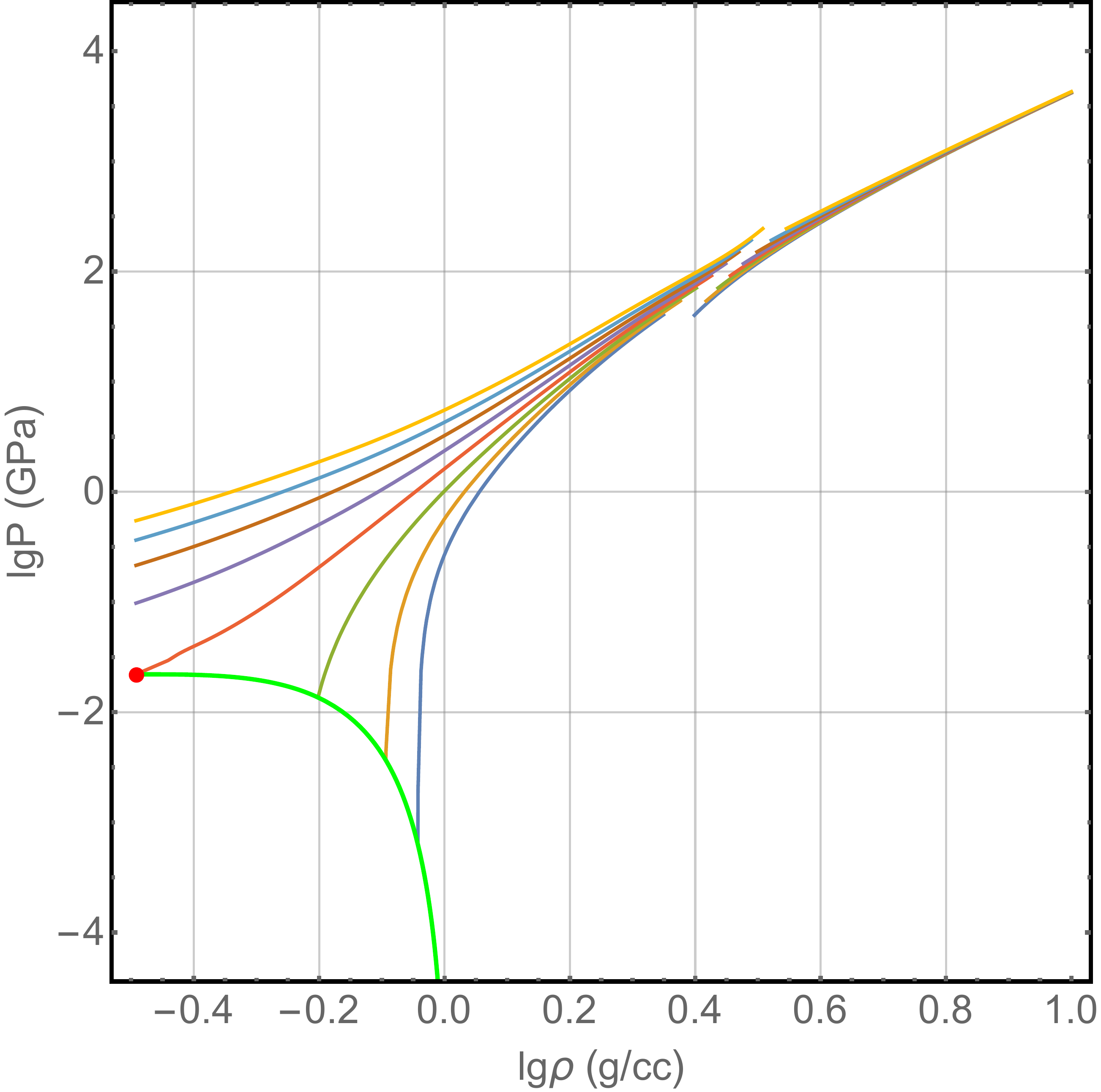}
\caption{Density-Pressure ($\rho$-$P$) plot of isentropes of H$_2$O. The discontinuity in each curve corresponds to the density jump at the phase boundary between fluidic and superionic. The color scheme of the curves are the same as Fig.~\ref{fig:H2O_phase_diagram_Mesh_20190428}. At high-pressure or high-density, all isentropes converge and have smooth behavior. This suggests that the EOS in this regime can be approximated by $\Gamma \equiv \bigg( \frac{\partial \ln{P}}{\partial \ln{\rho}} \bigg)_{s} \approx$ constant. The achieve higher fidelity, one could add a temperature-correction term. }
\label{fig:density_pressure_plot_eos_20190423}
\end{figure}

These eight isentropes can also be viewed in the $\rho$-$P$ (density-pressure) plot (Fig.~\ref{fig:density_pressure_plot_eos_20190423}), in order to calculated the mass-radius curve corresponding to each isentrope. 

\begin{figure}[h!]
\centering
\hspace*{-3cm}\includegraphics[scale=0.43]{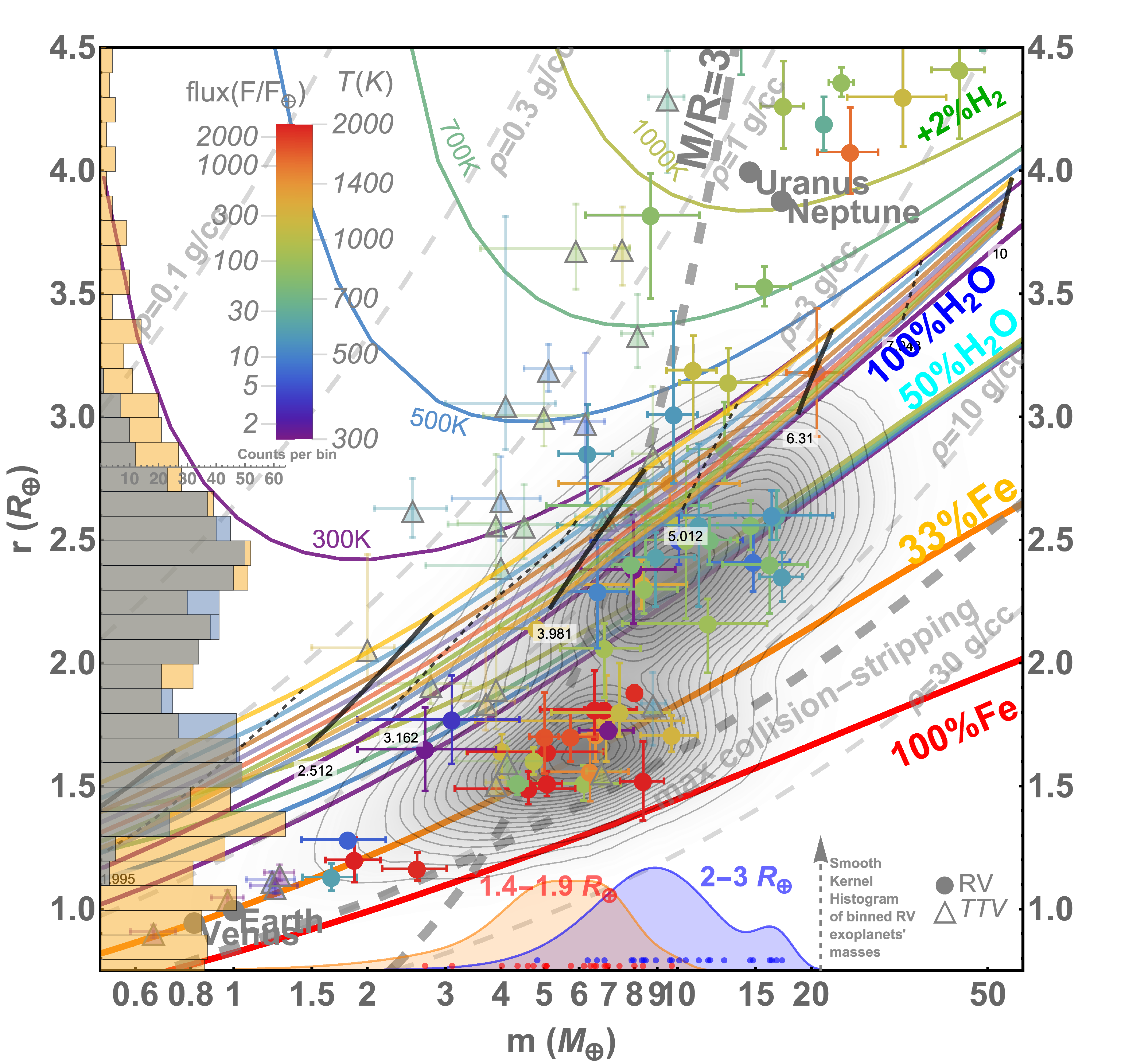}
\caption{Mass-radius curves of eight H$_2$O isentropes below 4 Earth radii (R$_\oplus$). These mass-radius curves suggest that with the appropriate consideration of interior temperature/entropy sustained by planet's own energy source, it could help explain some of the exoplanets observed in this regime as water-rich planets.  }
\label{fig:mrdiagram_small_20190423}
\end{figure}

\begin{figure}[h!]
\centering
\hspace*{-3cm}\includegraphics[scale=0.3]{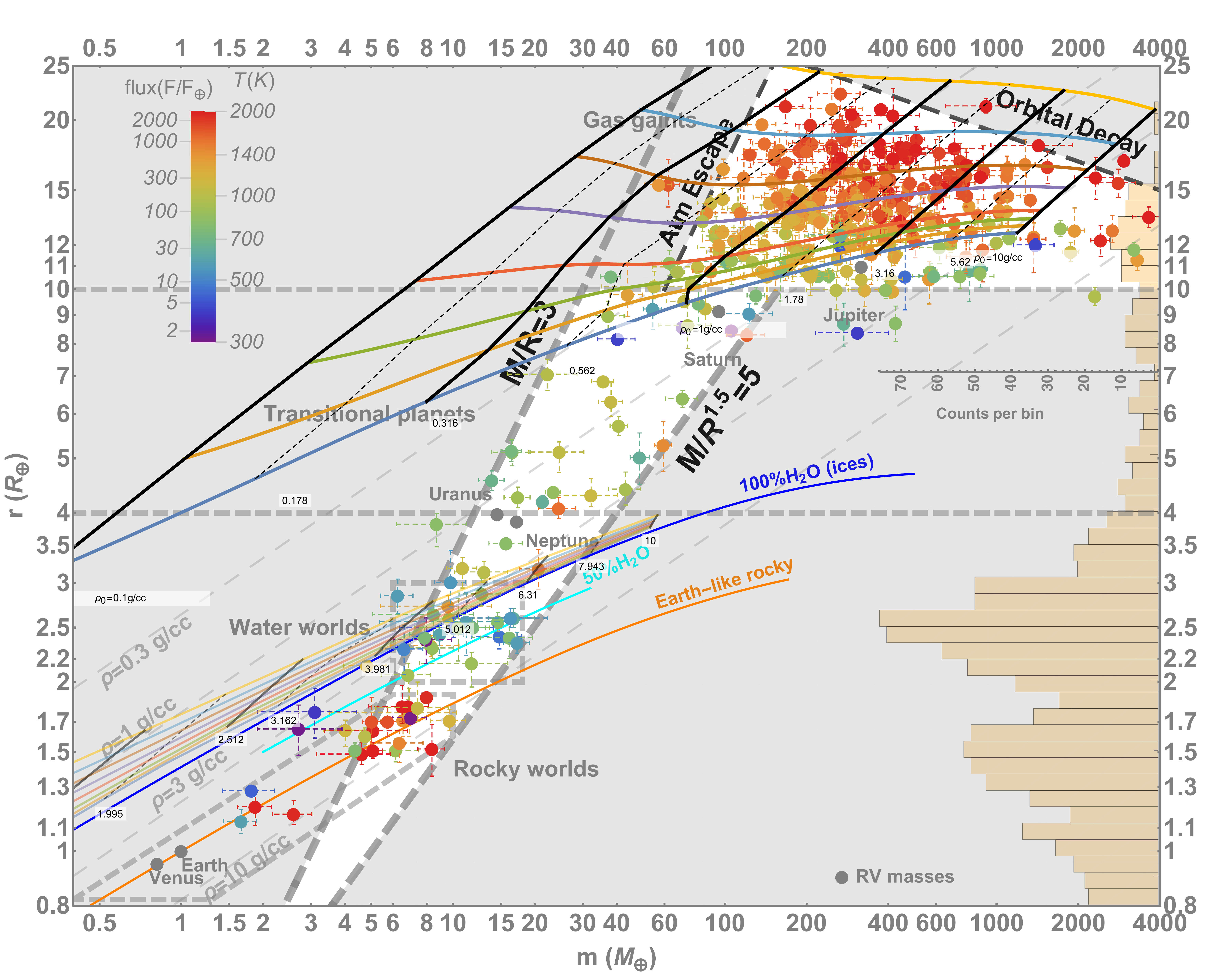}
\caption{Mass-radius curves of pure H$_2$O isentropes compared with the pure Hydrogen isentropes. The color of each mass-radius curve corresponds to different assumed specific entropy. The black mesh provides another set of contours of the central densities of planets.  }
\label{fig:mrdiagram_big_20190423}
\end{figure}

The mass-radius curves of these eight different isentropes, as a result, are shown in a small-scale mass-radius plot (Fig.~\ref{fig:mrdiagram_small_20190423}) and a large-scale mass radius plot (Fig.~\ref{fig:mrdiagram_big_20190423}). 

\begin{comment}%Compare with Mazevet 3D EOS
\begin{figure}[h!]
\centering
\hspace*{-2cm}\includegraphics[scale=0.8]{H2_eos_mazevet_2019_isentropes_3D.pdf}
\caption{Hydrogen EOS of Mazevet 2019 with isentropes visualized in the three dimensions of $lgT$, $lg\rho$, and $lgP$. The eight color curves corresponding to isentropes of specific entropy S (eV/1000Kelvin/atom) = {0.3, 0.4, 0.5, 0.6, 0.7, 0.8, 0.9, 1.0}. The bulge and the oscillation behaviors of the isentropes with lower specific entropies as a result of this bulge, are due to the Plasma Phase Transition (PPT). Recall that along an isentrope, $\gamma \equiv \bigg( \frac{\partial \ln{T}}{\partial \ln{\rho}} \bigg)_{s}$, and $\Gamma \equiv \bigg( \frac{\partial \ln{P}}{\partial \ln{\rho}} \bigg)_{s}$. Therefore, $\gamma$ and $\Gamma$ are just the tangential slopes of these 3-D isentropes projected onto the T-$\rho$ and P-$\rho$ plane correspondingly. }
\label{fig:H2EOSMazevet20193D}
\end{figure}
\end{comment}

\subsection{EOS of Ice VI and Ice VII}

According to~\citep{Bezacier2014}, the EOS parameters for Ice VI and Ice VII are as follows. 
Note that the underscript "$_{0}$" suggests extrapolation of the parameter under ambient conditions ($P_0=$1 bar and $T_0=$300 Kelvin).
These parameters are obtained by fit to BM2 EOS with temperature-corrections as: 

Density-correction:

\begin{equation}
    \rho(P,T) = \rho_{0} (P,T_0) \cdot \exp{\bigg( -\alpha_0 \cdot (T-T_0) \bigg)}
\end{equation}

Bulk-modulus correction:

\begin{equation}
    K(T) = K_0 + (T-T_0) \cdot \bigg( \frac{\partial K}{\partial T} \bigg)_{P}
\end{equation}

The melting curve of both Ice VI and Ice VII is fit to a \emph{Simon-Glatzel} form of Equation~\citep{Simon1929}. 

\begin{equation}
    P_M = P_0 + a \cdot \Bigg( \bigg( \frac{T_M}{T_0} \bigg)^c -1 \Bigg)
\end{equation}

where $a$ and $c$ are constants to be determined. More recent development involves the \emph{Kechin} Equation~\citep{Stishov1975}~\citep{Kechin1995}.

\subsubsection{Ice VI}

Thermodynamic parameters of Ice VI~\citep{Bezacier2014}: 
\begin{align}
\rho_{0} &\approx 1.27 \text{~g/cc} \nonumber\\
K_{0} &\approx 14 \text{~GPa} \nonumber\\
\alpha_{0} &\approx 14.6 \cdot 10^{-5} \text{K}^{-1} \nonumber\\
c_v &\approx 2.5~\frac{\text{J/K}}{g} \nonumber\\
\gamma &= \frac{\alpha \cdot K_T}{c_v \cdot \rho} \approx 0.65 \nonumber\\
\end{align}

Melting Curve of Ice VI~\citep{IAPWS:2011}: 

\begin{equation}
    P_M \approx 0.63 + 0.68 \cdot \Bigg( \bigg( \frac{T_M}{273\text{K}} \bigg)^{4.6} -1 \Bigg)~\text{GPa}
\end{equation}

\subsubsection{Ice VII}

Thermodynamic parameters of Ice VII~\citep{Bezacier2014}: 
\begin{align}
\rho_{0} &\approx 1.45 \text{~g/cc} \nonumber\\
K_{0} &\approx 20 \text{~GPa} \nonumber\\
\alpha_{0} &\approx 11.6 \cdot 10^{-5} \text{K}^{-1} \nonumber\\
c_v &\approx 2~\frac{\text{J/K}}{g} \nonumber\\
\gamma &= \frac{\alpha \cdot K_T}{c_v \cdot \rho} \approx 0.81 \nonumber\\
\end{align}

Melting Curve of Ice VII~\citep{Frank:2004}, fit to experimental data up to 60 GPa: 

\begin{equation}
    P_M \approx 2.17 + 0.764 \cdot \Bigg( \bigg( \frac{T_M}{355\text{K}} \bigg)^{4.32} -1 \Bigg)~\text{GPa}
\end{equation}

and the density of Ice VII along the melting curve is~\citep{Frank:2004}: 
\begin{equation}
    \rho_M \approx 1.45+ 0.4 \cdot \Bigg(1-\exp{\bigg(-0.0743 \frac{P_M}{\text{GPa}} \bigg)} \Bigg) + 2.8 \cdot \Bigg(1-\exp{\bigg(-0.0061 \frac{P_M}{\text{GPa}}\bigg)} \Bigg)~\text{g/cc}
\end{equation}

Thus, a good rule of thumb for the Gr\"{u}neisen parameter in the regime of both Ice VI and Ice VII is that $\gamma \approx 0.7\sim0.8$.

\subsection{3D EOS of H$_2$O}
\begin{figure}[h!]
\centering
\hspace*{-3.5cm}\includegraphics[scale=0.55]{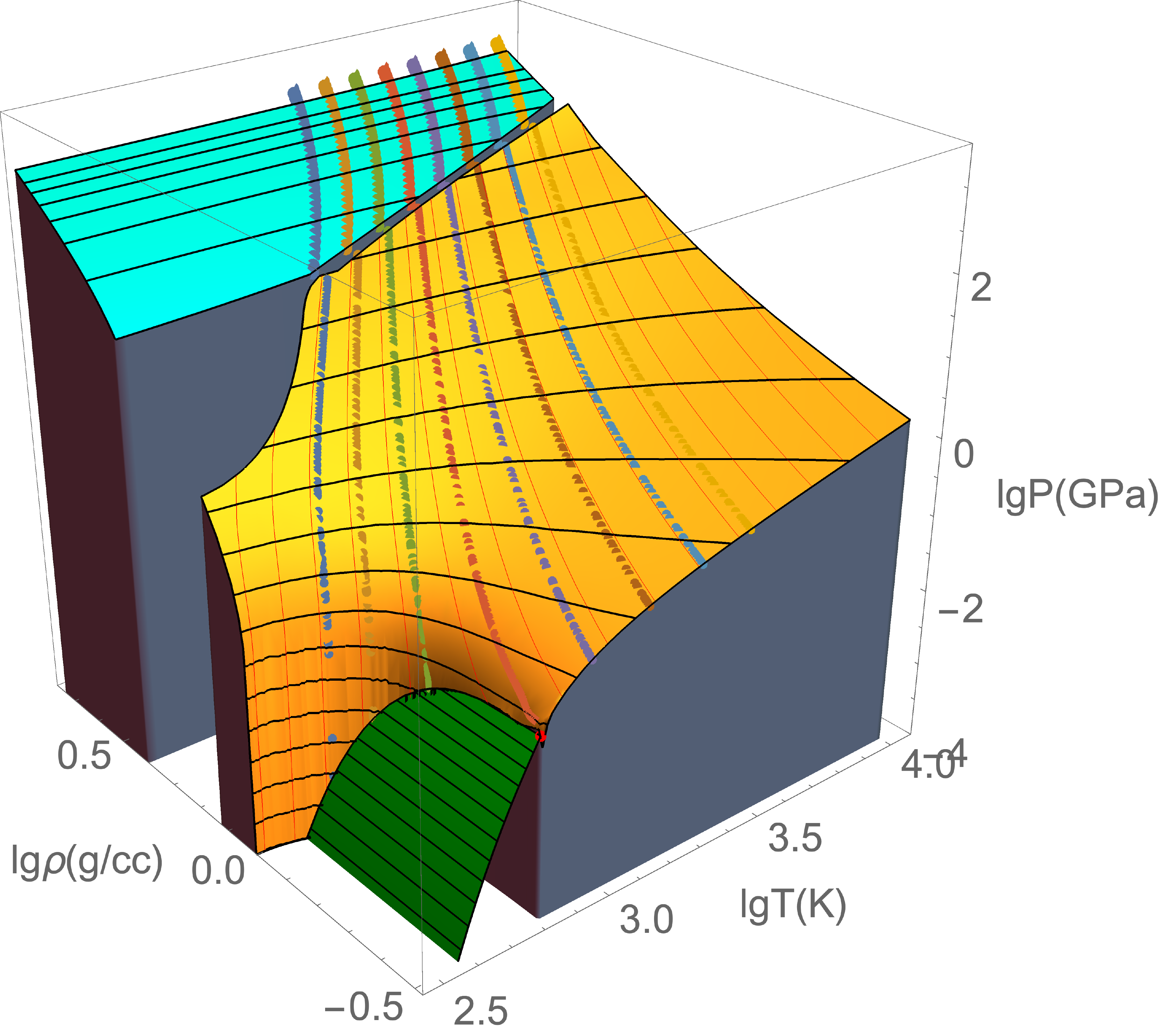}
\caption{A simplified 3-D $\rho-T-P$ EOS of H$_2$O EOS with isentropes (red-thin curves and colored dots) and isobars (black curves) visualized. The region of \emph{fluid} H$_2$O is colored golden. The cyan region above is \emph{super-ionic} H$_2$O, and green region below is the gas-liquid two-phase equilibrium (co-existence). }
\label{fig:H2OPhaseDiagram3D1}
\end{figure}

\begin{figure}[h!]
\centering
\hspace*{-3.5cm}\includegraphics[scale=0.5]{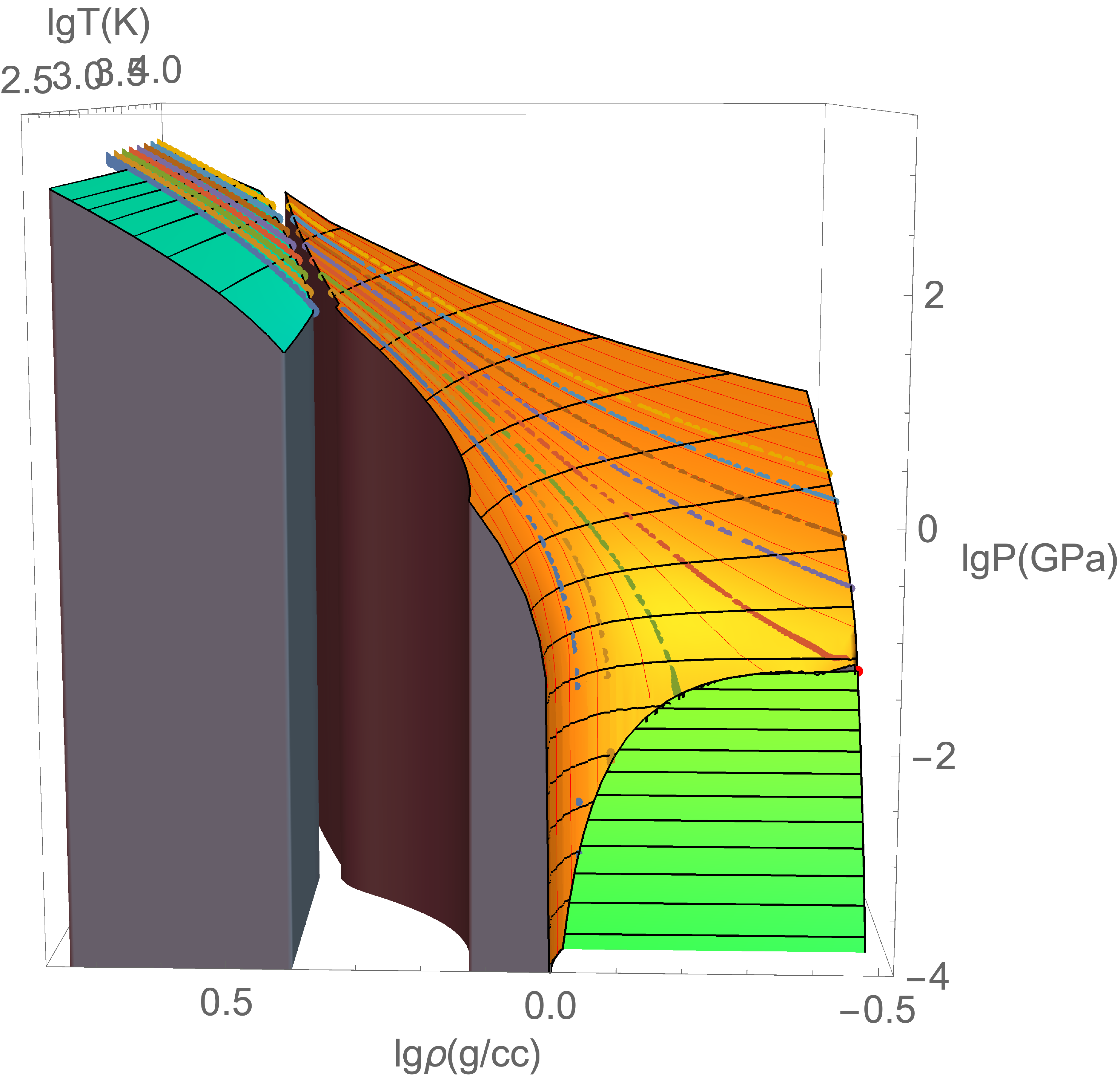}
\caption{3-D $\rho-T-P$ EOS of H$_2$O EOS with isentropes (red-thin curves and colored dots) and isobars (black curves) visualized. Recall that along an isentrope, $\Gamma \equiv \bigg( \frac{\partial \ln{P}}{\partial \ln{\rho}} \bigg)_{s}$, and thus, $\Gamma$ are just the tangential slopes of these 3-D isentropes when viewed from the perspective of the P-$\rho$ plane, c.f. Fig.~\ref{fig:density_pressure_plot_eos_20190423}.}
\label{fig:H2OPhaseDiagram3D2}
\end{figure}

\begin{figure}[h!]
\centering
\hspace*{-3.5cm}\includegraphics[scale=0.55]{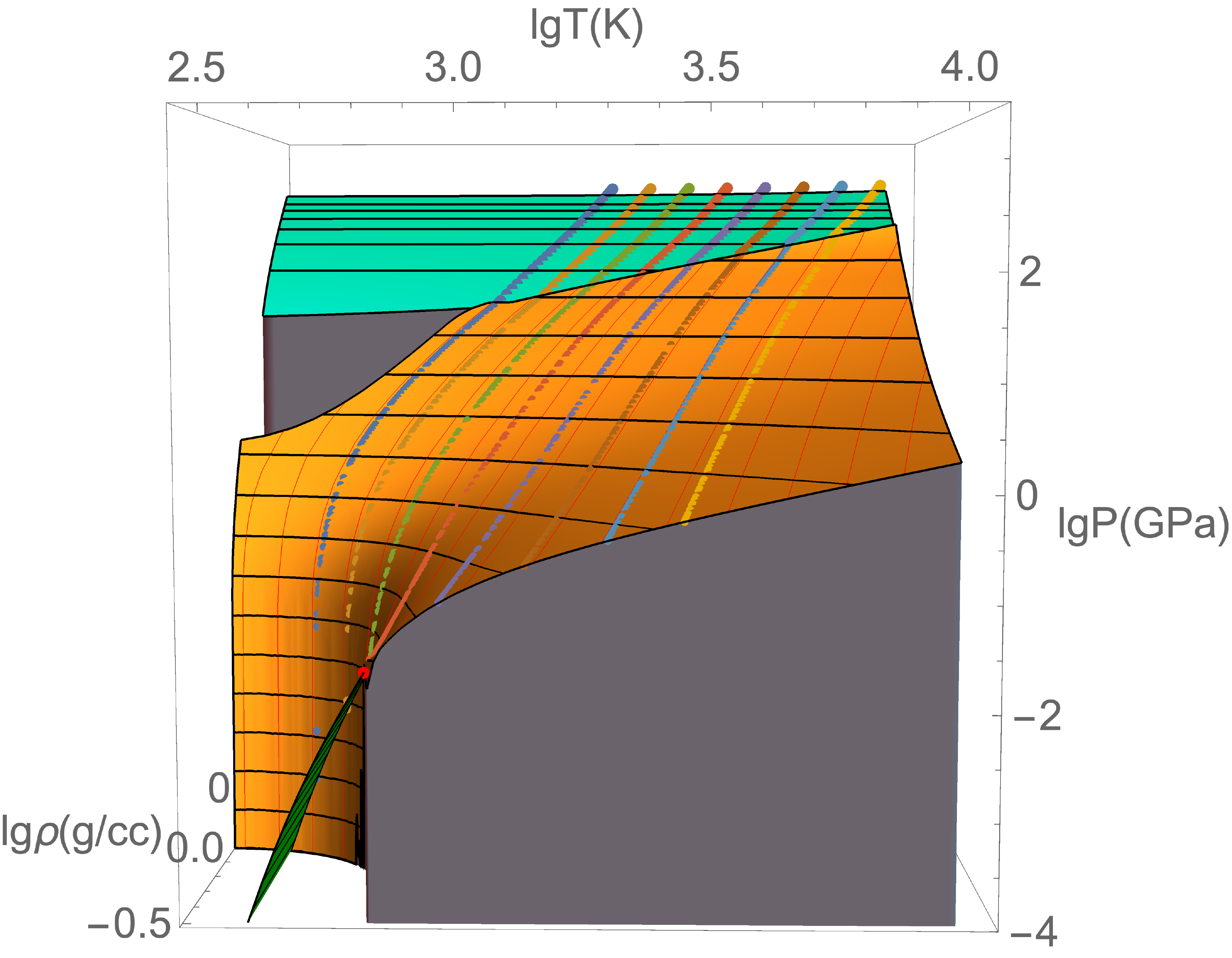}
\caption{3-D $\rho-T-P$ EOS of H$_2$O EOS with isentropes (red-thin curves and colored dots) and isobars (black curves) visualized. Viewed from the perspective of the T-P plane, the tangential slopes of isentropes are $\frac{\gamma}{\Gamma} = \bigg( \frac{\partial \ln{T}}{\partial \ln{P}} \bigg)_{s} = \bigg( \frac{\partial \ln{T}}{\partial \ln{\rho}} \bigg)_{s}  \Bigg/ \bigg( \frac{\partial \ln{P}}{\partial \ln{\rho}} \bigg)_{s} $. }
\label{fig:H2OPhaseDiagram3D3}
\end{figure}

\begin{figure}[h!]
\centering
\hspace*{-4.1cm}\includegraphics[scale=0.44]{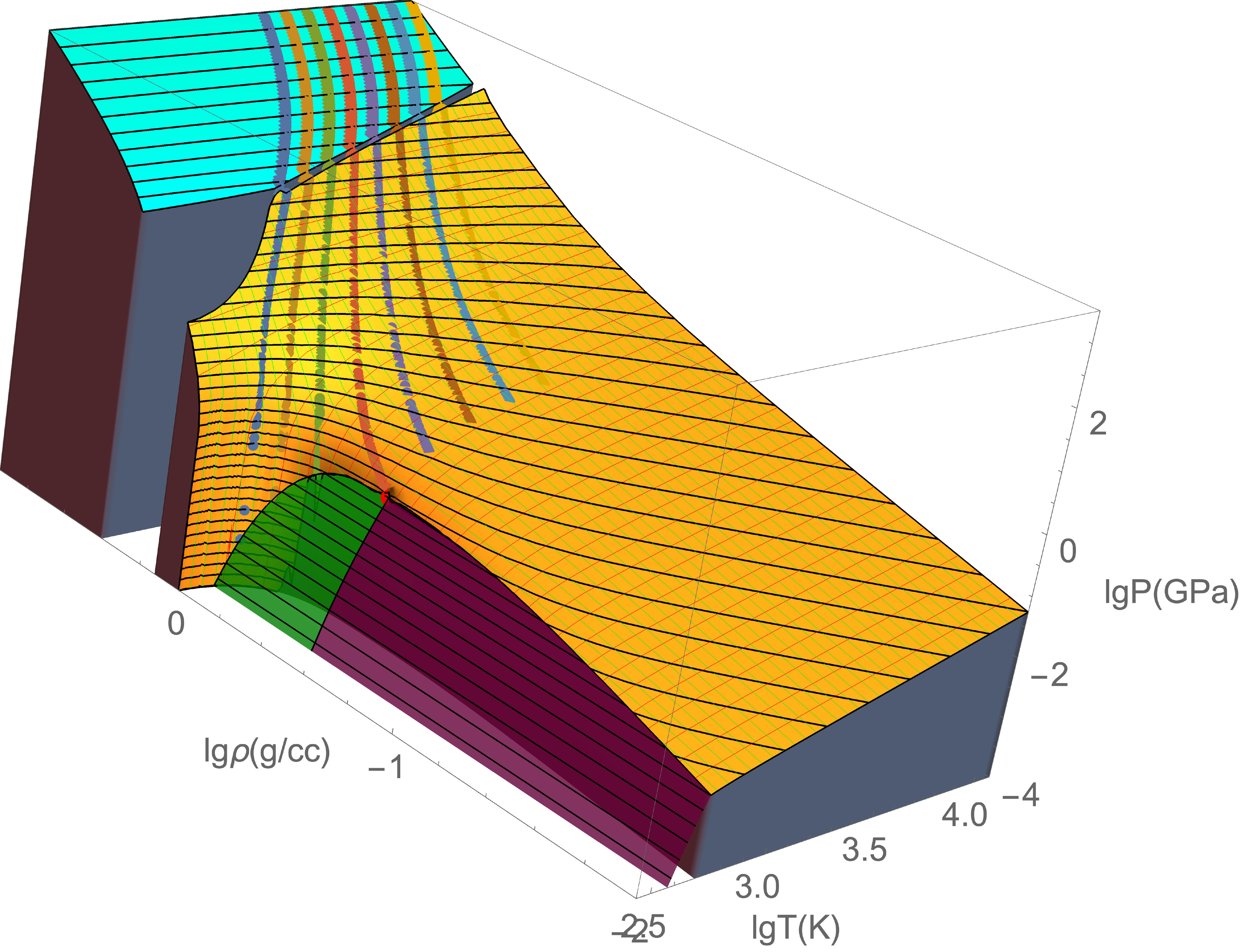}
\caption{3-D $\rho-T-P$ EOS of H$_2$O EOS extending down to lower density ($\rho$) below the critical point (big red dot). This shows how quickly it approaches the ideal gas EOS below the critical point: the contours of isochors (red thin curves), isobars (black thin curves), and isotherms (green thin curves) all become \emph{linear} in this low density regime in the lg-lg-lg plot. The gas-liquid equilibrium (co-existence) area below the critical density ($\rho_{c}$) is colored purple. This observation reinforces the idea to use a \emph{modified} ideal gas EOS to describe the entire \emph{fluid} regime of H$_2$O. }
\label{fig:H2OPhaseDiagram3D4}
\end{figure}

Finally, in order to aid our global understanding of the landscape of H$_2$O EOS, we \emph{visualize} the isentropes in a 3D $\rho T P$ EOS of H$_2$O in Fig.~\ref{fig:H2OPhaseDiagram3D1}, Fig.~\ref{fig:H2OPhaseDiagram3D2}, Fig.~\ref{fig:H2OPhaseDiagram3D3}, and Fig.~\ref{fig:H2OPhaseDiagram3D4}.

\clearpage
%%%%%%%%%%%%%%%%%%%%%%%%%%%%%%%
\section{EOS of NH$_3$, N$_2$, CH$_4$,...}

\subsection{Ammonia (NH$_3$)}

According to our earlier discussions, the Temperature-Density (T-$\rho$) phase diagram of Ammonia should resemble that of water. 
To the first-order approximation, we just need to re-scale the H$_2$O phase diagram \emph{properly} to obtain the one for Ammonia. 

Then, we will implement the ab initio calculation results from Rostock Group, in particular, Mandy Bethkenhagen~\citep{Bethkenhagen2013}~\citep{Bethkenhagen2015}. Their calculations show that NH$_3$ also contains a super-ionic phase. 

\subsubsection{Critical Point (C.P) for Ammonia}

The analytical formulae of NH$_3$ are anchored at its critical point ($c$)~\citep{NISTWebBookAmmonia}: 

\begin{align}
\rho_{c} &= 0.226 \text{~g/cc} \nonumber\\
T_{c} &= 405.5 \text{~K} = 132.3 ^{\circ}\text{C} \nonumber\\
P_{c} &= 1.128 \cdot 10^{7} \text{~Pa} = 112.8 \text{~bar} \nonumber\\
n_{c} &= 0.226/17 = 0.013 \text{~mol/cc} \nonumber\\
\end{align}

Here $n_{c}$ is the molar density of ammonia (NH$_3$) at its critical point. 

\subsubsection{Gas-Liquid-Ice Triple Point (t1) for Ammonia}
The Gas-Liquid-Ice triple point ($t1$) has the following properties~\citep{NISTWebBookAmmonia}~\citep{EngineeringtoolboxAmmonia}: 

\begin{align}
\rho_{t1} &= 0.733 \text{~g/cc for liquid} \nonumber\\
T_{t1} &= 195.4 \text{~K} = -77.75 ^{\circ}\text{C} \nonumber\\
P_{t1} &=6060 \text{~Pa} \approx 0.06 \text{~bar} \nonumber\\
n_{t1} &= 0.733/17 = 0.043 \text{~mol/cc for liquid} \nonumber\\
\end{align}

Here $n_{t1}$ is the molar density of ammonia (NH$_3$) at its triple point. 

\subsubsection{Re-Scaling}

Now, let's look at the \emph{ratio} of densities at the C.P.:
\begin{equation}
    \frac{\rho_{c,\text{NH}_3}}{\rho_{c,\text{H}_2\text{O}}} = \frac{0.226}{0.322} \approx 0.702
\end{equation}

and the \emph{ratio} of densities at the Gas-Liquid-Ice Triple Point (t1):

\begin{equation}
    \frac{\rho_{t1,\text{NH}_3}}{\rho_{t1,\text{H}_2\text{O}}} = \frac{0.733}{1.0} \approx 0.733
\end{equation}

These two \emph{ratios} are close to one another.

\subsection{Nitrogen (N$_2$)}

\subsubsection{Critical Point (C.P) for Nitrogen}

The analytical formulae of N$_2$ are anchored at its critical point ($c$)~\citep{NISTWebBookNitrogen}: 

\begin{align}
\rho_{c} &= 0.314 \text{~g/cc} \nonumber\\
T_{c} &= 126.19 \text{~K} = -146.96 ^{\circ}\text{C} \nonumber\\
P_{c} &= 3.4 \cdot 10^{6} \text{~Pa} = 34 \text{~bar} \nonumber\\
n_{c} &\approx 0.0112 \text{~mol/cc} \nonumber\\
\end{align}

Here $n_{c}$ is the molar density of nitrogen (N$_2$) at its critical point. 

\subsubsection{Gas-Liquid-Ice Triple Point (t1) for Nitrogen}
The Gas-Liquid-Ice triple point ($t1$) has the following properties~\citep{EngineeringtoolboxNitrogen}: 

\begin{align}
\rho_{t1} &= 0.867 \text{~g/cc for liquid} \nonumber\\
T_{t1} &= 63.14 \text{~K} = -210.01 ^{\circ}\text{C} \nonumber\\
P_{t1} &=1.25 \cdot 10^{4} \text{~Pa} = 0.125 \text{~bar} \nonumber\\
n_{t1} &\approx 0.031 \text{~mol/cc} \nonumber\\
\end{align}

Here $n_{t1}$ is the molar density of nitrogen (N$_2$) at its triple point. 

\subsubsection{Re-Scaling}

Now, let's look at the \emph{ratio} of densities at the C.P.:
\begin{equation}
    \frac{\rho_{c,\text{N}_2}}{\rho_{c,\text{H}_2\text{O}}} = \frac{0.314}{0.322} \approx 0.975
\end{equation}

and the \emph{ratio} of densities at the Gas-Liquid-Ice Triple Point (t1):

\begin{equation}
    \frac{\rho_{t1,\text{N}_2}}{\rho_{t1,\text{H}_2\text{O}}} = \frac{0.867}{1.0} \approx 0.867
\end{equation}

These two \emph{ratios} are close to one another. 

\subsection{Methane (CH$_4$)}

\subsubsection{Critical Point (C.P) for Methane}

The analytical formulae of CH$_4$ are anchored at its critical point ($c$)~\citep{NISTWebBookMethane}: 

\begin{align}
\rho_{c} &\approx 0.0101 \cdot 16 \approx 0.162 \text{~g/cc} \nonumber\\
T_{c} &= 190.6 \text{~K} = -82.55 ^{\circ}\text{C} \nonumber\\
P_{c} &= 4.61 \cdot 10^{6} \text{~Pa} = 46.1 \text{~bar} \nonumber\\
n_{c} &\approx 0.0101 \text{~mol/cc} \nonumber\\
\end{align}

Here $n_{c}$ is the molar density of methane (CH$_4$) at its critical point. 

\subsubsection{Gas-Liquid-Ice Triple Point (t1) for Methane}
The Gas-Liquid-Ice triple point ($t1$) has the following properties~\citep{EngineeringtoolboxMethane}: 

\begin{align}
\rho_{t1} &\approx 0.439 \text{~g/cc for liquid} \nonumber\\
T_{t1} &= 90.67 \text{~K} = -182.48 ^{\circ}\text{C} \nonumber\\
P_{t1} &=1.169 \cdot 10^{4} \text{~Pa} = 0.1169 \text{~bar} \nonumber\\
n_{t1} &\approx 0.439/16 \approx 0.0275 \text{~mol/cc} \nonumber\\
\end{align}

Here $n_{t1}$ is the molar density of methane (CH$_4$) at its triple point. 

\subsubsection{Re-Scaling}

Now, let's look at the \emph{ratio} of densities at the C.P.:
\begin{equation}
    \frac{\rho_{c,\text{CH}_4}}{\rho_{c,\text{H}_2\text{O}}} = \frac{0.162}{0.322} \approx 0.503
\end{equation}

and the \emph{ratio} of densities at the Gas-Liquid-Ice Triple Point (t1):

\begin{equation}
    \frac{\rho_{t1,\text{CH}_4}}{\rho_{t1,\text{H}_2\text{O}}} = \frac{0.439}{1.0} \approx 0.439
\end{equation}

These two \emph{ratios} are close to one another. 

\subsection{Discussion}

The comparison among all these gas species reveal their similarities in terms of their molar densities at their critical points and their triple points. 
Generally speaking, we have for all of them (H$_2$, H$_2$O, NH$_3$, N$_2$, CH$_4$,...):

\begin{align}
n_{c} &\sim 0.01 \text{~mol/cc} \nonumber\\
n_{t1} &\sim 0.03 \text{~mol/cc} \nonumber\\
n_{t1}/n_{c} &\sim 3 \nonumber\\
\end{align}

This approximation is generally good to within a factor of 2. Even more so for the ratio of $n_{t1}/n_{c}$. 

This supports the idea of the \emph{principle of corresponding states}~\cite{Karapetyants1978}, and will aid our future discussion of gaseous envelopes. We can think of $n_{c}\sim$0.01 \text{~mol/cc} as a general \emph{critical molar density} at which cohesion of molecules transition from being close to zero to non-zero as pointed out by D. Mendeleev (1870). %Any molar densities higher than this value will lead to condensation of 

$n_{t1}$ can be understood as another \emph{critical molar density} at which cohesion of molecules has exerted its full power, and any compression beyond this point is strongly resisted by the repulsive interactions of molecules, that is, the Pauli repulsion at shorter ranges due to overlapping electron orbitals. 

To go from $n_c$ to $n_{t1}$ only requires a three-fold compression. That means a mere a factor of $\sqrt[3]{3} = 1.44$ reduction in the linear dimension or linear spacing between neighboring molecules brings about the overturn in the inter-molecular interaction from cohesion to (strong) repulsion.

\subsection{An Approximate EOS}

Let's define the density of liquid at ambient conditions as $\rho_0 \sim 1$ g/cc. In contrast, the density at critical point is typically one-third of that. 

The density under other conditions as $\rho$. Their ratio can be defined as a dimensionless parameter: 

\begin{equation}
    \eta \equiv \rho/\rho_0
\end{equation}
Furthermore, we define another dimensionless parameter for temperature: 

\begin{equation}
    \tau \equiv T/T_c
\end{equation}

Then, approximately the pressure can be expressed as: 

\begin{equation}\label{Eq:EOSApprox}
    \frac{P}{\text{GPa}} \approx \eta^4 - \eta^2 + \tau \cdot \eta \cdot \exp{\bigg( \frac{\eta^2-1}{\eta^2+1} \bigg)}
\end{equation}

This EOS captures both the increase of pressure due to electronic Coulomb repulsion, and critical behavior, as well as thermal effects (including both ideal gas behavior at low-density high-temperature or atomic vibrations in crystal lattices at high-density). 

The reason it works is because at low-density: $P = n \cdot k_B \cdot T$, where $n$ is the number density of \emph{molecules}, where each molecule contains several ($\sim$ 3) atoms. At high-density: $P_{\text{thermal}} \approx 3 \cdot \Tilde{n} \cdot k_B \cdot T$, where $\Tilde{n}$ is the number of \emph{atoms}. It can be tuned to higher-accuracy by adding one or two extra parameters inside the exponential term.

Eq.~\ref{Eq:EOSApprox} can reproduce the approximate behavior of H$_2$O, NH$_3$, N$_2$, CH$_4$ from $\sim$one-tenth g/cc up to $\sim$5 g/cc, which covers the transition regime of fluid from gas-like behavior at lower density to more crystal-like behavior at higher density.

\begin{figure}[h!]
\centering
\hspace*{-2cm}\includegraphics[scale=0.7]{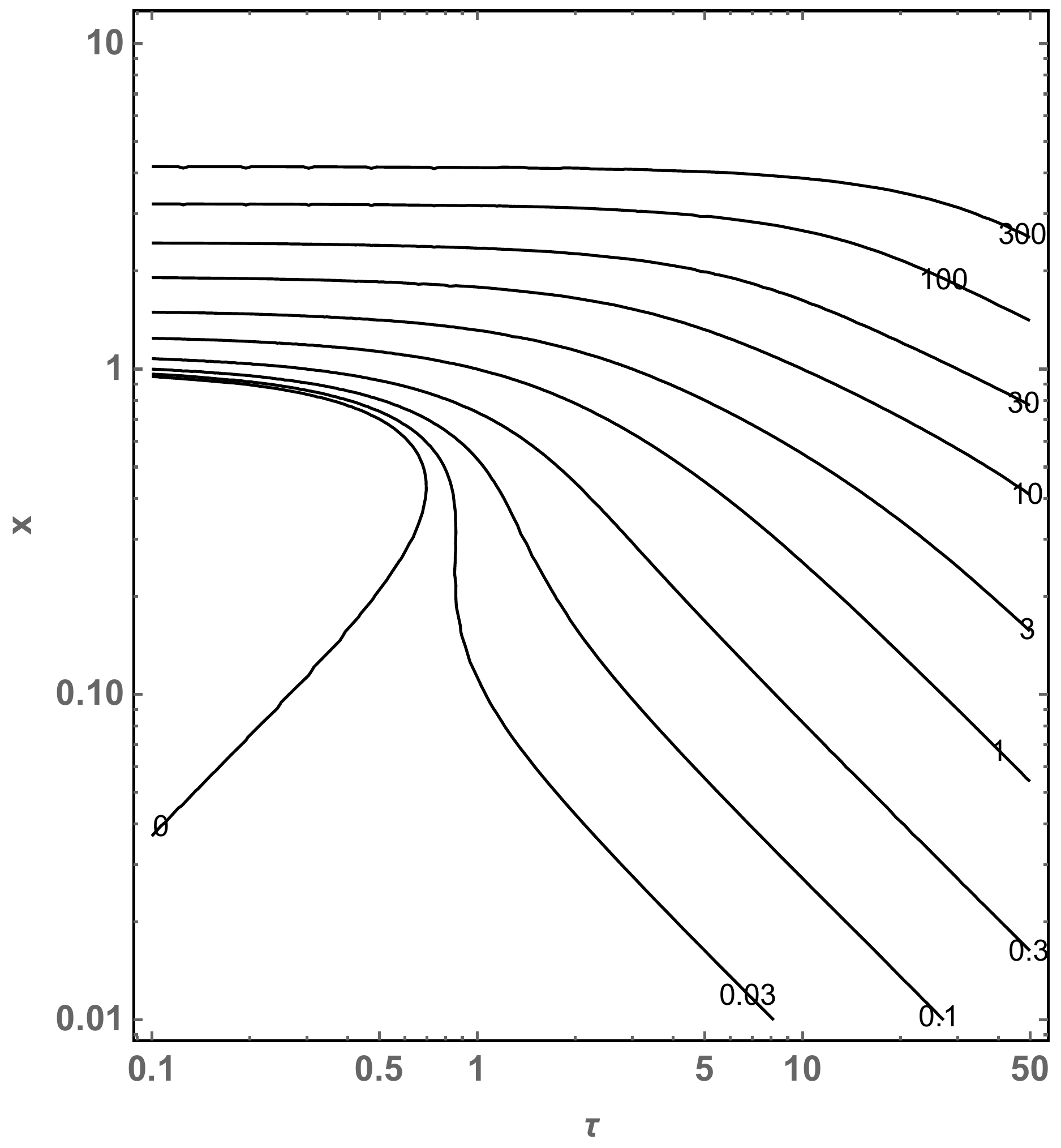}
\caption{Pressure contours (labelled in GPa), calculated from this approximate EOS (Eq.~\ref{Eq:EOSApprox}). }
\label{fig:PhaseDiagramH2OandNH3}
\end{figure}

\subsection{Approximate gas-liquid and gas-solid equilibrium\\and its relation to atmospheric escape}

On the other hand, the pressure of gas-liquid and gas-solid equilibrium can be crudely represented as: 

\begin{equation}\label{Eq:EOSApprox2}
    P_{\text{eq}} \sim P_{\infty} \cdot \exp{\bigg( -\frac{L}{k_B \cdot T} \bigg)}
\end{equation}

This equilibrium can be thought of an \emph{escape process}, similar to the escape process occurring in the planetary atmosphere. $L$ is the energy that each molecule has to overcome in order to go from the condensed phase (solid or liquid) into the dilute phase (gas) which is more or less close to ideal gas conditions. As a crude approximation, we treat $L$ as the same for gas-liquid and gas-solid equilibrium. 

If the system is close to thermal equilibrium, then the probability distribution of molecules (in both the condensed and the gas phases) satisfy the Boltzmann distribution. The probability of finding the molecules within the energy range of $E \sim E+\mathrm{d}E$ would be: 

\begin{equation}\label{Eq:BoltzmannDistribution1}
    f(E) \cdot \mathrm{d}E \propto \exp{\bigg( -\frac{E}{k_B \cdot T} \bigg)} \cdot \mathrm{d}E
\end{equation}

The probability of finding the molecules above a certain energy threshold $L$ would be: 

\begin{equation}\label{Eq:BoltzmannDistribution2}
    \int_{L}^{\infty} f(E) \mathrm{d}E \propto \exp{\bigg( -\frac{L}{k_B \cdot T} \bigg)}
\end{equation}

In fact, we realize this is exactly the scaling factor multiplying to pressure in Eq.~\ref{Eq:EOSApprox2}. 

On one hand, with regard to \emph{phase equilibrium}, $L$ is interpreted as the (nominal) latent heat of phase transition. 
$P_{\infty}$ is understood as an \emph{extrapolated} pressure as $T \rightarrow \infty$. 
For simplicity, $L/k_B$ defines a new Temperature $T_0 \equiv L/R$. Then, Eq.~\ref{Eq:EOSApprox2} can be re-written as: 

\begin{equation}\label{Eq:EOSApprox3}
    P_{\text{eq}} \sim P_{\infty} \cdot \exp{\bigg( -\frac{T_0}{T} \bigg)}
\end{equation}

For example, for H$_2$O, 

\begin{align}
P_{\text{eq}} &\sim 1 \text{~TPa} \nonumber\\
T_0 &\sim 6000 \text{~K} \nonumber
\end{align}

Interestingly enough, the critical point often occurs around $T_0/T \approx 10$ and the triple point often occurs around $T_0/T \approx 20$. Thus, for H$_2$O, 

\begin{align}
T_c &\sim 600 \text{~K} \nonumber\\
T_{t1} &\sim 300 \text{~K} \nonumber
\end{align}

From the magnitude of $P_{\text{eq}} \sim 1$ TPa, it indicates that it arises from electron degeneracy pressure, but as a manifestation of cohesion here. Thus, the factors of $10$ and $20$ within the exponential seem magical! 

\begin{align}
P_{\infty}/P_{\text{critical}} \sim \exp{(10)} &\approx 2.2 \times 10^4 \nonumber\\
P_{\infty}/P_{\text{triple}} \sim \exp{(20)} &\approx 0.5 \times 10^9 \nonumber
\end{align}

On the other hand, with regard to \emph{atmospheric escape}, $L$ can be understood as the depth of gravitational potential well at planet surface: 
\begin{equation}\label{Eq:AtmEscape1}
    L \equiv \frac{G \cdot M_p \cdot \mu}{R_p}
\end{equation}

where $\mu$ is the mass of each molecule or particle participating in the escaping process. We know that escape velocity $v_{\text{esc}}$ is defined as: 

\begin{equation}\label{Eq:AtmEscape2}
    v_{\text{esc}} \equiv \sqrt{\bigg(\frac{2 \cdot G \cdot M_p}{R_p} \bigg)}
\end{equation}

Therefore, $L$ can be re-written as: 

\begin{equation}\label{Eq:AtmEscape3}
    L = 1/2 \cdot \mu \cdot v_{\text{esc}}^2
\end{equation}

and, $L/k_B T$ is again dimensionless. 

\clearpage

%%%%%%%%%%%%%%%%%%%%%%%%%%%%%%%%%%%%%%%%%%%%%%%%
\section{Shock-wave Experiments on Ammonia and Methane: \\What to be expected}

We have put efforts into understanding the Temperature-Density Phase Diagram of ammonia and methane, in order to guide the shock wave experiments of these two species to be carried out on Z-machine at Sandia National Laboratories. These two chemical species are of particular interest in the current age of exoplanetary science, as they together with water may comprise a significant fraction of planetary interiors among exoplanets~\citep{PNAS:Zeng2019,Zeng2021}. Therefore, their physical and chemical properties in the high-pressure and high-temperature are interesting topic to pursue. 

There are existing ab initio calculations of the EOS of ammonia~\citep{Li2017}~\citep{Bethkenhagen2013}, an analytic EOS of Ammonia up to $\sim$3000 K and $\sim$20 GPa~\citep{Jahangiri2016}, and existing experimental data~\citep{Queyroux2019}, to be compared with, and guide the up-coming experiments. See also Fig.~\ref{fig:PhaseDiagramH2OandNH3}.

\begin{figure}[h!]
\centering
\hspace*{-2cm}\includegraphics[scale=0.7]{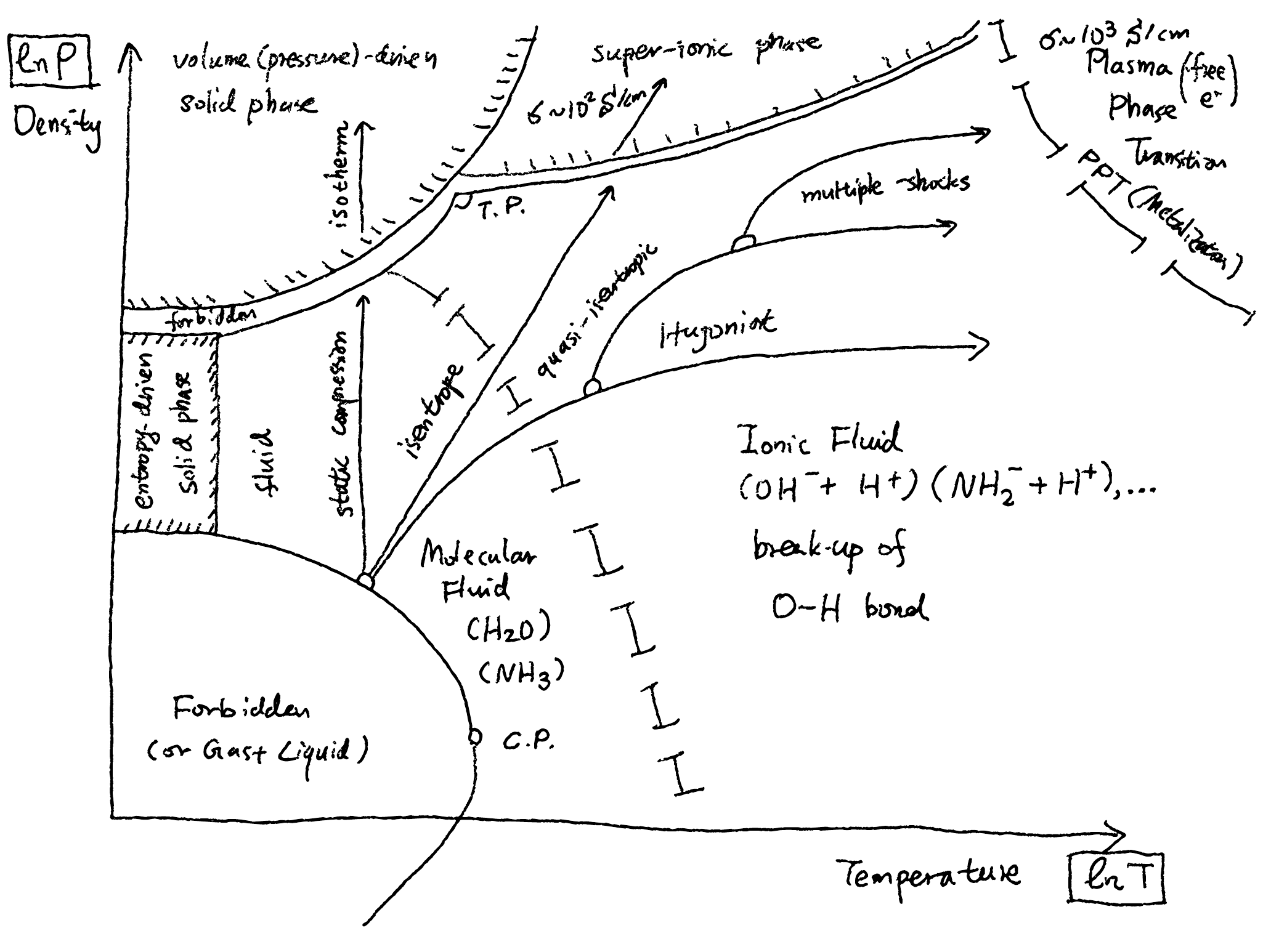}
\caption{Schematic phase diagram for water and ammonia, or their mixture. Also shown are trajectories of isotherm, isentropes, and single/multiple-shock Hugoniot. A single-shock Hugoniot will likely miss the super-ionic regime. The detailed differences among the various ice structures/phases have been ignored as they are all approaching close-packing at sufficient pressures. }
\label{fig:PhaseDiagramH2OandNH3}
\end{figure}

The Dissociation, Ionization, Super-Ionization, and Metallization of ammonia are key features to be expected in its phase diagrams. In particular, the \emph{super-ionic} phase has already been experimentally verified for water~\citep{Millot2018ExperimentalCompression}~\citep{Millot2019}. The \emph{super-ionic} phase of ammonia or water-ammonia mixture is expected to exist~\citep{Cavazzoni1999SuperionicConditions}~\citep{Bethkenhagen2015}~\citep{Jiang2017}.

In order to hit this \emph{super-ionic} phase of ammonia in terms of the shock-wave experiments, we need to either: 
\begin{itemize}
    \item use reverberating shocks technique starting with the liquid density at ambient conditions to approximate an isentrope, or, 
    \item use a single shock starting with a pre-compressed high-pressure ice phase, which is already 60\% or denser than the liquid
\end{itemize}
 
For methane, a simple analytic EOS was put forth by G. Kerley that goes up to $\sim$10$^{4}$ K and $\sim$2.5 g/cc~\citep{Kerley1980}, which agrees with the experimental data from W. Nellis. A detailed experimentally-verified methane EOS table for relatively low-temperature (up to 625 K) and low-pressure (up to 1 GPa) also exists~\citep{Setzmann1991}.

Recent ab initio simulations suggest that methane is expected to dissociate first into Hydro-carbons (above $\sim$ 100 GPa), and further into Carbon (Diamond) and Hydrogen (above $\sim$ 300 GPa)~\citep{Sherman2012}~\citep{Gao2010}~\citep{Benedetti1999}~\citep{Ancilotto1997}. 

For future purpose, in terms of their mixtures, ab initio calculations show that a \emph{linear mixing} model for water-ammonia system is good enough approximation for a wide range of pressure-temperature phase-space~\citep{Bethkenhagen2017}, because these two species are very soluble in each other, and behave chemically similar even under high-pressure conditions. However, ammonia ice and water ice may eventually separate above $\sim$ 500 GPa~\citep{Robinson2017}. 

In terms of the water-ammonia-methane triple mixture, the inclusion of methane (or carbon) into the water-ammonia mixture is more tricky at high-pressure high-temperature conditions. Methane/Carbon may separate out from the mixture, to either rain down as diamonds, or may float upwards as gas as it is somewhat lighter and more volatile. Some ab initio simulations are also available~\citep{Meyer2015}. 

\clearpage
%%%%%%%%%%%%%%%%%%%%%%%%%%%%%%%
\section{Light Gaseous Envelopes}

\subsection{Envelope-Core Radius Relation: \\Introducing \emph{fugacity} $\mathbf{f}$}
This section deals with planets covered with a not-so-massive gaseous envelope, and its effect on mass-radius relation. 
For a light (not-so-massive) envelope, typically less than about 20 percent ($\lesssim 20\%$) the planet core mass $M_c$ (defined as whatever underneath the envelope), the calculation is simplified with the following assumptions. 

\begin{enumerate}
    \item First, the gravitational field within the envelope depends only on the radial distance $r$ to the planet center, rather than any other conditions (such as mass distribution or scale height variation) within the envelope itself, because the planet core mass dominates. 
    \item Second, the energy transport process within the envelope is assumed to be understood and well-defined. There are three modes of energy transport process: radiative, convective, and conductive. They together define a unique path or trajectory that the thermal dynamical parameters $T$-$\rho$-$P$-entropy... corresponds to one another and to the envelope metric. It is exactly along this path that any differentiation is regarded in the following differential equations. It is implicitly path-independent.
\end{enumerate}

Under these two simplifications, the equation governing hydrostatic equilibrium is written as: 

\begin{equation}\label{Eq:HydrostaticEquilibriumInRadius}
  \frac{dP}{dr} = g \cdot \rho = -\frac{G \cdot M_c}{r^2} \cdot \rho
\end{equation}

By re-arrangement of terms, we have:

\begin{equation}\label{Eq:HydrostaticEquilibriumInRadius2}
  \frac{dP}{\rho} = -\frac{G \cdot M_c}{r^2} \cdot dr = G \cdot M_c \cdot d\Bigg( \frac{1}{r} \Bigg)
\end{equation}

The \textbf{RHS} of Eq.~\ref{Eq:HydrostaticEquilibriumInRadius2} can be integrated straightforwardly from the envelope top to the envelope bottom. Let's define this integral as parameter $x$: 

\begin{align}
x &\equiv \int_{\text{top}}^{\text{bottom}} G \cdot M_c \cdot d\Bigg( \frac{1}{r} \Bigg) = G \cdot M_c \cdot \Bigg(\frac{1}{r_{\text{bottom}}}-\frac{1}{r_{\text{top}}}  \Bigg) \\\nonumber
  &= G \cdot M_c \cdot \Bigg(\frac{1}{R_c}-\frac{1}{R_p}  \Bigg)  = \Bigg( \frac{G \cdot M_c}{R_c} \Bigg) \cdot \Bigg(1-\frac{R_c}{R_p} \Bigg)
\end{align}

Here $R_c$ is the radius measured at the core surface or the envelope bottom, and $R_p$ is the radius defined at the envelope top, also known as the total planet radius. We can re-express the relationship between $R_p$ and $R_c$ in terms of $x$ as: 

\begin{equation}\label{Eq:RadiusRelation}
  R_p = R_c \cdot \ddfrac{1}{1-\frac{x}{(G\cdot M_c/R_c)}}
\end{equation}

The cluster $GM_c/R_c$ has its physical meaning. It is the square of escape velocity at the core surface. It measures how deep inside a gravitational potential well does the core surface resides. It has the unit of specific energy (energy per unit mass). 

This Envelope-Core Radius-Relation (Eq.~\ref{Eq:RadiusRelation}) helps understand how the mass of a planet core $M_c$ may affect the extent of its envelope. Let's perform the following thought experiment: 
If $R_c$ stays constant, then, 
\begin{itemize}
    \item when $M_c \searrow$, $\frac{x}{(GM_c/R_c)} \nearrow$, $R_p \nearrow$
    \item  when $M_c \nearrow$, $\frac{x}{(GM_c/R_c)} \searrow$, $R_p \searrow$
\end{itemize}

Another corollary of Eq.~\ref{Eq:RadiusRelation} is that if $x$ becomes too large, then the denominator turns negative and situation becomes nonphysical. In reality, this suggests that the envelope becomes unbound.  

For the \textbf{LHS} of Eq.~\ref{Eq:HydrostaticEquilibriumInRadius2}, we need to consider the integral $\int \frac{dP}{\rho}$. Without any a priori knowledge of the thermal structure of the envelope, it is assumed to be at a constant temperature (isothermal) determined only by the host stellar radiation, i.e., the equilibrium temperature $T_{\text{eq}}$. This is similar to the situation on Earth surface where the heat flux from interior is orders-of-magnitude smaller than solar radiation. The mean temperature of fluids on Earth surface, i.e., the hydrosphere and atmosphere, is mostly determined by the average solar radiation. 

Then the integral $\int \frac{dP}{\rho}$ is a corollary function of \textbf{EOS} $P=P[\rho,T_{\text{eq}}]$, specifically for an isothermal process at constant temperature $T_{\text{eq}}$, with the proper upper and lower limits applied. This integral also has the unit of specific energy, that is, energy per unit mass. 

Alternatively, the same integral $\int \frac{dP}{\rho}$ can be derived by equating the \emph{chemical potential} $\mathbf{\mu}$ at any given height within this light isothermal envelope. In this manner, this integral is related to \emph{fugacity} $\mathbf{f}$, introduced by G. Lewis (1901). 

\begin{equation}
  x = \Bigg( \frac{R \cdot T_{\text{eq}}}{\mu} \Bigg) \cdot \ln{\Bigg( \frac{\mathbf{f_{\text{bottom}}}}{\mathbf{f_{\text{top}}}} \Bigg)} =  \int_{\text{top}}^{\text{bottom}} \frac{dP}{\rho}
\end{equation}

here the molar mass $\mu$ is needed to convert energy per mole to energy per unit mass. 

The \emph{fugacity} $\mathbf{f}$ is especially useful when considering an envelope consisting of multiple species, and chemical reactions occurring among the constituent species. When writing out an equilibrium equation for chemical reactions, \emph{fugacity} $\mathbf{f}$ is used often instead of partial pressure. $\mathbf{f}$ can be viewed as the pressure a given real system must produce to have the same action as an ideal system, given the proper asymptotic limit (constant of integration): $\lim_{P\to 0} \frac{\mathbf{f}}{P} = 1$. 

$\mathbf{f}$ is the corrected (effective) pressure which takes into account the attractive and repulsive intermolecular interactions. 

It should be emphasized that any $\mathbf{f}$ is only specific to a given temperature $T_{\text{eq}}$. For a different $T_{\text{eq}}$, $\int \frac{dP}{\rho}$ needs to be re-calculated. 

As regards to the limits of this integral:
\begin{itemize}
    \item The lower limit relates to what pressure-level or density-level a transit observation defines the rim of a planet. It is typically on the millibars (100 Pa) level or a few tens of millibars level, depending on the chemical and thermal conditions in the envelope (please see the detailed calculation of the slant-path optical depth which takes into account planet surface curvature in~\citep{DeWit2013} and its supplement). 
    \item The upper limit relates to the total mass or mass fraction of the envelope. This relation is elucidated by the next sub-section. 
\end{itemize}

Once $\int \frac{dP}{\rho}$, that is, $x$ is computed from an EOS, then we can use it to compute $R_p$ according to Eq.~\ref{Eq:RadiusRelation}. 
To aid our calculation, we define a dimensionless parameter $y$ as: 

\begin{equation}
    y \equiv \Bigg( \int_{\text{top}}^{\text{bottom}} \frac{dP}{\rho} \Bigg) \Bigg/ \Bigg( \frac{G \cdot M_c}{R_c} \Bigg) = x \Bigg/ \Bigg( \frac{G \cdot M_c}{R_c} \Bigg)
\end{equation}

Then the Envelope-Core Radius-Relation (Eq.~\ref{Eq:RadiusRelation}) becomes: 

\begin{equation}\label{Eq:RadiusRelation2}
  R_p = R_c \cdot \ddfrac{1}{1-y}
\end{equation}

In many cases, it is more convenient to express planet mass and radius in Earth units. Therefore,to aid our calculation, we introduce yet another dimensionless parameter $z$: 

\begin{equation}
    z \equiv \Bigg( \int_{\text{top}}^{\text{bottom}} \frac{dP}{\rho} \Bigg) \Bigg/ \Bigg( \frac{G \cdot M_{\oplus}}{R_{\oplus}} \Bigg) = x \Bigg/ \Bigg( \frac{G \cdot M_{\oplus}}{R_{\oplus}} \Bigg)
\end{equation}

The denominator equals a constant in unit of specific energy (erg/g): 

\begin{equation}\label{Eq:RadiusRelationConvert}
    \Bigg( \frac{G \cdot M_{\oplus}}{R_{\oplus}} \Bigg) = \frac{6.674 \cdot 10^{-8} \text{cc/g/s}^2 \cdot 5.972 \cdot 10^{27}\text{g}}{6.371 \cdot10^8 \text{cm}} = 6.256\cdot10^{11} \text{erg/g}
\end{equation}

Then the Envelope-Core Radius-Relation becomes: 

\begin{equation}\label{Eq:RadiusRelation3}
  R_p = R_c \cdot \ddfrac{1}{1- \Bigg( z \bigg/ \bigg( \frac{M_c/M_{\oplus}}{R_c/R_{\oplus}} \bigg) \Bigg)}
\end{equation}

Or, more simply if $M_c$ and $R_c$ are already expressed in Earth units: 

\begin{equation}\label{Eq:RadiusRelation4}
  R_p = R_c \cdot \ddfrac{1}{1- \Bigg( z \bigg/ \bigg( \frac{M_c}{R_c} \bigg) \Bigg)}
\end{equation}

This relation (Eq.~\ref{Eq:RadiusRelation4}) is useful when comparing to \emph{contours} of $M_{p}/R_{p}$ (planet mass divided by planet radius, in Earth units) in the  mass-radius plot. 

\subsection{Envelope-Core Pressure Relation}

In order to understand the relation between the mass of the envelope $M_{\text{env}}=(M_p-M_c)$ and the pressure $P_{\text{env}}$ it exerts onto the planet core, We now express the hydrostatic equilibrium Eq.~\ref{Eq:HydrostaticEquilibriumInRadius} in variable mass $m$ instead of radius $r$: 
\begin{equation}\label{Eq:HydrostaticEquilibriumInMass}
  \frac{dP}{dm} = -\frac{G \cdot m}{4 \pi \cdot r^4} = -\frac{1}{4 \pi \cdot G \cdot m} \cdot (g[m])^2
\end{equation}

Notice that here gravitational acceleration $g$ is a function dependent on $m$ or $r$. Its square occurs on the \textbf{RHS} of the above Eq.~\ref{Eq:HydrostaticEquilibriumInMass}. 
Through manipulation, we have: 

\begin{equation}\label{Eq:HydrostaticEquilibriumInMass2}
  dP = -\frac{g^2}{4 \pi  G} \cdot \frac{dm}{m}
\end{equation}

Eq.~\ref{Eq:HydrostaticEquilibriumInMass2} tells that Pressure $P$ and Gravity $g$ are essentially two sides of the same coin. Pressure arises from gravity from matter itself to compress itself, and thus, the square in the equation. 

It is very tempting to integrate Eq.~\ref{Eq:HydrostaticEquilibriumInMass2}: 

\begin{equation}
    P_{\text{env}} \approx \frac{\overbar{g^2}}{4 \pi G} \cdot \frac{M_{\text{env}}}{M_p}
\end{equation}

where $\overbar{g^2}$ is some kind of mass-averaged value within the envelope. It would be nice if we can somehow evaluate $g$ or $g^2$ within the envelope. In fact, this is feasible, as long as: 

\begin{enumerate}
    \item the envelope is light (not-so-massive) ($\lesssim 20\%~M_c$), compared to the planet core mass $M_c$ or the total planet mass $M_p=(M_c+M_{\text{env}})$. 
    \item the envelope mass is concentrated towards its bottom, so that gravity $g$ can be approximated by its value evaluated at the bottom of the envelope, or equivalently, on top of the planet core: $g_c$.
\end{enumerate}

If both (1) and (2) are satisfied, then the integral from the envelope top to the envelope bottom gives an approximate pressure $P_{\text{env}}$ that the envelope exerts onto the planet core due to gravitation: 

\begin{align}
  P_{\text{env}} \equiv P_{\text{bottom}} &\approx \bigg( P_{\text{bottom}}-P_{\text{top}} \bigg) \approx \frac{\abs{g_c}^2}{4 \pi  G} \cdot \ln{\Bigg( \frac{M_c+M_{\text{env}}}{M_c} \Bigg)} \\\nonumber &= \frac{\abs{g_c}^2}{4 \pi  G} \cdot \ln{\Bigg(1+ \frac{M_{\text{env}}}{M_c} \Bigg)} \approx \frac{\abs{g_c}^2}{4 \pi  G} \cdot \Bigg(\frac{M_{\text{env}}}{M_c} \Bigg)
\end{align}

and $M_c+M_{\text{env}}$ is no different than the total mass of the planet $M_p$. 
$g_c$ is of course related to the core mass $M_c$ and core radius $R_c$ as:

\begin{equation}
    g_c = \frac{G \cdot M_c}{(R_c)^2}
\end{equation}

Therefore, 

\begin{equation}
    P_{\text{env}} \approx \Bigg( \frac{G \cdot M_c^2}{4 \pi \cdot R_c^4} \Bigg) \cdot \Bigg(\frac{M_{\text{env}}}{M_c} \Bigg) \equiv P_c \cdot \Bigg(\frac{M_{\text{env}}}{M_c} \Bigg)
\end{equation}

Here we introduce $P_c$ as a \emph{characteristic} pressure in the deep interior of the planet core. For example, for Earth, $P_c=\frac{G \cdot M_{\oplus}}{4\pi R_{\oplus}^2} \approx$ 100 GPa or 1 megabar, which happens to be the pressure at the core-mantle boundary inside Earth, and about one-third of the pressure at the Earth's center. This is a \emph{fundamental} relation that relates the envelope pressure at its bottom to the characteristic pressure deep within the core, regardless of the detailed chemistry and physics of the envelope or the core themselves: 

\begin{equation}
    \Bigg(\frac{P_{\text{env}}}{P_c} \Bigg) \approx \Bigg(\frac{M_{\text{env}}}{M_c} \Bigg)
\end{equation}

\subsection{Real Gas EOS: \\Introducing \emph{Compression Factor} $\mathbf{Z}$}

The complexity of the fluid H$_2$O EOS benefits from the multitude of experimental data available and its application in industry. 
The accuracy achieved there is typically less than 0.1\% level. 
For other gases, let's start from scratch and \emph{solve a simpler problem first}. As for our purpose in planetary science, a 10\% accuracy is tolerable, so let's construct a \text{EOS} from a simpler standpoint. 

To aid our discussion, let's first define a dimensionless parameter: \emph{Compression Factor} $\mathbf{Z}$ as: 

\begin{equation}\label{Eq:CompressionFactor}
  \mathbf{Z} \equiv \frac{p \cdot V_m}{R \cdot T} = \frac{p}{(\rho/\mu) \cdot R \cdot T}
\end{equation}

Where $V_m$ is the molar volume, and $\mu$ is the molar weight of molecules. 
For Ideal Gas, $\mathbf{Z}=1$. Deviations of the \emph{Compression Factor} $\mathbf{Z}$ from unity are due to attractive and repulsive intermolecular interactions. By its definition, it is linked to the \emph{Fugacity} $\mathbf{f}$ as: 

\begin{equation}\label{Eq:CompressionFactor2}
  \mathbf{f} = \underbrace{ P \cdot  \exp{\Bigg(\int \frac{\mathbf{Z}-1}{P} \cdot dP} \Bigg)}_{\text{integral performed along an isotherm}}
\end{equation}

When $\mathbf{Z}=1$, Eq.~\ref{Eq:CompressionFactor2} immediately gives:

\begin{equation}\label{Eq:CompressionFactor2b}
  \mathbf{f} = P
\end{equation}

\subsubsection{Virial EOS}
$\mathbf{Z}$ can also be derived from the first principle through statistical mechanics, known as the Virial Equation~\citep{Karapetyants1978}: 

\begin{equation}\label{Eq:CompressionFactor3}
  \mathbf{Z} = 1 + B \cdot \Bigg( \frac{\rho}{\mu} \Bigg) + C \cdot \Bigg( \frac{\rho}{\mu} \Bigg)^2 + D \cdot \Bigg( \frac{\rho}{\mu} \Bigg)^3 + ...
\end{equation}

where $B$, $C$, and $D$, ... are the second, third, fourth, etc. \emph{virial coefficients}. These coefficients depend only on temperature. They account for interactions between successively larger groups of molecules. They reflect double ($B$), triple ($C$), quadruple ($D$), etc. interactions of molecules. The detailed calculations are based upon equations establishing the relationships between the energy of interaction of molecules, and distance between them, and their arrangement. 

For example, for spherical molecules, coefficient $B$ can be calculated as: 

\begin{equation}\label{Eq:CompressionFactor4}
  B = \frac{1}{2} \cdot N_A \cdot \int_{0}^{\infty} \Bigg(1-\exp{\bigg(-\frac{E_p(r)}{k_B \cdot T} \bigg)}\Bigg) \cdot (4\pi r^2) \cdot dr 
\end{equation}

where $E_p(r)$ is the intermolecular-force potential energy expressed as a function of distance $r$. $N_A$ is the Avogadro number and $k_B$ is the Boltzmann constant. $B$ has the unit of volume per mole. One example of $E_p(r)$-potential to plug in is the \emph{Lenard-Jones} 6-12 potential. 

%%%%%%%%%%%%%%%%%

For comparatively low pressures (in terms of the reduced pressure $\pi \equiv (P/{P_c})\lesssim0.5$), we can limit ourselves to the second virial coefficient $B$, which takes into account only double collisions: 

\begin{equation}\label{Eq:CompressionFactor5}
  \mathbf{Z} = 1 + B \cdot \Bigg( \frac{\rho}{\mu} \Bigg)
\end{equation}

If we include not only $B$, but also $C$, then the \emph{cubic} virial EOS gives satisfactory results up to reduced density $\delta \equiv (\rho/{\rho_c}) \lesssim2$ and \emph{inverse} reduced temperature $\tau \equiv (T_c/T) \lesssim2$: 

\begin{equation}\label{Eq:CubicVirialEOS}
  \mathbf{Z} = 1 + B \cdot \Bigg( \frac{\rho}{\mu} \Bigg) + C \cdot \Bigg( \frac{\rho}{\mu} \Bigg)^2
\end{equation}

The \emph{cubic} virial EOS (Eq.~\ref{Eq:CubicVirialEOS}) has all the nice characteristics of \emph{Van der Waals} EOS, but without its fatal singularity. The \emph{cubic} virial EOS accurately represents the gas-liquid equilibrium of most substance from the critical point down to the triple point, where solid phase starts to appear. 

\subsubsection{Berthelot EOS: \\For comparatively low pressures ($\lesssim$100 bar)\\or molar density up to $\sim$0.01 mol/cc}
D. Berthelot (1903) proposed the following EOS~\citep{Karapetyants1978}: 

\begin{equation}
  \mathbf{Z} = 1 + \frac{9 \cdot \pi \cdot \tau}{128} \cdot \Bigg( 1-6\cdot \tau^2 \Bigg)
\end{equation}

Here $\pi$ is the reduced pressure $\pi \equiv \frac{P}{P_c}$, and $\tau$ is the \emph{inverse} reduced temperature $\tau \equiv \frac{T_c}{T}$.

\subsection{Procedure to Calculate Mass-Radius Relation}

When we calculate the mass-radius curves for a fixed envelope-to-core mass ratio: $\bigg(\frac{M_{\text{env}}}{M_c}\bigg)$, we are essentially fixing $\bigg(\frac{P_{\text{env}}}{P_c}\bigg)$. However, recall the definition of $P_c$ that it depends on both $M_c$ and $R_c$, and thus, it varies along the mass-radius relation ($M_c$-$R_c$) of a given core composition. 

Therefore, we need to follow this procedure to calculate the mass-radius relation for a planet core with an envelope: 

\begin{itemize}
    \item First, we need to identify the mass-radius relation ($M_c$-$R_c$) of the core itself. 
    \item Second, we need to identify the mass $M_{\text{env}}$ or mass fraction $M_{\text{env}}/M_c$ of the envelope. Since it is a light envelope, we assume that it does not exert too much pressure onto the core, and change the core's mass-radius relation ($M_c$-$R_c$).  
    \item Third, we compute the pressure at the bottom of the light envelope $P_{\text{bottom}}$: 
        \begin{equation}
            P_{\text{bottom}} = \Bigg( \frac{G \cdot M_c^2}{4 \pi \cdot R_c^4} \Bigg) \cdot \Bigg(\frac{M_{\text{env}}}{M_c} \Bigg)
        \end{equation}
    \item Fourth, carry out the integral $\int \frac{dP}{\rho}$ from top to bottom of the envelope with the appropriate EOS, to find the dimensionless parameter $z$: 
        \begin{equation}
            z \equiv \Bigg( \int_{P_{\text{top}}}^{P_{\text{bottom}}} \frac{dP}{\rho} \Bigg) \Bigg/ \Bigg( \frac{G \cdot M_{\oplus}}{R_{\oplus}} \Bigg) = x \Bigg/ \Bigg( \frac{G \cdot M_{\oplus}}{R_{\oplus}} \Bigg)
        \end{equation}
    \item Fifth, we apply the Envelope-Core Radius-Relation, to calculate $R_p$:
        \begin{equation}
            R_p = R_c \cdot \ddfrac{1}{1- \Bigg( z \bigg/ \bigg( \frac{M_c}{R_c} \bigg) \Bigg)}
        \end{equation}
\end{itemize}

\begin{figure}[h!]
\centering
%\makebox[0pt]%
\hspace*{-2cm}\includegraphics[scale=0.38]{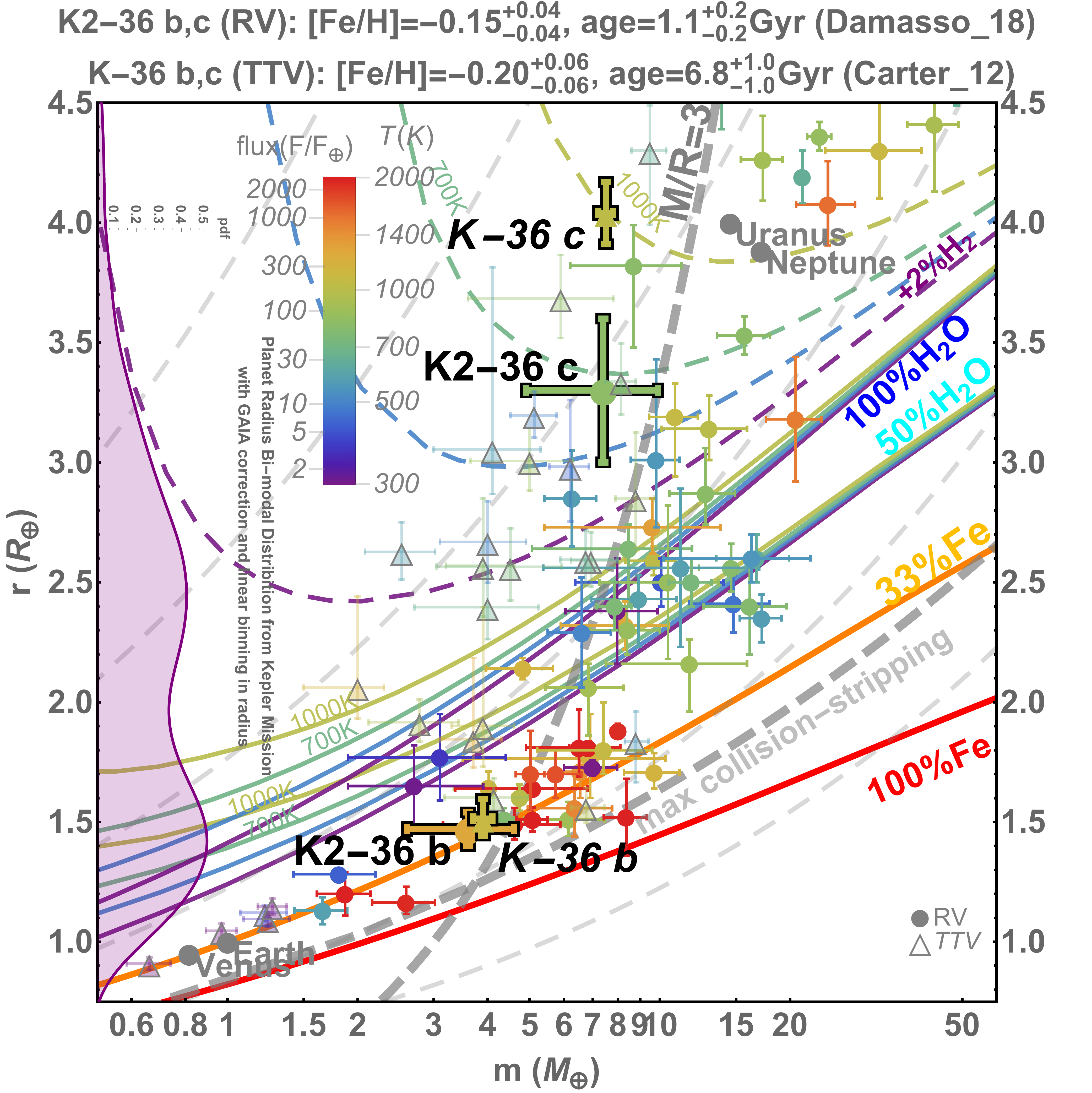}
\caption{The K-36 two-planet system and K2-36 two-planet system happen to resemble each other by mere coincidence of their designation. The two planetary systems, each has one rocky-super-Earth, and another planet with sufficient amount of volatiles. Their formation pathways may be similar~\cite{Damasso2019}~\cite{Carter:2012}. The planet occurrence rate with Gaia-correction is taken from~\citep{Berger2018Revised2}~\citep{Berger2020}. }
\label{fig:K2-36}
\end{figure}

Here we show an example of adding a 2-percent-by-mass pure-H$_2$-envelope onto a half-ice-half-Earth-like-rocky core and its effect on the mass-radius curves. We compare this set of isothermal-envelope mass-radius curves to the Kepler-36 system and K2-36 system (Fig.~\ref{fig:K2-36}).

A minimum point exists as a common feature for these mass-radius curves of fixed envelope-to-core mass ratio. It is due to the trade-off between the extensiveness of the envelope and the size of the core. Since in the mass-radius plot shown here (Fig.~\ref{fig:K2-36}), mass coordinate is in logarithmic scale, it suggests that the overall planet radius $R_p$ stays roughly constant over a significant mass-range. 

Using this minimum point, we could separate the envelope mass-radius curve into a \emph{left branch} and a \emph{right branch}. 

Within the \emph{left branch}, the planet radius $R_p$ increases with decreasing core mass $M_c$ or planet mass $M_p$. It quickly crosses over \emph{contours} of constant surface escape velocity or constant surface gravitational potential ($M_p/R_p$=const), contours of constant surface gravity ($M_p/R_p^2$=const), and contours of constant mean planet density ($M_p/R_p^3$=const), in the decreasing way. So it will quickly become unbound to the planet core and escape. 

Within the \emph{right branch}, the planet radius $R_p$ increases with increasing core mass $M_c$ or planet mass $M_p$. In the meantime, the envelope thickness ($R_p-R_c$) decreases and then tends towards constant with increasing core mass $M_c$ or planet mass $M_p$. The planet core holds the envelope tighter at higher mass. Thus, the envelope becomes more stably bound to the planet core, and less susceptible to escape. 

\subsection{Example: Separation of Variables}

Here we give an example to calculate the parameters $x$, $y$, and $z$ for a specific EOS--a Modified Ideal Gas EOS, and explain the implications of the results. 

For a generic fluid, including both gas and liquid, its pressure-EOS can often be treated with the technique of separation of variables: 

\begin{equation}\label{Eq:pressureEOS}
    P = \underbrace{P_{\text{thermal}}}_{\text{Part I}} + \underbrace{P_{\text{residual}}}_{\text{Part II}}
\end{equation}

\begin{equation}\label{Eq:pressureEOS2}
    P = \underbrace{P_{\text{thermal}}(T,\rho)}_{\text{Part I}} + \underbrace{P_{\text{residual}}(\rho)}_{\text{Part II}}
\end{equation}

\textbf{Part I} is, for an example, the Ideal Gas EOS. \textbf{Part II} results from interaction among molecules through effective pair potential. High-pressure shock wave experiments show that this effective pair potential is quite universal for fluid diatomic molecules such as N$_2$, CO, O$_2$ and H$_2$. It is independent of initial fluid densities at ambient pressure and potential parameters that scale with the critical points ($\rho_c$, $T_c$, $P_c$) of various fluids~\cite{Nellis2017}. 

Thus, it motivates us to re-express pressure $P$ in terms of molecular molar density $n$ (in mol/cc) instead of mass density $\rho$ (in g/cc), in order to make our derivations more generic. This is especially useful when treating a fluid which is a \emph{mixture} of multiple components rather than a single component. In order to convert $\rho$ into $n$ and vice versa, we need to introduce the mean molecular weight $\overbar{\mu}$ as:  

\begin{equation}\label{Eq:MeanMolecularWeight}
    n = \frac{\rho}{\overbar{\mu}}
\end{equation}

Suppose a generic fluid consists of Hydrogen Molecules H$_2$, Helium Atoms He, and other heavier species collectively called Z. Assume it is a \textbf{Linear Mixture} which obeys volume additive law, then the molar density $n$ follows the additive law as follows: 

\begin{equation}\label{Eq:VolumeAdditive}
    n_{\text{total}} = n_{\text{H$_2$}} + n_{\text{He}} + n_{\text{Z}}
\end{equation}

Eq.~\ref{Eq:VolumeAdditive} is of course violated if chemical reactions occur within or among species which result in association or dissociation (such as H$_2$ dissociates into two H-atoms) which change the total number of molecules. However, for now we assume this is not the case. This is approximately true for H$_2$ until pressures of the order of $\sim$100 GPa, for high pressure will prohibit the dissociation of H$_2$ when temperature $T$ is not too high. Furthermore, we assume each molecule behaves identically and uniformly in this mixture, and thus, their interaction can be described by a single effective pair potential. 

$\mu_{\text{H$_2$}}$=2 g/mol, and $\mu_{\text{He}}$= 4 g/mol. $\mu_{\text{Z}}$ is the mean molecular weight of the heavier species themselves, which is typically on the order of 20 or 30. 

\begin{align}
\frac{\rho_{\text{H$_2$}}}{\mu_{\text{H$_2$}}} &=n_{\text{H$_2$}} \leftarrow \text{molar density of H$_2$ (moles per unit volume)} \nonumber\\
\frac{\rho_{\text{He}}}{\mu_{\text{He}}} &=n_{\text{He}} \leftarrow \text{molar density of He (moles per unit volume)} \nonumber\\
\frac{\rho_{\text{Z}}}{\mu_{\text{Z}}} &=n_{\text{Z}} \leftarrow \text{molar density of heavier species (moles per unit volume)} \nonumber\\
\end{align}

However, in astronomy, people often like to component of a gas mixture in mass fraction instead of number fraction. The relationship can be expressed as follows: 

\begin{align}
X&\text{:(mass fraction of H$_2$)} \equiv \frac{\rho_{\text{H$_2$}}}{\rho_{\text{H$_2$}}+\rho_{\text{He}}+\rho_{\text{Z}}} = \frac{\mu_{\text{H$_2$}} \cdot n_{\text{H$_2$}}}{\mu_{\text{H$_2$}} \cdot n_{\text{H$_2$}}+\mu_{\text{He}} \cdot n_{\text{He}}+\mu_{\text{Z}} \cdot n_{\text{Z}}} \nonumber\\
Y&\text{:(mass fraction of He)} \equiv \frac{\rho_{\text{He}}}{\rho_{\text{He}}+\rho_{\text{He}}+\rho_{\text{Z}}} = \frac{\mu_{\text{H$_2$}} \cdot n_{\text{H$_2$}}}{\mu_{\text{H$_2$}} \cdot n_{\text{H$_2$}}+\mu_{\text{He}} \cdot n_{\text{He}}+\mu_{\text{Z}} \cdot n_{\text{Z}}} \nonumber\\
Z&\text{:(mass fraction of Z)} \equiv \frac{\rho_{\text{Z}}}{\rho_{\text{Z}}+\rho_{\text{He}}+\rho_{\text{Z}}} = \frac{\mu_{\text{H$_2$}} \cdot n_{\text{H$_2$}}}{\mu_{\text{H$_2$}} \cdot n_{\text{H$_2$}}+\mu_{\text{He}} \cdot n_{\text{He}}+\mu_{\text{Z}} \cdot n_{\text{Z}}} \nonumber\\
\end{align}

From Eq.~\ref{Eq:VolumeAdditive}, the relation between $\overbar{\mu}$ and $\mu_{\text{H$_2$}}$, $\mu_{\text{He}}$, $\mu_{\text{Z}}$ can be written as: 

\begin{equation}\label{Eq:VolumeAdditive2}
    \frac{1}{\overbar{\mu}} = \frac{X}{\mu_{\text{H$_2$}}} + \frac{Y}{\mu_{\text{He}}} + \frac{Z}{\mu_{\text{Z}}}
\end{equation}

Now, let's try the following generic form of pressure-EOS as motivated by real experimental data: 

\begin{equation}\label{Eq:pressureEOS2}
    P = \underbrace{R \cdot T \cdot n}_{\text{Part I}} + \underbrace{\lambda \cdot \Bigg(\frac{n}{\text{1 mol/cc}}\Bigg)^{\Gamma}}_{\text{Part II}}
\end{equation}

$\lambda$ is a constant in unit of pressure. It is about $10^3$ GPa or 1 TPa, or $10^{13}$dyn/cm$^2$ in cgs units, for H$_2$ and other molecules alike approximately. According to our demonstration earlier, we treat $\Gamma$ as constant, which is $\sim$2 for metallic fluid H$_2$,  and H$_2$-He Cosmic Mixture ($X=\frac{3}{4}$, $Y=\frac{1}{4}$), and other molecules alike approximately~\citep{Stevenson2017Ge131:PlanetaryEvolution}. 
At low number density $n$, the temperature-dependent \textbf{Part I} dominates. At high number density $n$, the temperature-independent \textbf{Part II} dominates. 
Then, the integral $\int \frac{dP}{\rho}$ for an isothermal envelope at constant temperature $T_{\text{eq}}$ and constant mean molecular weight $\overbar{\mu}$ can be performed: 

\begin{equation}
  x = \int_{\text{top}}^{\text{bottom}} \frac{dP}{\rho} = \frac{1}{\overbar{\mu}} \cdot \int_{\text{top}}^{\text{bottom}} \frac{dP}{n}
\end{equation}

Then, in cgs units, 

\begin{equation}
  dP = \underbrace{R \cdot T \cdot dn}_{\text{Part I}} + \underbrace{\lambda \cdot \Gamma \cdot n^{\Gamma-1} \cdot dn}_{\text{Part II}}
\end{equation}

Then,

\begin{equation}
  \frac{dP}{n} = \underbrace{R \cdot T \cdot \frac{dn}{n}}_{\text{Part I}} + \underbrace{\lambda \cdot \Gamma \cdot n^{\Gamma-2} \cdot dn}_{\text{Part II}}
\end{equation}

Then, 

\begin{equation}
  \int_{\text{top}}^{\text{bottom}} \frac{dP}{n} = \underbrace{R \cdot T \cdot \int_{\text{top}}^{\text{bottom}} \frac{dn}{n}}_{\text{Part I}} + \underbrace{\lambda \cdot \Gamma \cdot \int_{\text{top}}^{\text{bottom}} n^{\Gamma-2} \cdot dn}_{\text{Part II}}
\end{equation}

Then,

\begin{equation}
  \int_{\text{top}}^{\text{bottom}} \frac{dP}{n} = \underbrace{R \cdot T \cdot \ln{\Bigg( \frac{n_{\text{bottom}}}{n_{\text{top}}} \Bigg)}}_{\text{Part I}} + \underbrace{\lambda \cdot \frac{\Gamma}{\Gamma-1} \cdot \Bigg( n_{\text{bottom}}^{\Gamma-1} - n_{\text{top}}^{\Gamma-1} \Bigg)}_{\text{Part II}}
\end{equation}

Then, Parameter $x$, which is a represent of envelope thickness, is: 

\begin{equation}
  x = \underbrace{\frac{R \cdot T_{\text{eq}}}{\overbar{\mu}} \cdot \ln{\Bigg( \frac{n_{\text{bottom}}}{n_{\text{top}}} \Bigg)}}_{\text{thermal contribution to envelope thickness}} + \underbrace{\frac{\lambda}{\overbar{\mu}} \cdot \frac{\Gamma}{\Gamma-1} \cdot \Bigg( n_{\text{bottom}}^{\Gamma-1} - n_{\text{top}}^{\Gamma-1} \Bigg)}_{\text{intrinsic thickness of envelope}}
\end{equation}

Or, equivalently, since $n_{\text{bottom}} \gg n_{\text{top}}$, 

\begin{equation}
  x \approx \underbrace{\frac{R \cdot T_{\text{eq}}}{\overbar{\mu}} \cdot \ln{\Bigg( \frac{n_{\text{bottom}}}{n_{\text{top}}} \Bigg)}}_{\text{thermal contribution to envelope thickness}} + \underbrace{\frac{\lambda}{\overbar{\mu}} \cdot \frac{\Gamma}{\Gamma-1} \cdot \Bigg( n_{\text{bottom}}^{\Gamma-1} \Bigg)}_{\text{intrinsic thickness of envelope}}
\end{equation}

The interpretation of this result is that \textbf{Part I} gives rise to the \emph{thermal thickness} of an envelope, which is characterized by temperature-dependent scale-height. \textbf{Part II} gives rise to the \emph{intrinsic thickness} of an envelope, regardless of the thermal conditions and energy transport process. This result is very general. 

This Separation-of-Variable treatment allows us to understand how a fluidic envelope expands under increased stellar insolations, applicable to many short-period exoplanets, such as hot-Jupiters and hot-Saturns. 

\begin{equation}
  x \approx \underbrace{\frac{R \cdot T_{\text{eq}}}{\overbar{\mu}} \cdot \ln{\Bigg( \frac{n_{\text{bottom}}}{n_{\text{top}}} \Bigg)}}_{\text{Part I}} + \underbrace{\frac{\lambda}{\overbar{\mu}} \cdot \frac{\Gamma}{\Gamma-1} \cdot \Bigg( n_{\text{bottom}}^{\Gamma-1} \Bigg)}_{\text{Part II}}
\end{equation}

\subsubsection{Discussion}

Now, let's plug in the numbers and compare the actual results, in cgs units (erg/g): 

\begin{equation}
  x \approx \frac{1}{\overbar{\mu}} \cdot \Bigg( \underbrace{10^{11} \cdot \Bigg( \frac{T_{\text{eq}}}{10^3\text{K}} \Bigg) \cdot \ln{\Bigg( \frac{n_{\text{bottom}}}{n_{\text{top}}} \Bigg)}}_{\text{Part I}} + \underbrace{2 \cdot 10^{13} \cdot \Bigg( \frac{n_{\text{bottom}}}{\text{mol/cc}} \Bigg)}_{\text{Part II}} \Bigg)~\text{erg/g}
\end{equation}

$n_{\text{top}}$ corresponds to the number density of the top of the envelope that the transit observation probes. Let's treat it as $\sim10^{-8}$ mol/cc. The exact value shall depend on the photo-chemistry of the envelope top. This typically corresponds to pressures of millibar level (please see the detailed calculation of the slant-path optical depth which takes into account planet surface curvature in~\citep{DeWit2013} and its supplement). 

Note that \textbf{Part II} becomes comparable to \textbf{Part I} in magnitude when $n_{\text{bottom}}$ approaches $\sim10^{-2}$ mol/cc or $\sim$0.01 mol/cc. 

Recall that D. Mendeleev (1870) pointed out that it is useful to compare the molar densities among different chemical species, not at their boiling points, but at their critical points when the cohesion of the molecules transition from being close to zero to non-zero there~\citep{Karapetyants1978}. Indeed, in previous sections we see that: 

\begin{align}
n_{\text{c,H$_2$}} &\approx 0.015 \text{~mol/cc} \nonumber\\
n_{\text{c,H$_2$O}} &\approx 0.018 \text{~mol/cc} \nonumber\\
n_{\text{c,NH$_3$}} &\approx 0.013 \text{~mol/cc} \nonumber\\
n_{\text{c,N$_2$}} &\approx 0.0112 \text{~mol/cc} \nonumber\\
n_{\text{c,CH$_4$}} &\approx 0.010 \text{~mol/cc} \nonumber\\
\end{align}

Thus, near or above this \emph{critical} molar density of $\sim0.01$ mol/cc, the cohesive interactions among molecules start to take over and turn the substance into a \emph{condensed} state. They should be considered liquid or \emph{super-critical} fluid from there on, and they should not be considered \emph{gas-like} above this critical molar density. 

Using Eq.~\ref{Eq:RadiusRelationConvert}, we can obtain the dimensionless parameter $y$ from $x$: 
\begin{equation}
  y \approx \frac{1}{\overbar{\mu}} \cdot \Bigg( \underbrace{0.133 \cdot \Bigg( \frac{T_{\text{eq}}}{10^3\text{K}} \Bigg) \cdot \ln{\Bigg( \frac{n_{\text{bottom}}}{n_{\text{top}}} \Bigg)}}_{\text{Part I}} + \underbrace{16 \cdot \Bigg( \frac{n_{\text{bottom}}}{\text{mol/cc}} \Bigg)}_{\text{Part II}} \Bigg)\Bigg/ \Bigg( \frac{M_c}{R_c} \Bigg)~\text{,~$M_c$,$R_c$ in $\oplus$}
\end{equation}

and recall that: 

\begin{equation}
  R_p = R_c \cdot \ddfrac{1}{1-y}
\end{equation}

Now, let's focus on \textbf{Part I} which varies from $n_{\text{top}}\sim10^{-8}$ mol/cc to $n_{\text{bottom}}\sim10^{-2}$ mol/cc. This is approximately one-million-fold contrast in density. 

To \emph{visualize} this contrast, let's imagine a small volume box containing a certain number of molecules at the very top of the atmosphere defined by the transit observation. Now, when we move this box containing the same number of molecules down to the bottom, each side (linear dimension) of it shrinks by one-hundred fold, and this gives rise to one-million-fold increase in density. 

On the other hand, to go from the bottom of the atmosphere ($\sim0.01$ mol/cc) all the way to the center of a super-Earth exoplanet ($\sim0.1$ mol/cc) the density increase is only about one-order-of-magnitude (a factor of 10 or so). This shows how different and difficult it is to compress a condensed phase compared to a gas phase. 

Since $\ln{\Bigg(\frac{10^{-2}}{10^{-8}}\Bigg)} = \ln{(10^6)} \approx 13.8$, 

\begin{equation}
  y_{\text{I}} \approx \Bigg( \frac{2}{\overbar{\mu}} \Bigg) \cdot  \Bigg( \frac{T_{\text{eq}}}{10^3\text{K}} \Bigg)  \Bigg/ \Bigg( \frac{M_c}{R_c} \Bigg)
\end{equation}
%\lesssim
%\gtrsim

From Fig.~\ref{fig:K2-36}, one can see that most interesting super-Earth exoplanets have $M_p/R_p \gtrsim 3$. That suggests $M_c/R_c \gtrsim 3$ for most of these planets. 
Moreover, $\overbar{\mu} \gtrsim 2$ and it takes the minimum value only when it is a pure H$_2$ atmosphere. 
Moreover, most of these interesting super-Earth exoplanets are $\lesssim1000$ K equilibrium temperature. 
Therefore,

\begin{equation}
  y_{\text{I}} \lesssim 0.3 \cdot  \Bigg( \frac{T_{\text{eq}}}{10^3\text{K}} \Bigg)
\end{equation}

From this analysis, we show that though gas has low density, it is also very compressible under a gravitational potential gradient. 
It does not take a long distance for a gas-like species to build up its pressure and molar density until reaching its critical molar density, where the molecular cohesion takes over and turns it into a condensed phase. 

As a result of that, fundamentally speaking, a gaseous atmosphere can only account for a surface-layer of a planetary body, a mere skin-effect. This statement even applies to a gas giant like Jupiter or Saturn, because most of their interior is considered high-density \emph{super-critical} molecular fluid and then metallic fluid rather than a gaseous state, in terms of condensed matter science. 

The fractional boost gained from such a gaseous atmosphere is typically on the order of 10\% to 30\% (10-30 percent) of the planet core radius ($R_c$) if the atmosphere composition is dominated by H$_2$-He, and is typically on the order of 1\% to 3\% (1-3 percent) of the planet core radius ($R_c$) if the atmosphere composition is dominated by Oxygen-Nitrogen-Carbon-species since their molecular weight is about one order of magnitude higher. 

Then in Fig.~\ref{fig:K2-36}, the group of super-Earth exoplanets with $M_p/R_p \gtrsim 3$ and with lower-than-rocky density raise an interesting question. They may be explained by: 

\begin{itemize}
    \item \emph{intrinsic} lower-than-rocky density in the fluid-layer of these planets, or, 
    \item \emph{intrinsic} heat source or heat reservoir that maintains a high-entropy state in the interior of these planets. 
\end{itemize}

Last but not least, due to proximity to host star, the \emph{stability} of a such gaseous atmosphere on a short-period exoplanet needs to be examined carefully. Very plausibly, if such a gaseous atmosphere exists, then it is experiencing \emph{escape} and must be continuously replenished by a reservoir such as the fluid-layer underneath. 

\clearpage
%%%%%%%%%%%%%%%%%%%%%%%%%%%%%%%
\section{Conclusion}
I always thought something was fundamentally uniform with the universe. \citep{Vinet:1987}

\section{Acknowledgement}
The author Li Zeng would like to thank Dr. William J. Nellis for his help on the science of high-pressure shock wave experiments. 
The author Li Zeng would like to thank Dr. Dave Latham for his continuous pursuit of exoplanets through the \emph{TESS} mission. The author Li Zeng would like to thank Dr. Ronald Redmer and his research group, Dr. Stephane Mazevet, Dr. Amit Levi, and Dr. Morris Podolak for discussion on Hydrogen EOS and H$_2$O EOS. %The author Li Zeng would like to thank Dr. Juan Perez-Mercader for discussion on ...

\bibliographystyle{abbrv}
%\bibliographystyle{apj}
%\bibliography{mendeley_v6}
%\bibliography{references}

\begin{thebibliography}{10}


\bibitem{EncyclopediaBritannica_bandtheory}
{Band theory | physics | Britannica}.

\bibitem{EncyclopediaBritannica_crystalbonding}
{Chemical bonding (chemistry) - Videos and Images | Britannica}.

\bibitem{EncyclopediaBritannica_electricalconductor}
{Electrical conductor | physics | Britannica}.

\bibitem{EncyclopediaBritannica_electricity}
{Electricity - Conductors, insulators, and semiconductors | Britannica}.

\bibitem{EngineeringtoolboxAmmonia}
{Engineeringtoolbox: Ammonia - Properties at Gas-Liquid Equilibrium
  Conditions}.

\bibitem{EngineeringtoolboxHydrogen}
{Engineeringtoolbox: Hydrogen - Density and Specific Weight}.

\bibitem{EngineeringtoolboxMethane}
{Engineeringtoolbox: Methane - Density and Specific Weight}.

\bibitem{EngineeringtoolboxNitrogen}
{Engineeringtoolbox: Nitrogen - Density and Specific Weight}.

\bibitem{NISTWebBookAmmonia}
{NIST webbook: Ammonia}.

\bibitem{NISTWebBookHydrogen}
{NIST webbook: Hydrogen}.

\bibitem{NISTWebBookMethane}
{NIST webbook: Methane}.

\bibitem{NISTWebBookNitrogen}
{NIST webbook: Nitrogen}.

\bibitem{IAPWS:2016}
The international association for the properties of water and steam revised
  release on the iapws formulationd 1995 for the thermodynamic properties of
  ordinary water substance for general and scientific use.
\newblock Technical report, The International Association for the Properties of
  Water and Steam, 2016.

\bibitem{Ancilotto1997}
F.~Ancilotto, G.~L. Chiarotti, S.~Scandolo, and E.~Tosatti.
\newblock {Dissociation of methane into hydrocarbons at extreme (planetary)
  pressure and temperature}.
\newblock {\em Science}, 275(5304):1288--1290, feb 1997.

\bibitem{Ashcroft1976_2}
N.~W. Ashcroft and N.~D. Mermin.
\newblock {Table for Fermi Energies, Fermi Temperatures, and Fermi Velocities}.

\bibitem{Ashcroft1976}
N.~W. Ashcroft and N.~D. Mermin.
\newblock {\em {Solid State Physics}}.
\newblock 1976.

\bibitem{Becker2014AbDwarfs}
Andreas Becker, Winfried Lorenzen, Jonathan~J. Fortney, Nadine Nettelmann,
  Manuel Sch{\"{o}}ttler, and Ronald Redmer.
\newblock {ab initio equations of state for hydrogen (H-REOS.3) and helium
  (He-REOS.3) and their implications for the interior of brown dwarfs}.
\newblock {\em The Astrophysical Journal Supplement Series}, 215(2):21, 12
  2014.

\bibitem{Becker2013IsentropicInteriors}
Andreas Becker, Nadine Nettelmann, Bastian Holst, and Ronald Redmer.
\newblock {Isentropic compression of hydrogen: Probing conditions deep in
  planetary interiors}.
\newblock {\em Physical Review B}, 88(4):045122, 7 2013.

\bibitem{Benedetti1999}
Laura~Robin Benedetti, Jeffrey~H. Nguyen, Wendell~A. Caldwell, Hongjian Liu,
  Michael Kruger, and Raymond Jeanloz.
\newblock {Dissociation of CH4 at high pressures and temperatures: Diamond
  formation in giant planet interiors?}
\newblock {\em Science}, 286(5437):100--102, oct 1999.

\bibitem{Berger2018Revised2}
Travis~A. Berger, Daniel Huber, Eric Gaidos, and Jennifer~L. van Saders.
\newblock {Revised Radii of Kepler Stars and Planets using Gaia Data Release
  2}.
\newblock {\em The Astrophysical Journal}, 5 2018.

\bibitem{Berger2020}
Travis~A. Berger, Daniel Huber, Eric Gaidos, Jennifer~L. van Saders, and
  Lauren~M. Weiss.
\newblock {The Gaia-Kepler Stellar Properties Catalog. II. Planet Radius
  Demographics as a Function of Stellar Mass and Age}.
\newblock {\em The Astronomical Journal}, may 2020.

\bibitem{Bethkenhagen2017}
M.~Bethkenhagen, E.~R. Meyer, S.~Hamel, N.~Nettelmann, M.~French, L.~Scheibe,
  C.~Ticknor, L.~A. Collins, J.~D. Kress, J.~J. Fortney, and R.~Redmer.
\newblock {Planetary Ices and the Linear Mixing Approximation}.
\newblock {\em The Astrophysical Journal}, 848(1):67, oct 2017.

\bibitem{Bethkenhagen2015}
Mandy Bethkenhagen, Daniel Cebulla, Ronald Redmer, and Sebastien Hamel.
\newblock Superionic phases of the 1:1 water-ammonia mixture.
\newblock {\em The Journal of Physical Chemistry A}, 119(42):10582--10588,
  2015.
\newblock PMID: 26390374.

\bibitem{Bethkenhagen2013}
Mandy Bethkenhagen, Martin French, and Ronald Redmer.
\newblock Equation of state and phase diagram of ammonia at high pressures from
  ab initio simulations.
\newblock {\em The Journal of Chemical Physics}, 138(23):234504, 2013.

\bibitem{Bezacier2014}
Lucile Bezacier, Baptiste Journaux, Jean~Philippe Perrillat, Herv{\'{e}}
  Cardon, Michael Hanfland, and Isabelle Daniel.
\newblock {Equations of state of ice VI and ice VII at high pressure and high
  temperature}.
\newblock {\em Journal of Chemical Physics}, 141(10):104505, sep 2014.

\bibitem{Bolmatov_FrenkelFrequency}
D.~Bolmatov, V.~V. Brazhkin, and K.~Trachenko.
\newblock The phonon theory of liquid thermodynamics.
\newblock {\em Scientific Reports}, 2(1):421, 2012.

\bibitem{Carter:2012}
Joshua~A Carter, Eric Agol, William~J Chaplin, Sarbani Basu, Timothy~R Bedding,
  Lars~A Buchhave, JÃžrgen Christensen-Dalsgaard, Katherine~M Deck, Yvonne
  Elsworth, Daniel~C Fabrycky, Eric~B Ford, Jonathan~J Fortney, Steven~J Hale,
  Rasmus Handberg, Saskia Hekker, Matthew~J Holman, Daniel Huber, Christopher
  Karoff, Steven~D Kawaler, Hans Kjeldsen, Jack~J Lissauer, Eric~D Lopez,
  Mikkel~N Lund, Mia Lundkvist, Travis~S Metcalfe, Andrea Miglio, Leslie~A
  Rogers, Dennis Stello, William~J Borucki, Steve Bryson, Jessie~L
  Christiansen, William~D Cochran, John~C Geary, Ronald~L Gilliland, Michael~R
  Haas, Jennifer Hall, Andrew~W Howard, Jon~M Jenkins, Todd Klaus, David~G
  Koch, David~W Latham, Phillip~J MacQueen, Dimitar Sasselov, Jason~H Steffen,
  Joseph~D Twicken, and Joshua~N Winn.
\newblock {Kepler-36: A Pair of Planets with Neighboring Orbits and Dissimilar
  Densities}.
\newblock {\em Science}, 337(6094):556--559, 2012.

\bibitem{Cavazzoni1999SuperionicConditions}
C.~Cavazzoni, G.~L. Chiarotti, S.~Scandolo, E.~Tosatti, M.~Bernasconi, and
  M.~Parrinello.
\newblock {Superionic and Metallic States of Water and Ammonia at Giant Planet
  Conditions}.
\newblock {\em Science}, 283(5398):44, 1999.

\bibitem{Chabrier_2019}
G.~Chabrier, S.~Mazevet, and F.~Soubiran.
\newblock A new equation of state for dense hydrogen{\textendash}helium
  mixtures.
\newblock {\em The Astrophysical Journal}, 872(1):51, feb 2019.

\bibitem{ChaplinSuperIonic}
Martin Chaplin.
\newblock {Ice-eighteen: Superionic Ice}.

\bibitem{ChaplinSuperCritical}
Martin Chaplin.
\newblock {Superionic Water}.

\bibitem{Collins1978}
George~W. Collins.
\newblock {\em {The Virial Theorem in Stellar Astrophysics}}.
\newblock 1978.

\bibitem{Damasso2019}
{Damasso, M.}, {Zeng, L.}, {Malavolta, L.}, {Mayo, A.}, {Sozzetti, A.},
  {Mortier, A.}, {Buchhave, L. A.}, {Vanderburg, A.}, {Lopez-Morales, M.},
  {Bonomo, A. S.}, {Cameron, A. C.}, {Coffinet, A.}, {Figueira, P.}, {Latham,
  D. W.}, {Mayor, M.}, {Molinari, E.}, {Pepe, F.}, {Phillips, D. F.}, {Poretti,
  E.}, {Rice, K.}, {Udry, S.}, and {Watson, C. A.}
\newblock So close, so different: characterization of the k2-36 planetary
  system with harps-n.
\newblock {\em A\&A}, 624:A38, 2019.

\bibitem{DeWit2013}
Julien {De Wit} and Sara Seager.
\newblock {Constraining exoplanet mass from transmission spectroscopy}.
\newblock {\em Science}, 342(6165):1473--1477, dec 2013.

\bibitem{Dewar1899}
James Dewar.
\newblock {"Sur la solidification de l'hydrog{\`{e}}ne" (English: On the
  solidification of hydrogen)}.
\newblock {\em Annales de chimie et de physique}, 1899.

\bibitem{Douce2011}
Alberto~Pati{\~{n}}o Douce.
\newblock {\em {Thermodynamics of the earth and planets}}, volume
  9780521896214.
\newblock Cambridge University Press, jan 2011.

\bibitem{Edwards2010Metallization}
P~P Edwards, M~T~J Lodge, F~Hensel, and R~Redmer.
\newblock '... a metal conducts and a non-metal doesn't'.
\newblock {\em Philosophical transactions. Series A, Mathematical, physical,
  and engineering sciences}, 368(1914):941--965, 03 2010.

\bibitem{Hensel_Metallic_Oxygen}
Peter~P. Edwards and Friedrich Hensel.
\newblock Metallic oxygen.
\newblock {\em ChemPhysChem}, 3(1):53--56, 2002.

\bibitem{Edwards1978}
Peter~Phillip Edwards and Michell~J. Sienko.
\newblock {Universality aspects of the metal-nonmetal transition in condensed
  media}.
\newblock {\em Physical Review B}, 17(6):2575--2581, mar 1978.

\bibitem{Feynman1949}
R.~P. Feynman, N.~Metropolis, and E.~Teller.
\newblock {Equations of state of elements based on the generalized fermi-thomas
  theory}.
\newblock {\em Physical Review}, 75(10):1561--1573, may 1949.

\bibitem{Frank:2004}
Mark~R Frank, Yingwei Fei, and Jingzhu Hu.
\newblock {Constraining the equation of state of fluid H2O to 80 GPa using the
  melting curve, bulk modulus, and thermal expansivity of Ice VII}.
\newblock {\em Geochimica et Cosmochimica Acta}, 68(13):2781--2790, 2004.

\bibitem{Gao2010}
Guoying Gao, Artem~R. Oganov, Yanming Ma, Hui Wang, Peifang Li, Yinwei Li,
  Toshiaki Iitaka, and Guangtian Zou.
\newblock {Dissociation of methane under high pressure}.
\newblock {\em Journal of Chemical Physics}, 133(14):144508, oct 2010.

\bibitem{Goncharov:2005}
Alexander~F Goncharov, Nir Goldman, Laurence~E Fried, Jonathan~C Crowhurst,
  I-Feng~W Kuo, Christopher~J Mundy, and Joseph~M Zaug.
\newblock {Dynamic Ionization of Water under Extreme Conditions}.
\newblock {\em Phys. Rev. Lett.}, 94(12):125508, 4 2005.

\bibitem{Goncharov:2009}
Alexander~F. Goncharov, Chrystele Sanloup, Nir Goldman, Jonathan~C. Crowhurst,
  Sorin Bastea, W.~M. Howard, Laurence~E. Fried, Nicolas Guignot, Mohamed
  Mezouar, and Yue Meng.
\newblock {Dissociative melting of ice VII at high pressure}.
\newblock {\em The Journal of Chemical Physics}, 130(12):124514, mar 2009.

\bibitem{Grimvall1999}
G.~Grimvall.
\newblock {\em {Thermophysical Properties of Materials}}.
\newblock Elsevier, 1999.

\bibitem{Hansen2004}
Carl~J. Hansen, Steven~D. Kawaler, and Virginia. Trimble.
\newblock {\em {Stellar Interiors : Physical Principles, Structure, and
  Evolution}}.
\newblock Springer New York, 2004.

\bibitem{Hensel2015Metallization}
Friedrich Hensel, Daniel~R. Slocombe, and Peter~P. Edwards.
\newblock On the occurrence of metallic character in the periodic table of the
  chemical elements.
\newblock {\em Philosophical Transactions: Mathematical, Physical and
  Engineering Sciences}, 373(2037):1--12, 2015.

\bibitem{Holzapfel2018}
Wilfried~B. Holzapfel.
\newblock {Coherent thermodynamic model for solid, liquid and gas phases of
  elements and simple compounds in wide ranges of pressure and temperature}.
\newblock {\em Solid State Sciences}, 80:31--34, jun 2018.

\bibitem{IAPWS:2011}
IAPWS.
\newblock {IAPWS R14-08(2011), Revised Release on the Pressure along the
  Melting and Sublimation Curves of Ordinary Water Substance}.
\newblock Technical report, The International Association for the Properties of
  Water and Steam, 2011.

\bibitem{Jahangiri2016}
Soran Jahangiri and Hassan Behnejad.
\newblock {An analytical equation of state for ammonia at high temperatures and
  high pressures}.
\newblock {\em Journal of Molecular Liquids}, 222:733--738, oct 2016.

\bibitem{BNLCryogenicDataNotebook}
J.E. Jensen, W.A. Tuttle, R.B. Stewart, H.~Brechna, and A.G. Prodell.
\newblock {Cryogenic Data Notebook from Brookhaven National Laboratory (BNL
  10200-R, Revised August 1980)}.

\bibitem{Jiang2017}
Xue Jiang, Xue Wu, Zhaoyang Zheng, Yingying Huang, and Jijun Zhao.
\newblock {Ionic and superionic phases in ammonia dihydrate N
  H3{\textperiodcentered}2 H2 O under high pressure IONIC and SUPERIONIC PHASES
  in AMMONIA ... JIANG, WU, ZHENG, HUANG, and ZHAO}.
\newblock {\em Physical Review B}, 95(14):144104, apr 2017.

\bibitem{Karapetyants1978}
M.~Kh. {Karapetyants}.
\newblock {\em {Chemical Thermodynamics}}.
\newblock Mir Publishers Moscow, 1978.

\bibitem{Kechin1995}
V~V Kechin.
\newblock {Thermodynamically based melting-curve equation}.
\newblock {\em Journal of Physics: Condensed Matter}, 7(3):531--535, jan 1995.

\bibitem{Kerley1980}
G.~I. Kerley.
\newblock {A theoretical equation of state for methane}.
\newblock {\em Journal of Applied Physics}, 51(10):5368--5374, 1980.

\bibitem{Kireev1978}
P.S. {Kireev}.
\newblock {\em {Semiconductor Physics}}.
\newblock Mir Publishers Moscow, 1978.

\bibitem{Kittel2004}
Charles Kittel.
\newblock {\em {Introduction to Solid State Physics}}.
\newblock 2004.

\bibitem{Ledoux1958}
P.~Ledoux.
\newblock {Stellar Stability}.
\newblock In {\em Encyclopedia of Physics}, pages 605--688. Springer, Berlin,
  Heidelberg, 1958.

\bibitem{Li2017}
Dafang Li, Cong Wang, Jun Yan, Zhen~Guo Fu, and Ping Zhang.
\newblock {Structural and transport properties of ammonia along the principal
  Hugoniot}.
\newblock {\em Scientific Reports}, 7(1):1--12, dec 2017.

\bibitem{NISTWebBookWater}
P.J. Linstrom and W.G. Mallard, editors.
\newblock {\em NIST Chemistry WebBook, NIST Standard Reference Database Number
  69: Water}, page~1.
\newblock National Institute of Standards and Technology, 2020.

\bibitem{Lorenzen2009}
Winfried Lorenzen, Bastian Holst, and Ronald Redmer.
\newblock {Demixing of hydrogen and helium at megabar pressures}.
\newblock {\em Physical Review Letters}, 102(11):115701, mar 2009.

\bibitem{Meyer2015}
Edmund~R. Meyer, Christopher Ticknor, Mandy Bethkenhagen, Sebastien Hamel,
  Ronald Redmer, Joel~D. Kress, and Lee~A. Collins.
\newblock {Bonding and structure in dense multi-component molecular mixtures}.
\newblock {\em Journal of Chemical Physics}, 143(16):164513, oct 2015.

\bibitem{Millot2019}
Marius Millot, Federica Coppari, J.~Ryan Rygg, Antonio Correa~Barrios,
  Sebastien Hamel, Damian~C. Swift, and Jon~H. Eggert.
\newblock Nanosecond x-ray diffraction of shock-compressed superionic water
  ice.
\newblock {\em Nature}, 569(7755):251--255, 2019.

\bibitem{Millot2018ExperimentalCompression}
Marius Millot, Sebastien Hamel, J.~Ryan Rygg, Peter~M. Celliers, Gilbert~W.
  Collins, Federica Coppari, Dayne~E. Fratanduono, Raymond Jeanloz, Damian~C.
  Swift, and Jon~H. Eggert.
\newblock {Experimental evidence for superionic water ice using shock
  compression}.
\newblock {\em Nature Physics}, page~1, 2 2018.

\bibitem{Mott1982}
N~F Mott.
\newblock {Review lecture: Metal-insulator transitions}.
\newblock {\em Proceedings of the Royal Society of London. A. Mathematical and
  Physical Sciences}, 382(1782):1--24, jul 1982.

\bibitem{Needham1997}
Tristan. Needham.
\newblock {\em {Visual Complex Analysis}}.
\newblock Clarendon Press, 1997.

\bibitem{Nellis2006}
W~J Nellis.
\newblock Dynamic compression of materials: metallization of fluid hydrogen at
  high pressures.
\newblock {\em Reports on Progress in Physics}, 69(5):1479--1580, apr 2006.

\bibitem{Nellis2017}
William {Nellis}.
\newblock {\em {Ultracondensed Matter by Dynamic Compression}}.
\newblock Cambridge University Press, 2017.

\bibitem{Nettelmann2012JUPITERH-REOS.2}
N.~Nettelmann, A.~Becker, B.~Holst, and R.~Redmer.
\newblock {JUPITER MODELS WITH IMPROVED AB INITIO HYDROGEN EQUATION OF STATE
  (H-REOS.2)}.
\newblock {\em The Astrophysical Journal}, 750(1):52, 5 2012.

\bibitem{Queyroux2019}
Jean~Antoine Queyroux, Sandra Ninet, Gunnar Weck, Gaston Garbarino, Thomas
  Plisson, Mohamed Mezouar, and Fr{\'{e}}d{\'{e}}ric Datchi.
\newblock {Melting curve and chemical stability of ammonia at high pressure:
  Combined x-ray diffraction and Raman study}.
\newblock {\em Physical Review B}, 99(13):134107, apr 2019.

\bibitem{Robinson2017}
Victor~Naden Robinson, Yanchao Wang, Yanming Ma, and Andreas Hermann.
\newblock {Stabilization of ammonia-rich hydrate inside icy planets}.
\newblock {\em Proceedings of the National Academy of Sciences},
  114(34):9003--9008, aug 2017.

\bibitem{Setzmann1991}
U.~Setzmann and W.~Wagner.
\newblock {A New Equation of State and Tables of Thermodynamic Properties for
  Methane Covering the Range from the Melting Line to 625 K at Pressures up to
  1000 MPa}.
\newblock {\em Journal of Physical and Chemical Reference Data},
  20(6):1061--1155, nov 1991.

\bibitem{Sherman2012}
Benjamin~L. Sherman, Hugh~F. Wilson, Dayanthie Weeraratne, and Burkhard
  Militzer.
\newblock {Ab initio simulations of hot dense methane during shock
  experiments}.
\newblock {\em Physical Review B - Condensed Matter and Materials Physics},
  86(22):224113, dec 2012.

\bibitem{SilveraDias2018MetallicHydrogen}
Isaac~F Silvera and Ranga Dias.
\newblock Metallic hydrogen.
\newblock {\em Journal of Physics: Condensed Matter}, 30(25):254003, may 2018.

\bibitem{Simon1929}
Franz Simon and Gunther Glatzel.
\newblock {Bemerkungen zur Schmelzdruckkurve}.
\newblock {\em Zeitschrift f{\"{u}}r anorganische und allgemeine Chemie},
  178(1):309--316, jan 1929.

\bibitem{Sommerfeld1933}
A.~Sommerfeld, H.~Bethe, and A.~Smekal.
\newblock {Elektronentheorie der Metalle}.
\newblock In {\em Aufbau Der Zusammenh{\"{a}}ngenden Materie}, pages 333--622.
  Springer Berlin Heidelberg, 1933.

\bibitem{Stacey:1977}
F.~D. {Stacey} and R.~D. {Irvine}.
\newblock {A simple dislocation theory of melting}.
\newblock {\em Australian Journal of Physics}, 30:641, November 1977.

\bibitem{Stacey:2019}
Frank~D. Stacey and Jane~H. Hodgkinson.
\newblock Thermodynamics with the gruneisen parameter: Fundamentals and
  applications to high pressure physics and geophysics.
\newblock {\em Physics of the Earth and Planetary Interiors}, 286:42 -- 68,
  2019.

\bibitem{Stevenson2017Ge131:PlanetaryEvolution}
D.~J. Stevenson.
\newblock {Ge131:Planetary Structure and Evolution}, 2017.

\bibitem{Stishov1975}
S.~M. Stishov.
\newblock {The Thermodynamics of Melting of Simple Substances}.
\newblock {\em Soviet Physics - Uspekhi}, 17(5):625--643, may 1975.

\bibitem{Tassoul1968}
Monique Tassoul and Jean~Louis Tassoul.
\newblock {Asymptotic Approximations for Stellar Pulsations}.
\newblock {\em The Astrophysical Journal}, 153:127, jul 1968.

\bibitem{Vinet:1987}
P~Vinet, J~Ferrante, J~H Rose, and J~R Smith.
\newblock {Compressibility of Solids}.
\newblock {\em Journal of Geophysical Research}, 92(B9):9319--9325, 1987.

\bibitem{Wahl2017JunoCoreFuzzy}
S.~M. Wahl, W.~B. Hubbard, B.~Militzer, T.~Guillot, Y.~Miguel, N.~Movshovitz,
  Y.~Kaspi, R.~Helled, D.~Reese, E.~Galanti, S.~Levin, J.~E. Connerney, and
  S.~J. Bolton.
\newblock Comparing jupiter interior structure models to juno gravity
  measurements and the role of a dilute core.
\newblock {\em Geophysical Research Letters}, 44(10):4649--4659, 2017.

\bibitem{Yoo2020}
Choong~Shik Yoo.
\newblock {Chemistry under extreme conditions: Pressure evolution of chemical
  bonding and structure in dense solids}, jan 2020.

\bibitem{Zeng2016VARIATIONALINTERIORS}
L.~Zeng and S.B. Jacobsen.
\newblock {VARIATIONAL PRINCIPLE for PLANETARY INTERIORS}.
\newblock {\em Astrophysical Journal}, 829(1), 2016.

\bibitem{Zeng:2016c}
Li~Zeng and Stein~B Jacobsen.
\newblock {A Simple Analytical Model for Rocky Planet Interior}.
\newblock {\em ApJ}.

\bibitem{Zeng2021}
Li~Zeng, Stein~B. Jacobsen, Eugenia Hyung, Amit Levi, Chantanelle Nava, James
  Kirk, Caroline Piaulet, Gaia Lacedelli, Dimitar~D. Sasselov, Michail~I.
  Petaev, Sarah~T. Stewart, Munazza~K. Alam, Mercedes L{\'{o}}pez-Morales,
  Mario Damasso, and David~W. Latham.
\newblock {New Perspectives on the Exoplanet Radius Gap from a Mathematica Tool
  and Visualized Water Equation of State}.
\newblock {\em The Astrophysical Journal}, 923(2):247, dec 2021.

\bibitem{PNAS:Zeng2019}
Li~Zeng, Stein~B. Jacobsen, Dimitar~D. Sasselov, Michail~I. Petaev, Andrew
  Vanderburg, Mercedes Lopez-Morales, Juan Perez-Mercader, Thomas~R. Mattsson,
  Gongjie Li, Matthew~Z. Heising, Aldo~S. Bonomo, Mario Damasso, Travis~A.
  Berger, Hao Cao, Amit Levi, and Robin~D. Wordsworth.
\newblock Growth model interpretation of planet size distribution.
\newblock {\em Proceedings of the National Academy of Sciences},
  116(20):9723--9728, 2019.

\end{thebibliography}

\end{document}